\documentclass[12pt]{iopart}
\usepackage[margin=3cm]{geometry}

\newcommand{\volume}{{\ooalign{\hfil$V$\hfil\cr\kern0.08em--\hfil\cr}}}
\usepackage{threeparttable}    
\usepackage{dcolumn}           
\newcolumntype{d}{D{.}{.}{-1}}
\usepackage{nomencl}           

\makenomenclature
\usepackage{graphicx}          
\usepackage{caption}          
\usepackage{fancyvrb}          
\fvset{fontsize=\footnotesize,xleftmargin=2em}  
\usepackage{lettrine}          
\expandafter\let\csname equation*\endcsname\relax
\expandafter\let\csname endequation*\endcsname\relax
\usepackage{amsmath}           
\usepackage{hyperref}          
\graphicspath{{figs/}}         
\usepackage{float}             
\usepackage{upquote}
\usepackage{placeins}

\usepackage[retainorgcmds]{IEEEtrantools}
\usepackage{subfig}
\usepackage{siunitx}
\usepackage[export]{adjustbox}
\usepackage[title]{appendix}
\newcommand{\be}{\begin{equation}}
\newcommand{\ee}{\end{equation}} 

\usepackage{cite}

\usepackage[nameinlink,capitalise]{cleveref}
\usepackage{tikz}
\usetikzlibrary{positioning,shadows,arrows,shapes.multipart,decorations.pathmorphing}

\newcommand{\colormarkerempty}[1]{\raisebox{0.5pt}{\protect\tikz{\protect\node[draw,scale=0.3,circle,fill=white](){};}}}

\definecolor{greenF}{rgb}{0.55, 0.71, 0.0}
\definecolor{greenD}{rgb}{0.0, 0.5, 0.0}
\definecolor{capri}{rgb}{0.0, 0.75, 1.0}
\definecolor{violetM}{rgb}{0.73, 0.2, 0.52}
\definecolor{orange}{rgb}{1, 0.375, 0}

\begin{document}

\title[]{From Coils to Surface Recession: Multiphysics Simulation of Ablation in ICP Wind Tunnels}

\author{Sanjeev Kumar$^{1}$, Alessandro Munaf\`{o}$^{1,2}$, Blaine Vollmer$^{1}$, Massimo Franco$^{1}$, Kelly A. Stephani$^{1,2}$, Daniel J. Bodony$^{1}$, and Marco Panesi$^{1,2}$\footnote{\label{footnote_2}Corresponding author (mpanesi@uci.edu).}}

\address{$^1$Center for Hypersonics and Entry Systems Studies (CHESS), \\ University of Illinois Urbana-Champaign, Urbana, IL 61801, USA}

\address{$^2$Department of Mechanical and Aerospace Engineering, \\ University of California Irvine, Irvine, CA 92697, USA}

\ead{mpanesi@uci.edu}
\vspace{10pt}


\begin{abstract}
This work presents a multi-solver, coupled computational framework for predicting the thermo-chemical material response of thermal protection systems in inductively coupled plasma (ICP) wind tunnels. The framework integrates a high-fidelity Navier-Stokes plasma solver, an electromagnetic field solver, and a discontinuous-Galerkin material response solver using a partitioned coupling strategy. This enables an \textit{ab initio}, end-to-end simulation of the \SI{350}{kW} Plasmatron X facility at the University of Illinois Urbana-Champaign (UIUC), including plasma generation, electromagnetic heating, near-wall thermochemistry, and time-accurate material ablation. The model captures key ICP physics such as vortex-mode recirculation, Joule-heating-driven plasma formation, and Lorentz-force-induced flow confinement, and accurately predicts the transition from subsonic to supersonic jet behavior at low pressures. Validation against cold-wall calorimetry shows that predicted stagnation-point cold-wall heat fluxes fall well within experimental uncertainty, while coupled ablation simulations accurately reproduce measured stagnation temperature histories and recession rates with errors below 12\% and 10\%, respectively. Remaining discrepancies are attributed to uncertainties in power-coupling efficiency, equilibrium ablation modeling, and material property datasets. Sensitivity analyses reveal that a $\pm10\%$ variation in system efficiency can induce changes of up to 11\% in steady-state surface temperature and 17\% in recession rate, whereas an equivalent variation in material thermal conductivity results in comparatively minor deviations of about 1.5\% and 0.5\%, respectively. Overall, the framework demonstrates strong predictive capability for ICP wind tunnel environments and provides a foundation for improved design, interpretation, and planning of hypersonic material testing campaigns.

\end{abstract}

\section{Introduction}\label{sec:intro}
In the hypersonic regime, a vehicle traveling at extreme velocities generates a strong bow shock that severely compresses the surrounding gas, resulting in very high post-shock temperatures and imposing stringent thermal loads on the thermal protection system (TPS) \cite{duffa2013ablative}. The maintenance of the vehicle’s safety and the achievement of mission success depend heavily on the ability of the TPS to withstand intense aerodynamic heating, chemical reactions, and mechanical loads encountered during flight. Given the high cost of testing new thermal protection materials during flight, the aerospace community is heavily relying on specialized ground-based facilities. High-temperature wind tunnels have been developed specifically for this purpose, enabling researchers to expose material samples to controlled, flight-like aerothermal environments and evaluate their performance well before they are ever integrated into a real mission \cite{calomino2010evaluation,loehle2022assessment}. Within the spectrum of ground-based testing facilities, inductively coupled plasma (ICP) wind tunnels occupy a unique role. These systems generate high-enthalpy plasma flows through electromagnetic induction, allowing researchers to recreate key aspects of hypersonic flight within a controlled laboratory setting. Because the plasma is produced without electrodes, the test environment remains exceptionally clean and chemically well-defined, avoiding contamination that could alter material response. ICP facilities can operate in a near-continuous mode while precisely tuning flow properties to match specific mission requirements, including targeted stagnation-point cold-wall heat fluxes and enthalpy levels. This capability makes them valuable for the assessment of thermal protection materials under realistic and high-temperature conditions \cite{boulos2023handbook,capponi2023aerothermal}.

When exposed to extreme temperatures, the surrounding gas strongly interacts with the vehicle’s TPS, leading to substantial surface heating and material loss. This material removal process, known as ablation, can arise from chemical reactions as well as mechanical processes such as erosion or spallation, and may eventually alter the external shape of the TPS. In addition, the gas particles striking the surface undergo catalytic reactions that may play a major role in the overall heat load \cite{gnoffo1999planetary}. Because these coupled chemical and thermal processes strongly influence the performance of a TPS, an accurate modeling of gas–surface chemistry is paramount for the reliable design of hypersonic vehicles. However, simulating the response of a TPS sample to a reacting flow is inherently challenging. The difficulty stems not only from the intricate physics and chemistry governing both the flow and the material surface, but also from the significant uncertainties associated with the predicted flow conditions and the models used to represent these processes. A foundational and still widely adopted approach for modeling thermal response and ablation of a material is the film coefficient method \cite{anderson1968further,moyer1968analysis,bartlett1968analysis,cooper2023numerical,de2011stagnation,padovan2024extended}. In this framework, surface heating is represented by the enthalpy difference between the boundary layer edge and the wall, scaled by effective heat and mass transfer coefficients that capture conductive and diffusive fluxes. Ablation rates are calculated assuming surface chemical equilibrium and a specified mass transfer coefficient. The method offers reliable accuracy near stagnation regions and benefits from a decoupled formulation, enabling rapid simulations for preliminary heatshield design. Additionally, it exhibits low sensitivity to wall temperature assumptions, with cold walls and radiative equilibrium conditions yielding comparable heat transfer coefficients. However, the accuracy of the film coefficient approach diminishes away from the stagnation region, primarily due to the emergence of non-equilibrium effects and its limited capability to capture variations in heat and mass transfer coefficients induced by ablation gas blowing \cite{milos2012nonequilibrium,zibitsker2023deviation}. Geometric shape changes further complicate the method, as boundary condition profiles become increasingly dependent on the evolution of surface topology \cite{driver2010understanding,zibitsker2022study}. Under such conditions, a coupled flow-material solver offers a significant advantage by enabling accurate resolution of time-varying boundary conditions and incorporating non-equilibrium chemistry effects within the boundary layer. As a result, developing robust and efficient flow–material coupling strategies has become an important and active research topic in the hypersonics community \cite{martin2015strongly,chen2010loosely,chen2003graphite,bianchi2010navier,bersbach2022ablative,zibitsker2023finite,mcclernan2025finite,le2025numerical}. These considerations form the basis for the present study, whose objective is to establish a multiphysics coupled numerical framework capable of predicting the response of thermal protection system materials within the environment of an ICP wind tunnel. In particular, this work aims to model the response of a TPS sample in the \SI{350}{kW} Plasmatron X ICP facility at UIUC and validate the results against recent graphite ablation experiments performed in this facility. Achieving this goal necessitates a comprehensive multiphysics formulation, one that simultaneously models the plasma flow, the electromagnetic field generated by the induction coils, and the complex thermo-chemical and mechanical behavior of the material itself. By integrating these interacting physical processes within a unified computational platform, the framework aims to provide a high-fidelity representation of TPS material behavior under realistic high-enthalpy testing conditions. 

Numerical modeling of ICPs involves solving the coupled Navier–Stokes and Maxwell’s equations to capture the plasma’s physico-chemical and electromagnetic behavior. Early studies in the 1960s–1970s modeled the torch as an infinite solenoid under local thermodynamic equilibrium (LTE), simplifying the problem to the coupled solution of the gas energy and induction equations \cite{freeman1968energy,keefer1973electrodeless,eckert1970analysis,eckert1970analytical,eckert1972analysis,eckert1977two}. Advances in computational fluid dynamics (CFD) subsequently enabled multi-dimensional magnetohydrodynamic (MHD) simulations \cite{Boulos_1976,mostaghimi1984parametric,mostaghimi1985analysis,mostaghimi1987two,proulx1987heating,mostaghimi1990effect,chen1991modeling,panesi2007analysis,abeele2000efficient,utyuzhnikov2004simulation}. During the 1990s and early 2000s, researchers at the Institute for Problems in Mechanics (IPM), Moscow, developed a combined numerical–experimental methodology to determine the catalytic coefficients of TPS samples \cite{kolesnikov1999combined,kolesnikov1995aerothermodynamic,gordeev2000methodology,vasil1996mathematical}. As part of this approach, which required both detailed experiments and full-facility numerical simulations, IPM researchers were among the first to develop comprehensive simulation frameworks for ICP facilities, modeling the system from the induction coils to the plasma jet without relying on simplifying assumptions \cite{kolesnikov2016heat,bykova2004effect}. Nevertheless, these simulations still depended on experimental data to characterize the TPS material response. The present work overcomes this limitation by introducing a loosely coupled ICP–material simulation framework. To the author’s knowledge, this study represents the first \textit{ab initio}, end-to-end numerical simulation of a high-enthalpy inductively coupled plasma facility used for hypersonic testing, spanning from the electromagnetic field generation in the induction coils to the plasma flow and material response, without reliance on experimentally measured input data for parameter tuning.

Most existing works on ICP modeling, as highlighted above, still assume LTE conditions, which remain valid at high pressures ($\approx 10^4$ Pa or above) where frequent particle collisions maintain equilibrium, thereby reducing computational cost. However, more recent investigations have revealed that non-local thermodynamic equilibrium (NLTE) effects can become significant at lower pressures \cite{zhang2016analysis,kumar2024electronic,kumar2025numerical}, highlighting the need for more detailed plasma modeling in ICP facilities. The state-to-state approach provides the most accurate treatment of non-equilibrium by resolving each internal energy state as a separate pseudo-species, enabling non-Boltzmann population distributions \cite{laux2012state,munafo2015tightly,panesi2013rovibrational,kumar2024investigation,colonna2015non,laporta2016electron,capitelli2013plasma}. Despite its fidelity, this method is computationally expensive and depends on extensive atomic and molecular datasets that are not always available. Consequently, simplified models assuming Maxwell–Boltzmann internal state distributions have been developed, among which the two-temperature (2T) formulation is the most widely used for ICP discharge simulations \cite{park1989nonequilibrium,gnoffo1989conservation}. This model assumes rapid equilibration between translational and rotational degrees of freedom of heavy particles (\emph{i.e.}, atoms and molecules) while electronic and vibrational modes equilibrate with translation of free electrons. In this work, the two-temperature NLTE formulation is adopted for the low-pressure simulations where non-equilibrium effects prevail, whereas the high-pressure cases, such as the coupled flow–material response simulations presented in this paper, are modeled under the LTE assumption to describe the plasma within the facility.

The paper is organized as follows: \cref{sec:ICP_framework} provides a concise overview of the multi-solver coupled ICP framework used in this work. \cref{sec:coupling_framework} describes how the material solver is integrated into the ICP framework, enabling end-to-end simulation of the entire facility, from the torch and induction coils through the chamber and up to the test sample. \cref{sec:problem_setup} provides a detailed description of the Plasmatron X facility and the corresponding simulation setup. \cref{sec:Results} then presents the overall MHD characteristics within the facility, followed by an analysis of the material response and a comparison with experiments. Conclusions and future work are summarized in \cref{sec:conclusions}.
 
\section{ICP framework}\label{sec:ICP_framework}
A complete description of an inductively coupled plasma requires modeling the flow and the electromagnetic field created by the inductor coils. The simulations presented in this work employ a multi-solver coupled numerical framework for ICPs, detailed in ref. \cite{kumar2024investigation}. Here, only a concise overview of the framework is provided to contextualize the novel contributions of this work to its advancement. The flow governing equations are solved in a block-structured finite volume (FV) solver, \textsc{HEGEL} (High fidElity tool for maGnEtogas-dynamic appLications) \cite{munafo2024hegel}, and the electromagnetic (EM) field is solved in a mixed finite element solver,
\textsc{FLUX} (Finite-element soLver for Unsteady electromagnetiX) \cite{kumar2022self}. \textsc{HEGEL} is coupled with the physico-chemical library \textsc{PLATO} \cite{munafo2025plato} for all plasma-related quantities (\emph{e.g.}, transport, thermodynamics), given the flow properties at a given location. The plasma and the electromagnetic field are coupled through the Lorentz force and Joule heating in the fluid momentum equation and energy equations, respectively, and the electrical conductivity in the induction equation. This interaction between the plasma and the electromagnetic solver enables the accurate simulation of intricate magnetohydrodynamic phenomena within ICP wind tunnels. The coupling between \textsc{HEGEL} and \textsc{FLUX} occurs throughout the CFD domain and is called volume coupling. Throughout this paper, this framework is referred to as the \textquotedblleft base\textquotedblright $\ $ICP framework, emphasizing that it comprises the bare minimum set of solvers required to simulate the Plasmatron X facility.

\begin{figure}[hbt!]
\centering
\includegraphics[scale=0.4]{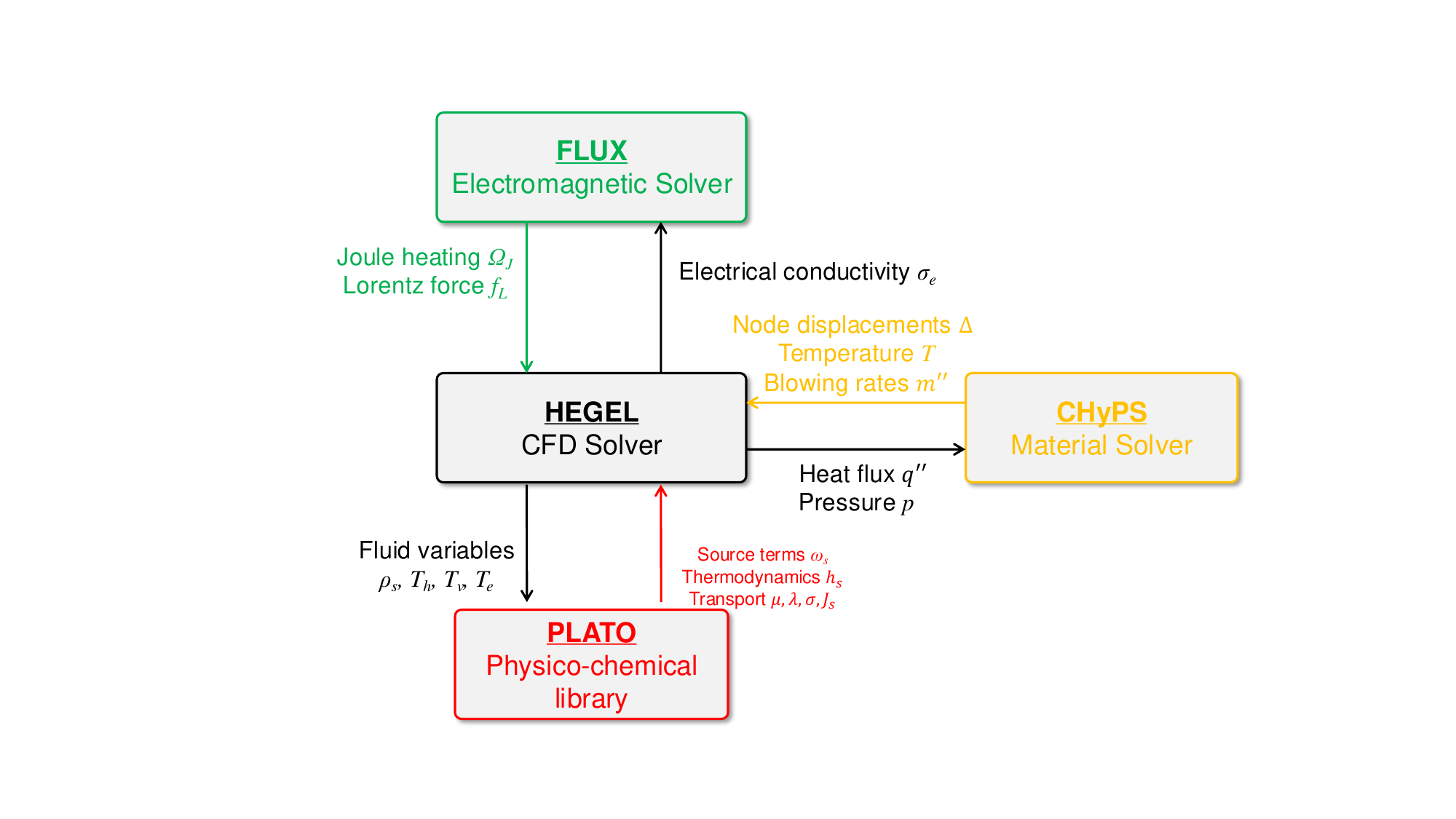}
\caption{Flowchart of the multiphysics coupled numerical framework used in this work.}
\label{fig:overall_coupling}
\end{figure}

\begin{figure}[hbt!]
\centering
\includegraphics[trim={0 4cm 0 4cm},clip,scale=0.4]{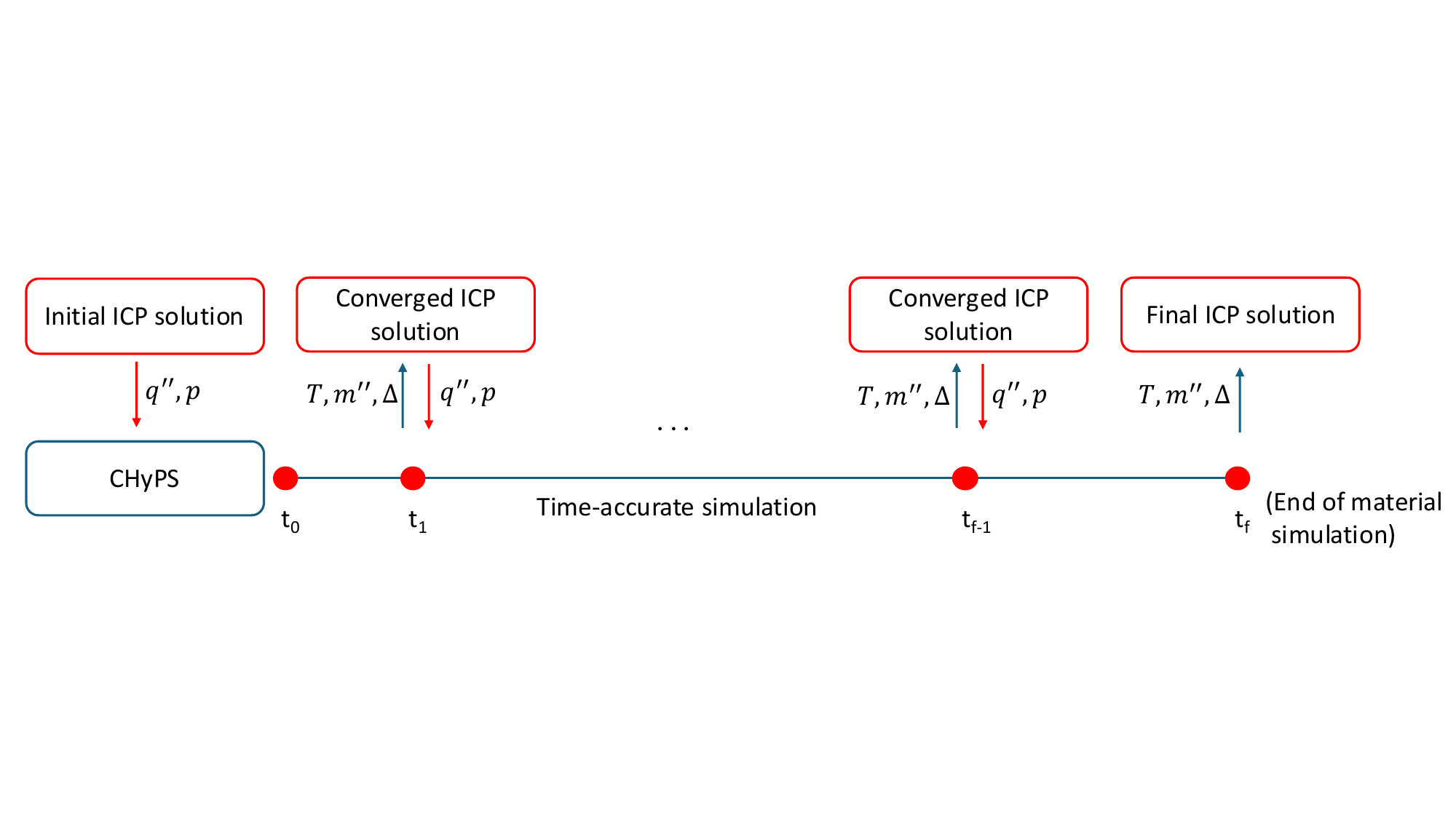}
\caption{Flowchart of the surface coupling between plasma and material solvers.}
\label{fig:surface_coupling}
\end{figure}

\section{Coupling of the base ICP framework with material response solver }\label{sec:coupling_framework}
In this work, the response of the TPS sample in the Plasmatron X facility is modeled with the Coupled Hypersonic Protection System (\textsc{CHyPS}) Simulator \cite{chiodi2022chyps}. \textsc{CHyPS} is a high-order material response solver based on the discontinuous Galerkin formulation, built on top of the finite-element capabilities of the open-source library \textsc{MFEM} \cite{mfem}. More details on the governing equations, numerical approach, various capabilities, \emph{etc.,} can be found in ref. \cite{chiodi2022chyps}. 

The coupling of the base ICP framework with the material response solver in this work uses a loosely coupled procedure, where the boundary conditions at the sample surface between \textsc{HEGEL} and \textsc{CHyPS} are exchanged at every coupling window. Further, the surface coupling assumes an equilibrium thermo-chemistry, where the plasma simulations are run assuming LTE conditions, while the surface recession modeling in the material solver uses the well-known $B^{\prime}$-table approach. This formulation has been implemented in several existing ablation codes (\emph{e.g.,} PATO \cite{lachaud2014porous} and KATS \cite{weng2014multidimensional}), where the dimensionless char ablation mass flux $B^{\prime}_c$ and wall enthalpy $h_w$ are tabulated as a function of surface pressure, temperature and
dimensionless pyrolysis gas mass flux $B^{\prime}_g$. More details on the $B^{\prime}$ ablation modeling in \textsc{CHyPS} can be found in ref. \cite{chiodi2022chyps}. The TPS sample in the Plasmatron X wind tunnel is generally tested for several seconds. Hence, the numerical simulation of the material needs to run for several seconds to reproduce the material response measured during the experiments. However, running a time-accurate CFD simulation for several seconds is impractical due to the time-step restriction for the simulation. Hence, this work uses a pseudo-steady-state approach for the surface coupling, where the ICP (\emph{i.e.}, \textsc{HEGEL} + \textsc{FLUX}) simulation is run till convergence for the given surface quantities from the material solver at every coupling window. The coupling window in this work is based on the simulation time of the material solver, \emph{i.e.}, the material solver exchanges the surface data with the plasma solver at a fixed simulation time interval. The time interval for the coupling window is decided based on the gradient of the sample temperature and is changed dynamically during the simulation. \cref{fig:overall_coupling} shows the flowchart of the multiphysics coupled numerical framework consisting of the base ICP framework and the material solver, highlighting the various quantities being exchanged. 
All data exchange between the solvers is managed through \textsc{preCICE} \cite{bungartz2016precice}, an open-source coupling library for partitioned multi-physics simulations. The communication between solvers occurs within designated coupling windows, which define when and how data is transferred. These coupling windows are generally distinct and non-overlapping across solvers, as their configurations are tailored to the specific temporal and numerical requirements of each participating solver.

\cref{fig:surface_coupling} illustrates the algorithm for coupling between the base ICP framework and the material solver. First, an uncoupled steady-state ICP solution is obtained. \textsc{HEGEL} then passes the surface heat flux ($q^{\prime\prime}$) and pressure ($p$) to \textsc{CHyPS}, which are used as boundary conditions at the sample$^{\prime}$s active surface to advance the time-accurate material simulation. \textsc{CHyPS} communicates the displacement ($\Delta$) resulting from ablation, surface temperature ($T$) and blowing rate ($m^{\prime\prime}$) back to \textsc{HEGEL} at the next coupling window. \textsc{HEGEL} then makes use of the displacement data to modify the surface boundary of the mesh. The updated coordinates of the surface are then used to modify those of the inner mesh points by means of Linear Transfinite Interpolation (LTI) \cite{spekreijse2002simple,eriksson1982generation}. The surface temperature and blowing rate data are used in a blowing wall boundary condition \cite{thompson2008implementation}. The implementation of this boundary condition relies on the surface normal momentum balance (SMoB) obtained by equating the fluxes normal to the surface from the gas phase to those from the material. When diffusive contributions are neglected, the SMoB relation reads  
\begin{equation}
p_\eta=p_{f}+\rho_{f} v_{f}^2=p_{\mathrm{w}}+\rho_{\mathrm{w}} v_{\mathrm{w}}^2 \label{eq:smob}
\end{equation}
where $p$ is the gas pressure, and the subscripts w, $f$, and $\eta$ denote the wall, flow, and net conditions, respectively. Under LTE conditions, the wall density is retrieved by numerically solving \cref{eq:smob}, where the dependence of the wall pressure on the temperature and the sought density is given by the equation of state (\emph{i.e.}, $p = p(\rho, \, T)$),
\begin{equation}
\rho_{\mathrm{w}}=\frac{m^{\prime\prime}_{\mathrm{w}}{}^2}{p_\eta-p_{\mathrm{w}}\left(\rho_{\mathrm{w}}, T_{\mathrm{w}}\right)}.
\end{equation}

Once the wall density is computed, the blowing velocity follows from
\begin{equation}
    v_{\mathrm{w}} = \frac{m^{\prime\prime}_{\mathrm{w}}}{\rho_{\mathrm{w}}}.
\end{equation}
\begin{figure}[hbt!]
\centering
\includegraphics[scale=0.8]{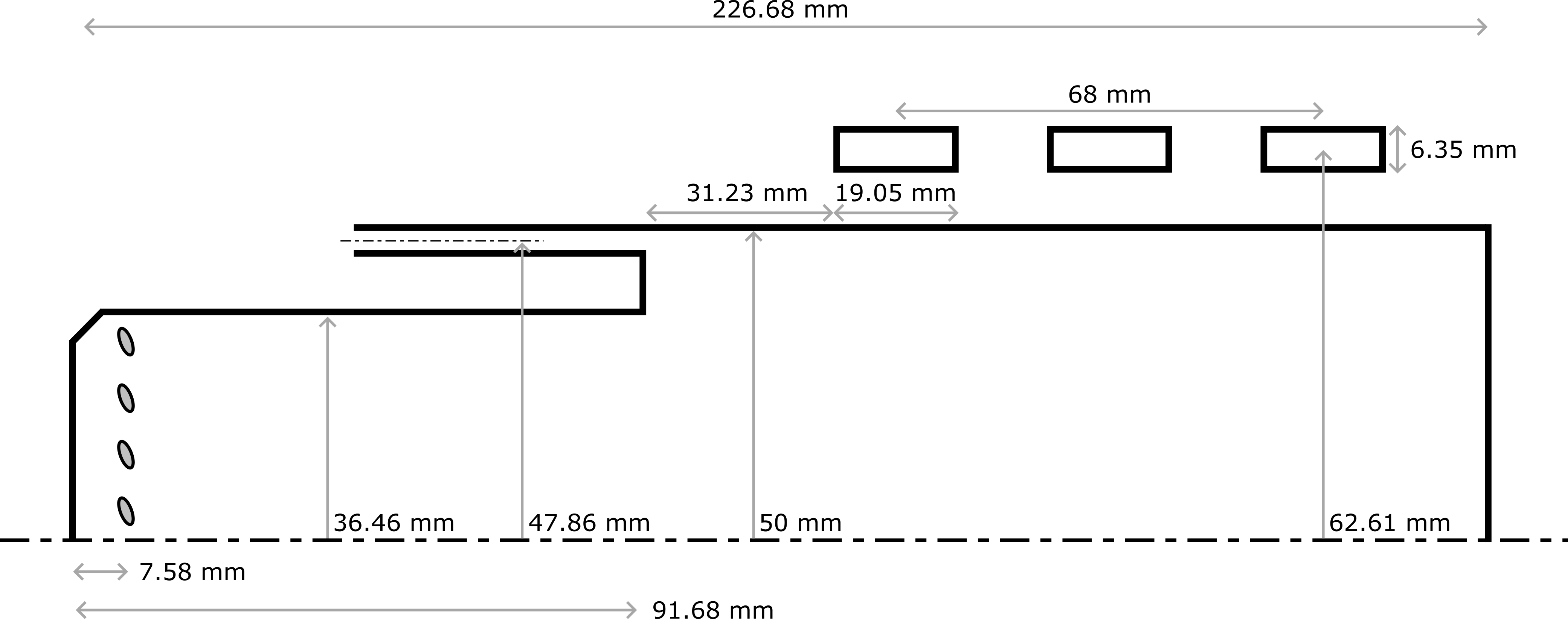}
\caption{Schematic of the Plasmatron X torch \cite{oruganti2023modeling}.}
\label{fig:pX_torch}
\end{figure}
\section{Problem description and simulation setup}\label{sec:problem_setup}
The multiphysics computational framework discussed above simulates the \SI{350}{\kilo\watt} Plasmatron X facility at UIUC. All the simulations discussed here are 2D axisymmetric. The schematic of the torch section is shown in \cref{fig:pX_torch}. The torch has two injectors: central injection consisting of \num{15} holes, which are inclined \SI{15}{\degree} downward and have \SI{24}{\degree} swirl angle, and sheath gas injection consisting of \num{72} holes and inject gas straight in the axial direction. For simplicity, both injectors are assumed to be continuous annular injectors. The inductor coil has three turns with a rectangular cross-section. In the adopted configuration, the inductor coils are assumed to be parallel to enforce symmetry about the torch axis. The rings are assumed infinitely thin and are located at the mid-point of the innermost part of the coil cross-section, which is a good approximation since most of the current is concentrated there due to the skin-effect \cite{abeele2000efficient,belevitch1971lateral,blackwell2020demonstration}. The current running through the set of $N_{\mathrm{c}}$ coils is modeled based on a point-source model as described in ref. \cite{kumar2024investigation}. The frequency of the coils for the facility is \SI{2.1}{\mega\hertz}. For all simulations, the central and sheath mass flows are kept fixed at \SI{0.86}{\gram/\second} and \SI{7.13}{\gram/\second}, respectively. The only quantities that vary in the experiments and simulations presented in this work are the operating pressure and power. It is important to note that during the facility$^\prime$s operation, only a portion of the operating power is delivered to the plasma due to inefficiencies in the electrical circuit of the RF generator. The efficiency is defined considering only the electrical losses in the RF generator, which are estimated from the power dissipated as Joule heating in the copper components of the RF circuit. The generator efficiency is defined as:
   \begin{equation}
       \eta_{\text {gen }}=\frac{P_{A C}-P_{\text {gen }}^{\text {loss }}}{P_{A C}} 100,
   \end{equation}
    where $P_{AC }$ is the input electrical power from the AC supply, and $P_{\text {gen }}^{\text {loss }}$ is the power dissipated as heat in the copper components of the RF circuit. These losses are absorbed by the deionized (DI) water cooling lines and are estimated from the measured mass flow rate and temperature rise of the coolant as:
    \begin{equation}
           P_{\text {loss }}=\dot{m}_{\text {water }} \cdot c_{p_{\text {water }}} \cdot\left[\left(T_{\text {out }}-T_{\text {in }}\right)_{\text {plasma ON }}-\left(T_{\text {out }}-T_{\text {in }}\right)_{\text {plasma OFF }}\right]. 
    \end{equation}
 The term $\left(T_{\text {out }}-T_{\text {in }}\right)_{\text {plasma OFF }}$ represents the baseline temperature rise of the DI water in the absence of plasma, accounting for heat generated by non-plasma sources, including viscous dissipation due to pressure losses along the cooling lines. Subtracting this baseline isolates the additional heat associated with plasma operation. The power values used in the simulations are obtained by multiplying the experimental operating power by the corresponding experimentally measured efficiencies (reported for all cases). 

\begin{figure}[hbt!]
\centering
\includegraphics[trim={0 3cm 0 3cm},clip,scale=0.4]{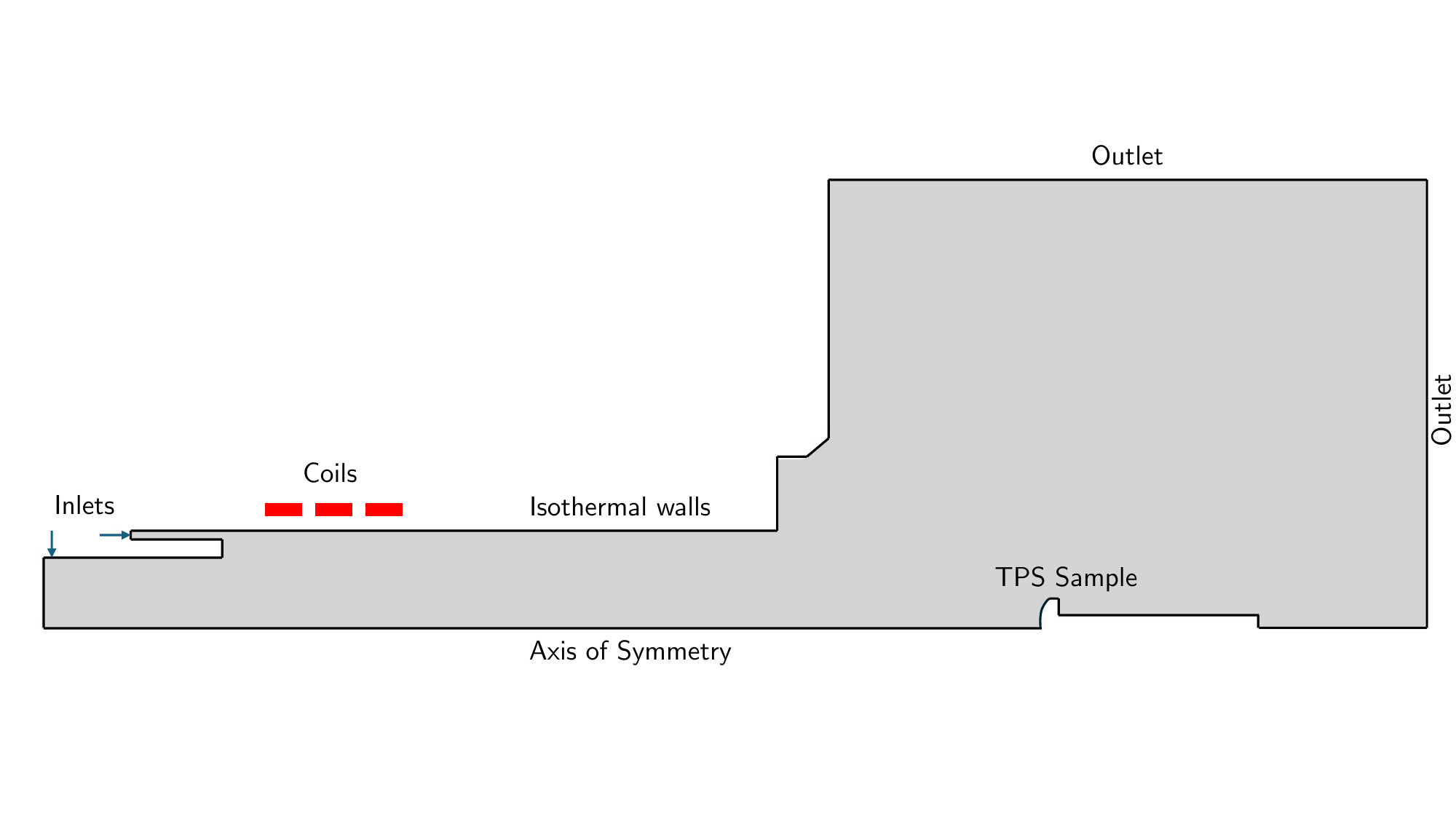}
\caption{Schematic of the adopted simplified geometry of the Plasmatron X facility consisting of: torch, nozzle, chamber, and TPS sample.}
\label{fig:domain}
\end{figure}

\begin{figure}[hbt!]
\centering
\includegraphics[trim={0 6cm 0 4cm},clip,scale=0.45]{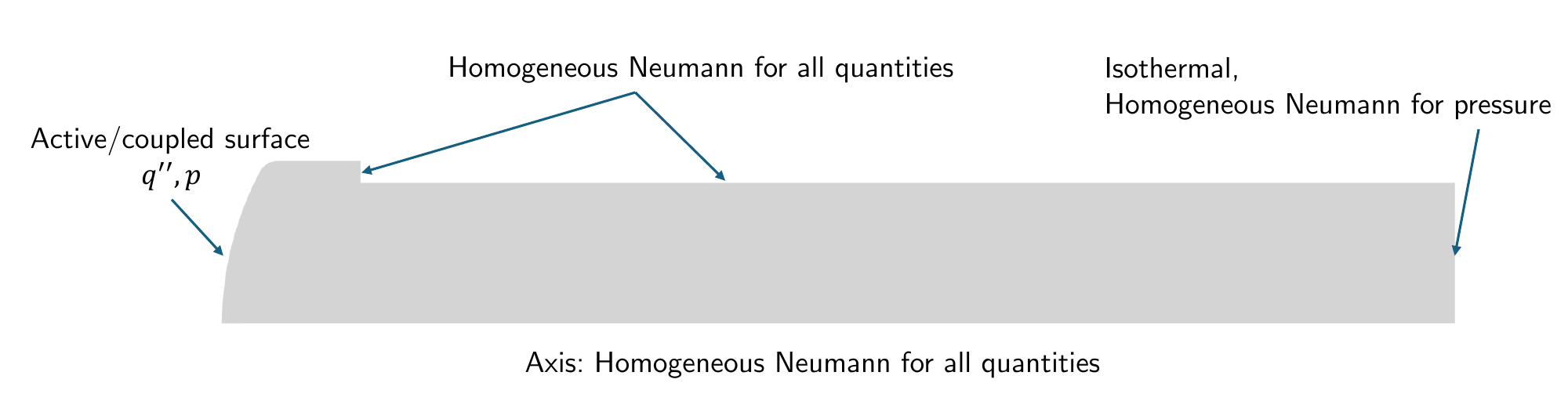}
\caption{Schematic of the TPS sample geometry (including the holder) with the boundary conditions applied on all surfaces.}
\label{fig:sample_BC}
\end{figure}

The Plasmatron X facility has three kinds of nozzles (attached at its outlet): one straight and two converging-diverging. In this work, only the straight nozzle (of length \SI{129.54}{\milli\meter} and diameter \SI{100}{\milli\meter}) has been considered. \cref{fig:domain} shows the simplified geometry of the entire facility, including the torch, nozzle, chamber, and TPS sample. The TPS sample used in the present calculations is a \SI{30}{\milli\meter} diameter isoQ graphite sample placed at an axial distance of \SI{108}{\milli\meter} from the torch exit as in the experiments. The shape has a radius of curvature of the front face of \SI{46}{mm} and a corner radius of \SI{1.875}{mm}. The sample is threaded onto a \SI{100}{mm} hollow graphite holder, which is threaded onto the water-cooled arm of the displacement system, which positions it into the plasma jet (see \cref{fig:sample_BC}).

The gas mixture considered for all the simulations is a conventional eleven species air mixture: $\mathcal{S} = \left\{\mathrm{e}^-, \, \mathrm{N}_2, \, \mathrm{O}_2, \, \mathrm{NO},\mathrm{N},\, \mathrm{O}, \, \mathrm{N^+_2}, \, \mathrm{O^+_2},\, \mathrm{NO^+},\, \mathrm{N^+},\, \mathrm{O^+}\right\}$. To account for NLTE effects, a conventional two-temperature model is adopted, where the heavy particle temperature ($T_{\mathrm{h}}$) is assigned to the roto-translational degrees of freedom of heavy particles, whereas the \emph{vibronic} temperature ($T_{\mathrm{ve}}$) characterizes the translational energy of free electrons as well as those of the vibrational modes and excited electronic states of heavy particles. 

\subsection{Boundary conditions}\label{sec:boundary_conditions}
\subsubsection{Plasma}
The fluid governing equations are solved in \textsc{HEGEL} by imposing the following boundary conditions.
\begin{itemize}
\item injectors (subsonic inflow):
$$
\rho u=\frac{\dot{m_x}}{A}, \quad \rho v=\frac{\dot{m_y}}{A}, \quad \rho w=\frac{\dot{m_z}}{A}, \quad  y_s = y_{s\, \mathrm{in}}, \quad \frac{\partial p}{\partial n}=0 \quad \text{and} \quad T_{\mathrm{h}}= T_{\mathrm{ve}} = T_{\mathrm{in}},
$$
where the $\mathrm{in}$ lower-script denotes the injection conditions. In the above formulas, $\dot{m}$ is the mass flow; $A$ the area of the injector; $u$, $v$, and $w$ the axial, radial, and swirl velocity components, respectively; $T$ the temperature; and $y_s$ the mass fraction of species $s \in \mathcal{S}$. 
\item centerline (symmetry):
$$
\frac{\partial \rho_s}{\partial r}=\frac{\partial u}{\partial r}=\frac{\partial p}{\partial r}=0 \quad \text{and} \quad v=w=0.
$$
\item walls (isothermal):
$$
u=v=w=0 \quad \text{and} \quad T_{\mathrm{h}}=T_{\mathrm{ve}} = T_{\mathrm{w}}.
$$

\item Sample surface (blowing wall): as detailed in \cref{sec:coupling_framework}.

\item outlet (subsonic/supersonic outflow): if the local normal Mach number at the boundary is less than 1, a constant ambient pressure is imposed, $p=p_{\mathrm{a}}$. Conversely, if the local Mach number is greater than 1, the flow properties are extrapolated from the interior. 
\end{itemize} 
In all simulations, the temperature at the inlet is set to \SI{300}{\kelvin}. The walls are also assumed at \SI{300}{\kelvin}, consistent with the experimental conditions in which the quartz tube is actively cooled. In practice, the wall temperature may not be maintained exactly at \SI{300}{\kelvin}. However, the plasma field is generally insensitive to modest deviations in the wall temperature.
\\

\subsubsection{Electromagnetic field}
In \textsc{FLUX}, the induction equation is solved by setting the electric field to zero on the symmetry axis and far from the coils: $E(x, 0)=0$, $E(x, r\rightarrow\infty)=0$, and $E(x\rightarrow \pm \infty, r)=0$. 
\\

\subsubsection{Material}
\cref{fig:sample_BC} illustrates the boundary conditions applied in \textsc{CHyPS} on the various surfaces of the TPS sample. At the front (\emph{i.e.}, active/coupled) surface, the heat flux ($q^{\prime\prime}$) and pressure ($p$) obtained from \textsc{HEGEL} are prescribed. Along the axis, homogeneous Neumann boundary conditions are imposed for all variables, including temperature and pressure. The same goes for the surface of the sample holder. The back face of the sample holder is instead modeled as an isothermal wall maintained at \SI{300}{K}, with a homogeneous Neumann boundary condition for pressure.  

\subsection{Mesh}\label{sec:plasma_mesh}
\begin{figure}[hbt!]
\centering
\includegraphics[trim={0 1cm 0 1cm},clip,scale=0.7]{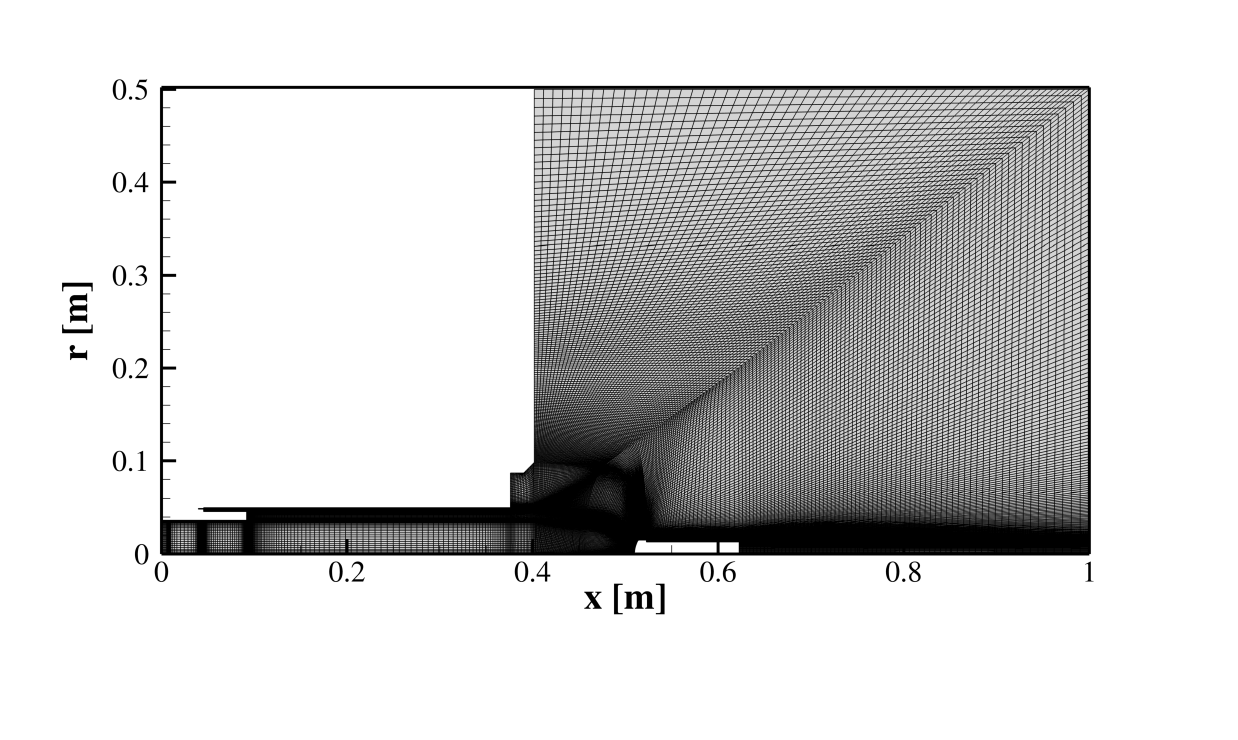}
\caption{Multi-block structured mesh used by \textsc{HEGEL}.}
\label{fig:hegel_mesh}
\end{figure}

All simulations in this study employ a point-match, block-structured mesh for \textsc{HEGEL}. The mesh for the torch-chamber-only domain (without the sample) is made of \num{62700} cells (with 128 across the torch diameter), while the configuration including the sample contains approximately \num{58000} cells (with 90 cells across the torch diameter) (see \cref{fig:hegel_mesh}). These grid resolutions were selected based on a grid-convergence analysis of the plasma flowfield as presented in \ref{appendix:grid_convergence_torch}. At the sample surface, the cell thickness was set to \SI{0.8}{\micro\meter}, which ensured a grid-converged heat flux as shown in \ref{appendix:grid_convergence_with_sample}.

\begin{figure}[hbt!]
\centering
\includegraphics[scale=0.4]{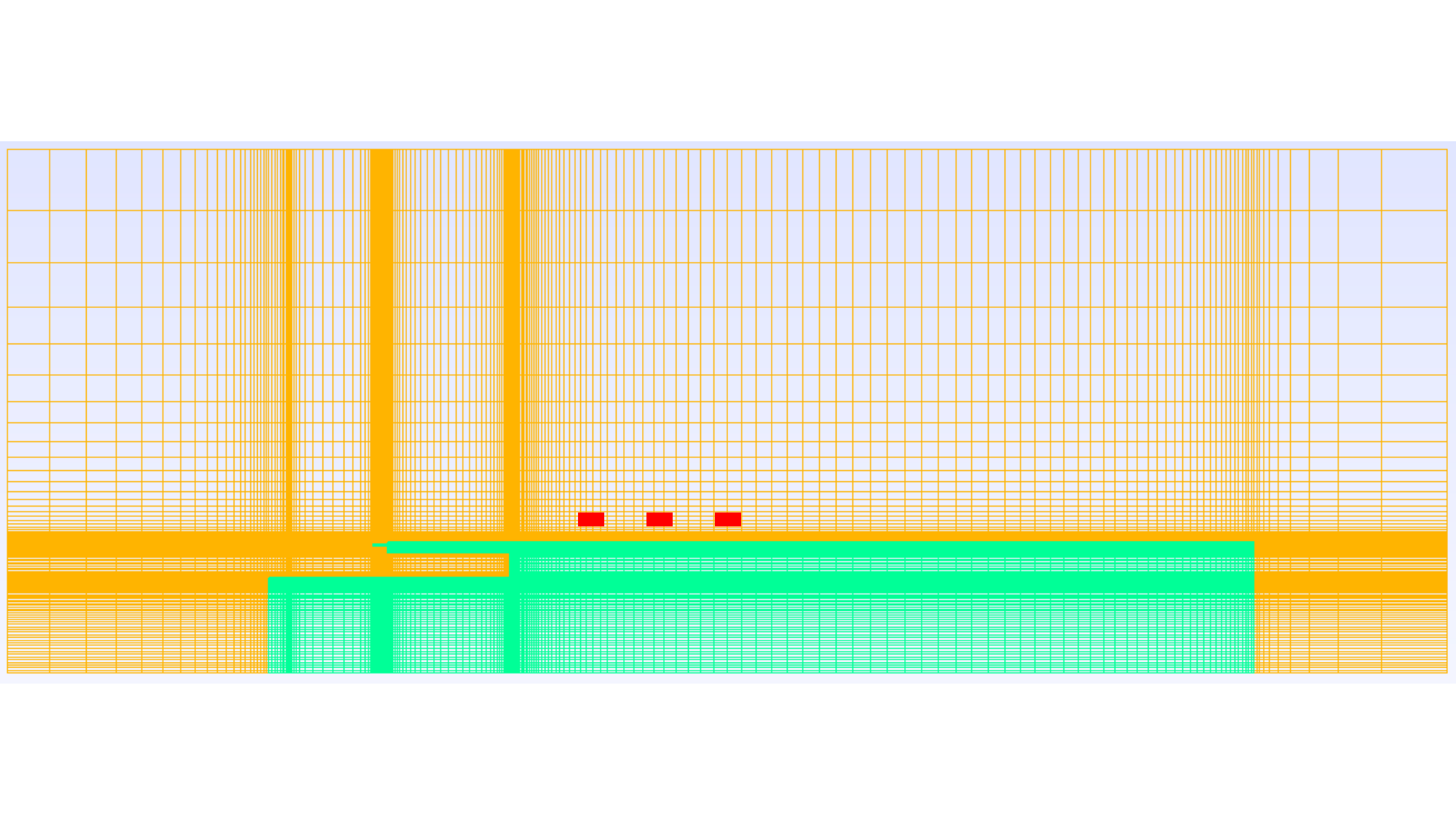}
\caption{\textsc{FLUX} mesh. The portion of the mesh highlighted in green overlaps with that used in \textsc{HEGEL} (coil locations are marked in red).}
\label{fig:flux_mesh}
\end{figure}

For \textsc{FLUX}, a structured mesh consisting of around \num{40000} rectangular elements was used, as shown in \cref{fig:flux_mesh}. For the sake of robustness and simplicity, the vertices of the mesh used by \textsc{FLUX} were taken coincident with those of \textsc{HEGEL} in the coupling region (\emph{i.e.}, torch) \cite{kumar2024investigation}. Hence, a separate grid-convergence study for the FLUX mesh is not needed, as it changes according to the flow solver$^{\prime}$s mesh.

\begin{figure}[hbt!]
\centering
\includegraphics[scale=0.8]{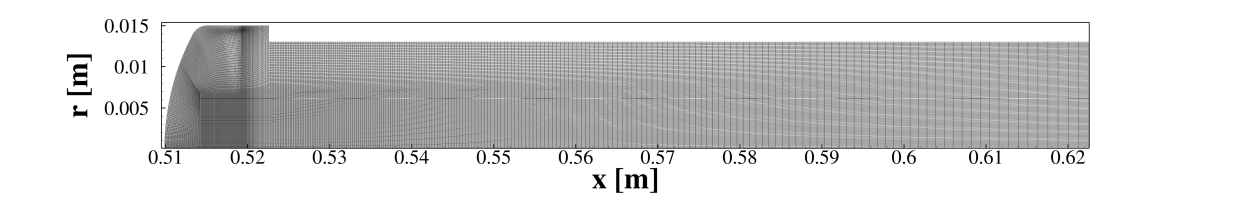}
\caption{TPS sample (\SI{30}{mm} isoQ) mesh used by \textsc{CHyPS}. The $x$–$y$ coordinates are defined with respect to the global reference frame employed for the complete computational domain.}
\label{fig:sample_mesh}
\end{figure}

\cref{fig:sample_mesh} shows the isoQ \SI{30}{mm} TPS sample mesh used for the material solver. The mesh contains approximately \num{33500} elements, with 168 and 88 elements oriented parallel and perpendicular to the active surface within the sample domain, and 200 and 94 elements parallel and perpendicular to the holder surface, respectively. This mesh size was found to be sufficient to attain a grid-converged solution as demonstrated in \ref{appendix:grid_convergence_material}.

\subsection{Numerical method}\label{sec:plasma_numerical_method}
The steady-state plasma simulations presented in this work have been performed by integrating the FV discretized flow governing equations by means of the backward Euler method, with CFL-based local time-stepping to accelerate convergence. The simulations were considered converged when the $L_2$ norm of the solution variation (in terms of conservative variables) decreased by at least 6 orders of magnitude. The plasma solver uses a linear MUSCL reconstruction to achieve second-order accuracy in space . Inviscid fluxes were evaluated using the all-speed AUSM\textsuperscript{+}-up flux function \cite{liou2006sequel} to handle the low Mach number stiffness often associated with ICPs. Diffusive fluxes are evaluated using Green-Gauss’ theorem to compute face-averaged gradients.

At each coupling window, \textsc{HEGEL} exchanges data with \textsc{FLUX}, which computes the steady-state electromagnetic field based on the given plasma electrical conductivity distribution. To solve the resulting linear system, \textsc{FLUX} employs the Flexible Generalized Minimal Residual (FGMRES) method, leveraging Algebraic Multigrid (AMG) as a preconditioner to enhance convergence. 

The governing equations (\emph{i.e.,} gas pressure and temperature equations) in \textsc{CHyPS} are advanced in time using the backward Euler method with a fixed time step of \SI{1}{\milli\second}. Non-linearities are addressed via Picard iterations.

\section{Results}\label{sec:Results}
\subsection{General features of the plasma inside the Plasmatron X facility}\label{sec:general_flow_features}
To introduce the main features of the plasma inside the Plasmatron X facility, an LTE simulation of the torch-chamber geometry (without the sample) at \SI{20}{\kilo\pascal}, \SI{55}{\kilo\watt} and \SI{63.75}{\percent} efficiency ($\eta$) is presented in this section.

\begin{figure}[H]
\centering
\includegraphics[scale=0.75]{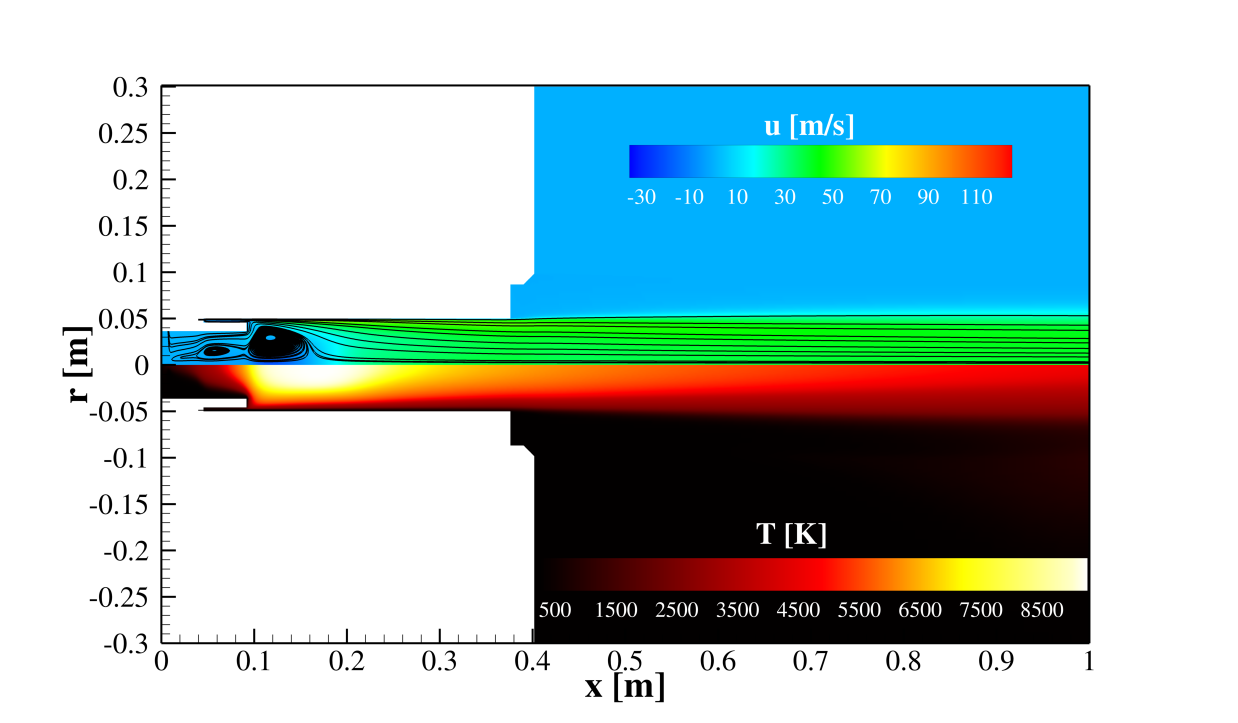}
\caption{Plasma field inside the Plasmatron X facility. Top: axial velocity with streamlines, bottom: temperature ($p_{\mathrm{a}} = \SI{20}{\kilo\pascal}$, $P = \SI{55}{\kilo\watt}$, $\eta = \SI{63.75}{\percent}$). }
\label{fig:20kPa_35kW_u_top_T_bottom}
\end{figure}

\cref{fig:20kPa_35kW_u_top_T_bottom} shows the axial velocity and temperature distributions inside the facility. The streamlines added on top of the velocity field highlight a large recirculation bubble in the torch near the inlet, representing a typical vortex-mode discharge. This phenomenon plays a crucial role in sustaining the hot plasma within the torch. The recirculation bubble carries the hot plasma from the inductor zone back towards the inlet, preventing the plasma from getting swept away by the flow. The plasma is suspended in the middle of the inductor zone, shielded from the tube by a layer of cold fluid coming from the sheath gas injector. 

\begin{figure}[H]
\hspace{-1cm}
\subfloat[][]{\includegraphics[trim={0 0.75cm 0 0},clip,scale=0.25]{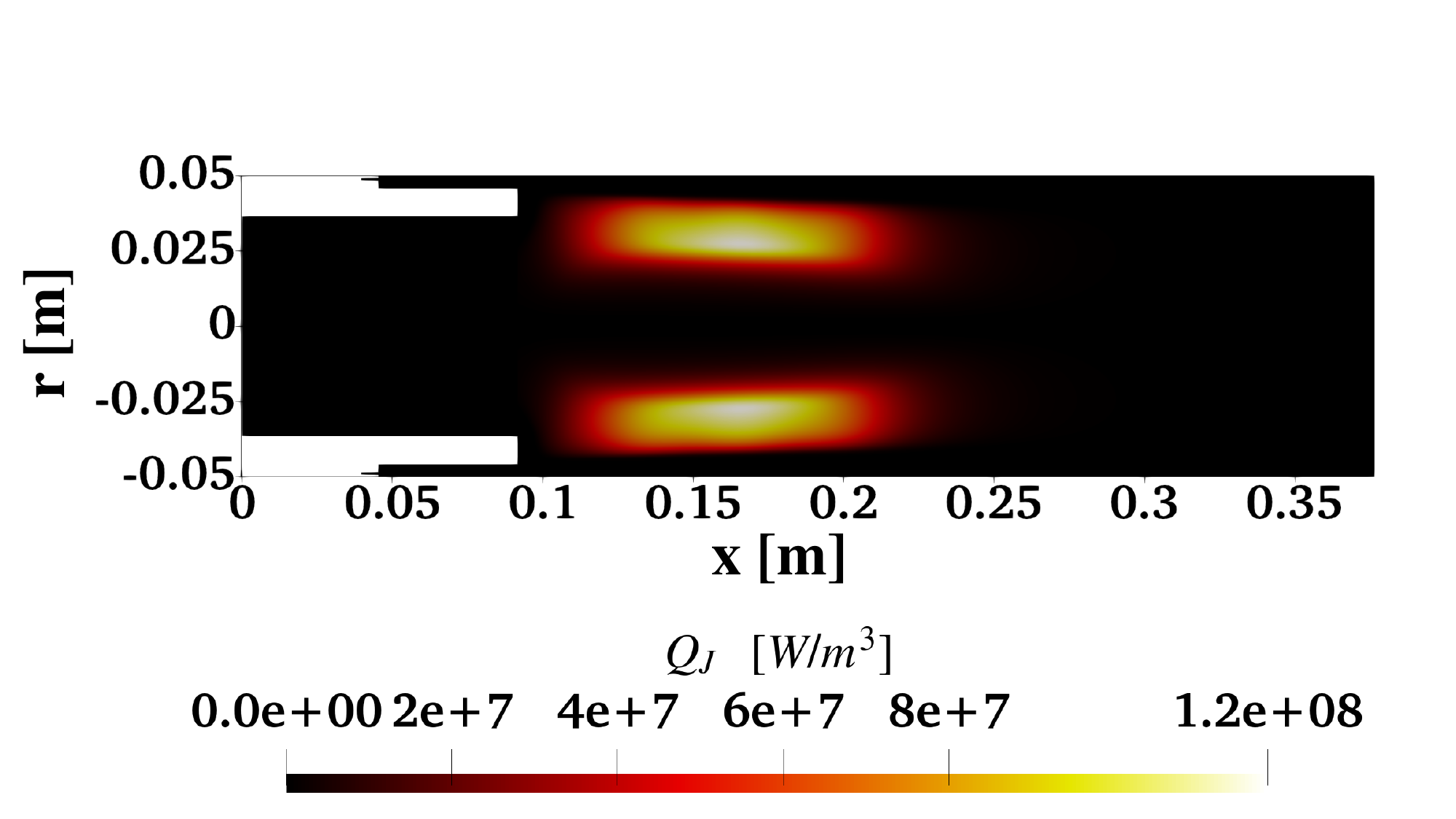}}
\subfloat[][]{\includegraphics[scale=0.24]{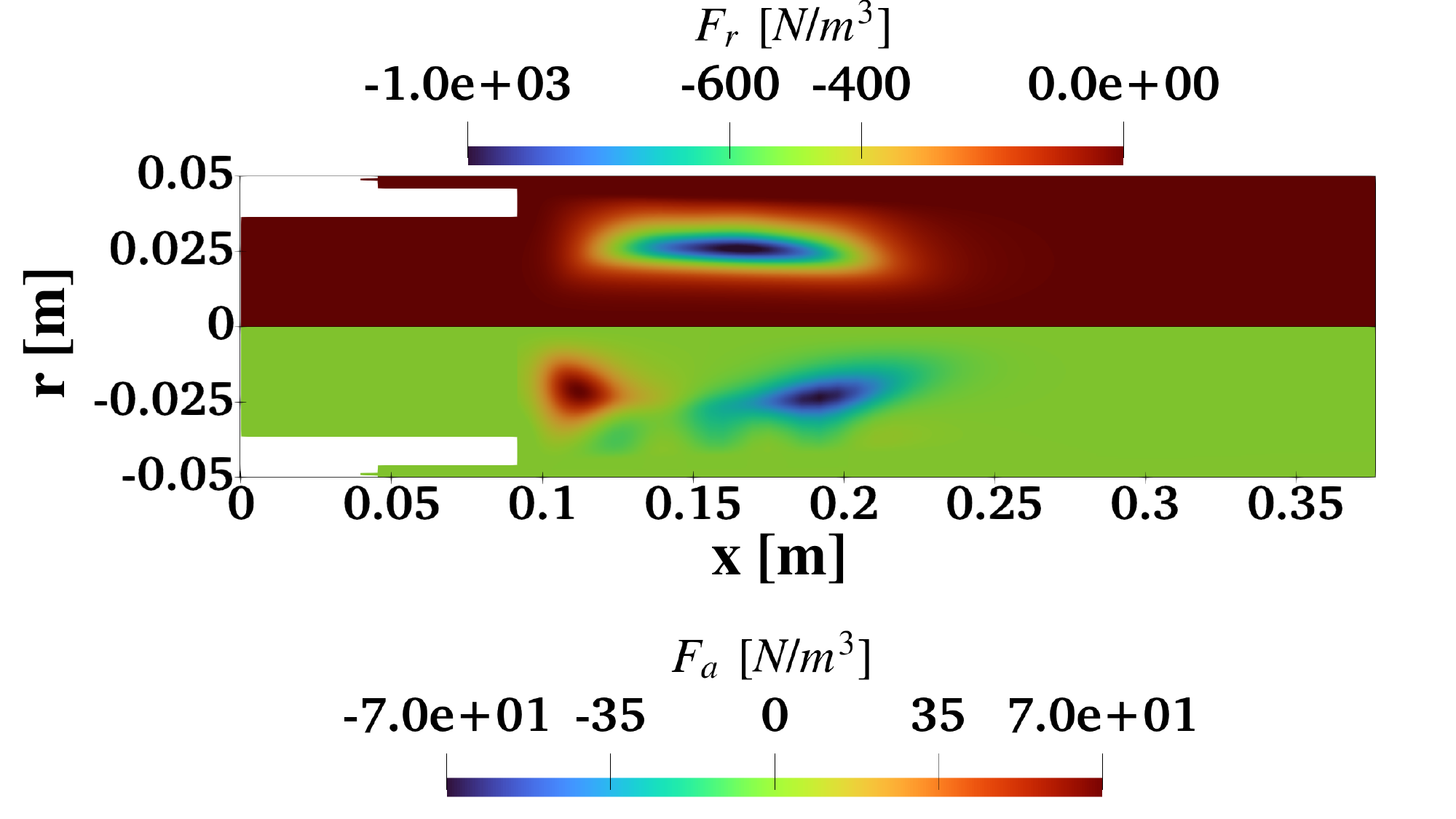}}
\caption{Distribution of the electromagnetic quantities inside the torch. (a) Joule heating, (b) Lorentz force (top: radial component, bottom: axial component) ($p_{\mathrm{a}} = \SI{20}{\kilo\pascal}$, $P = \SI{55}{\kilo\watt}$, $\eta = \SI{63.75}{\percent}$).}
\label{fig:em_quantities}
\end{figure}

\cref{fig:em_quantities} displays the distribution of Joule heating and Lorentz forces within the torch, predominantly concentrated in the coil region. The large Joule heating in this region raises the temperature of the incoming cold gas, forming a hot plasma core. Additionally, as shown in \cref{fig:em_quantities}(b), the radial component of the Lorentz force is directed towards the axis, ensuring that the hot plasma is pushed away from the walls, whereas the axial component, acting in the negative $x$-direction around $x = \SI{0.2}{m}$, pushes the plasma towards the inlet. The interplay between the swirling injector and Lorentz forces ultimately generates the large recirculation bubble shown in \cref{fig:20kPa_35kW_u_top_T_bottom}(a), thereby stabilizing the plasma discharge.

Next, a series of simulations was performed at a very low pressure of \SI{600}{Pa} by varying the power only. This was done to qualitatively validate the current ICP framework by investigating its ability to accurately replicate the transition of the plasma jet in the facility from subsonic to supersonic at very low pressures, a feature observed in the experiments \cite{capponi2024multi}. This observation aligns with Rayleigh flow theory, which states that heat addition in a constant-area duct drives the flow toward, but not past, sonic conditions. Therefore, at a fixed (very low) operating pressure at which the flow velocities are much higher, a continuous increase in power eventually chokes the nozzle exit, resulting in a supersonic plasma jet in the suddenly expanding chamber region. Hence, a qualitative agreement with the plasma jet transition experiments would not only validate the correct implementation of the models, geometries, etc., but also indicate the correctness of the translation of various operating parameters (\emph{e.g.}, power and efficiency) from the experiments to the simulations. To predict the transition using the current numerical framework, simulations were conducted at five different powers: 100, 150, 200, 250, and 300 \si{\kilo\watt}, assuming an efficiency of \SI{50}{\percent}. Due to the low pressure values, simulations were conducted using the two-temperature NLTE model from Park \emph{et al.} \cite{park2001chemical}.  

     \begin{table}[!hbt]
    
    \centering
    \begin{tabular}{lccc}
    \hline\hline
    Power & l [Measured] \cite{capponi2024multi} & l [Predicted] & Relative error \\
    (\si{\kilo\watt}) & (\si{\milli\meter}) & (\si{\milli\meter}) & (\si{\percent}) \\
    \hline
    150 & \SI{68}{} & \SI{79}{} & 16.2 \\
    200 & \SI{85}{} & \SI{103}{} & 21.2\\
    250 & \SI{110}{} & \SI{122}{} & 10.9\\
    300 & \SI{120}{} & \SI{136}{} & 13.3\\
    \hline\hline
    \end{tabular}
    \caption{\label{tab:shock_distance} Average distance between shocks in the supersonic plasma jet cases ($p_{\mathrm{a}} = \SI{650}{Pa}$).}
    \end{table}
    
\begin{figure}[H]
\centering
\subfloat[][\SI{100}{\kilo\watt}]{\includegraphics[trim={0 0.75cm 0 0.75cm},clip,scale=0.35]{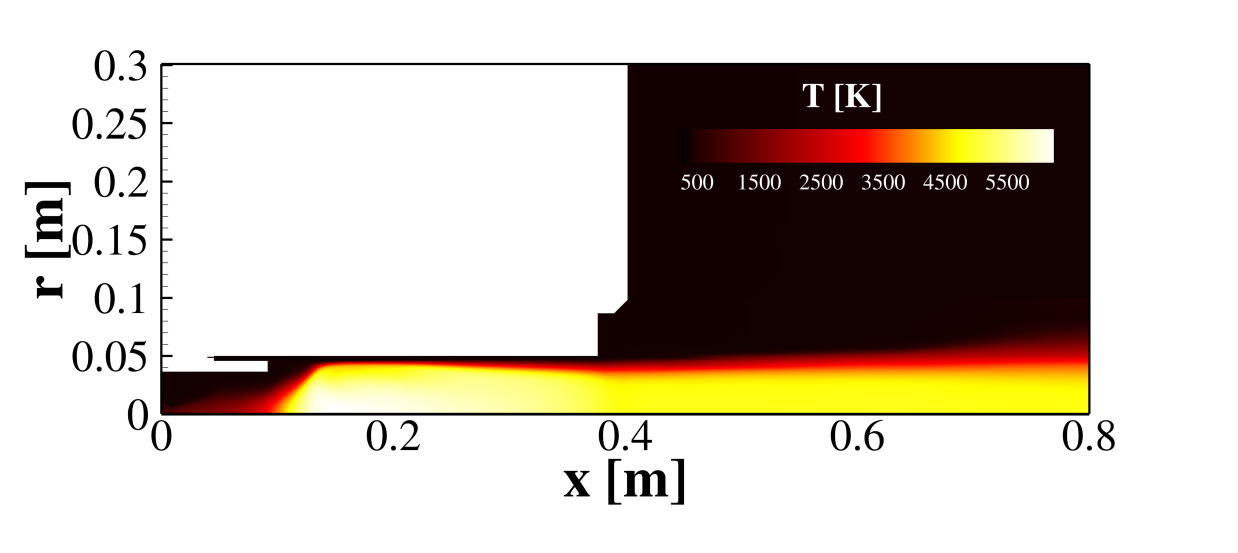}}
\subfloat[][\SI{100}{\kilo\watt}]{\includegraphics[trim={0 0.75cm 0 0},clip,scale=0.35]{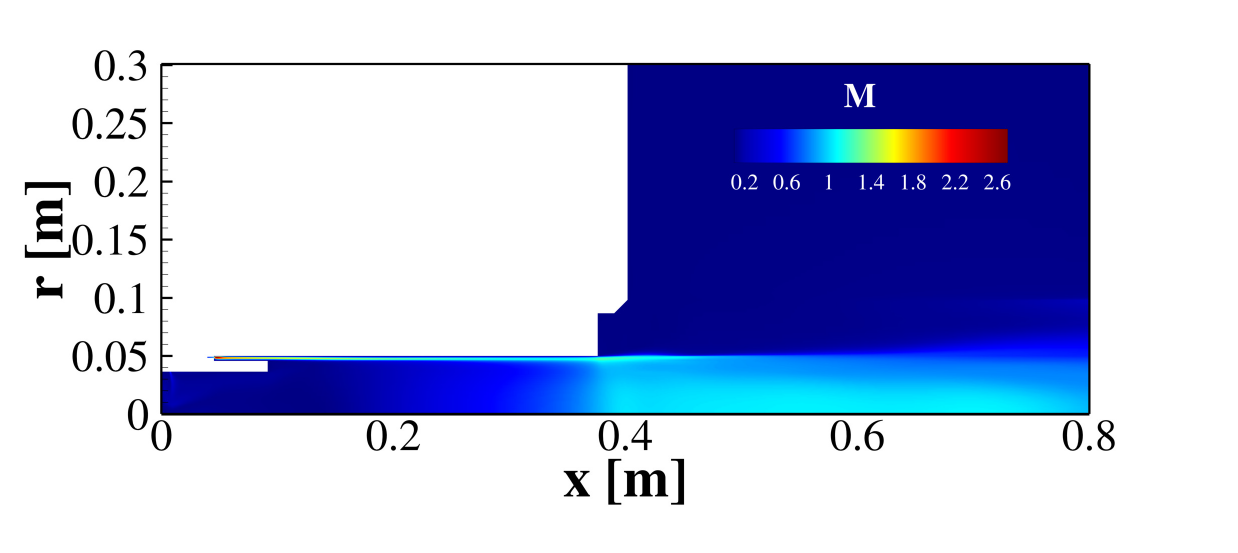}}\\
\subfloat[][\SI{150}{\kilo\watt}]{\includegraphics[trim={0 0.75cm 0 0},clip,scale=0.35]{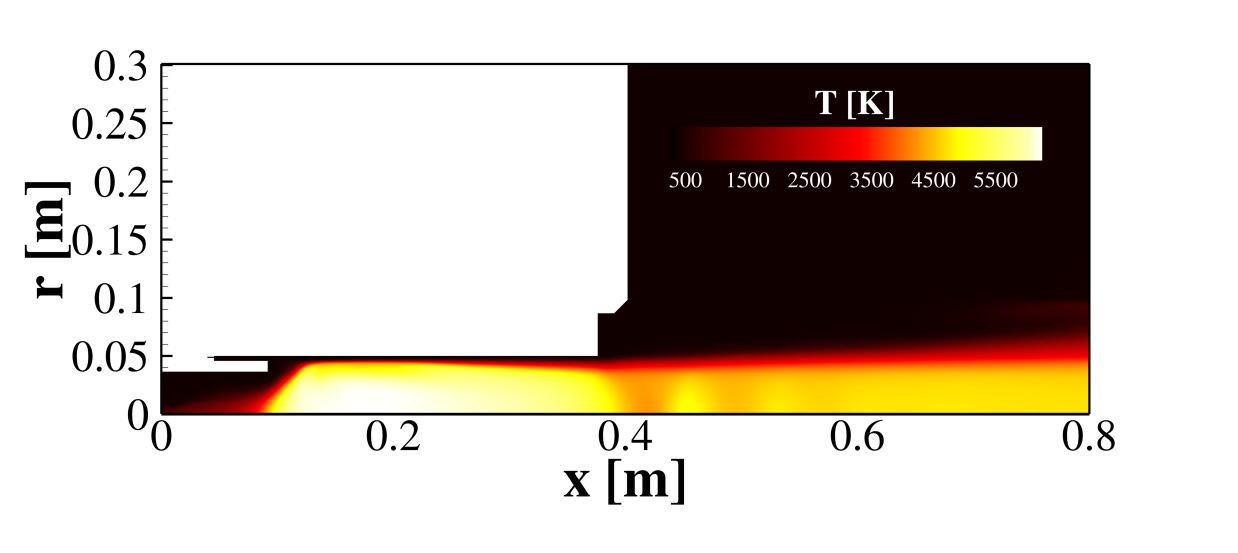}}
\subfloat[][\SI{150}{\kilo\watt}]{\includegraphics[trim={0 0.75cm 0 0},clip,scale=0.35]{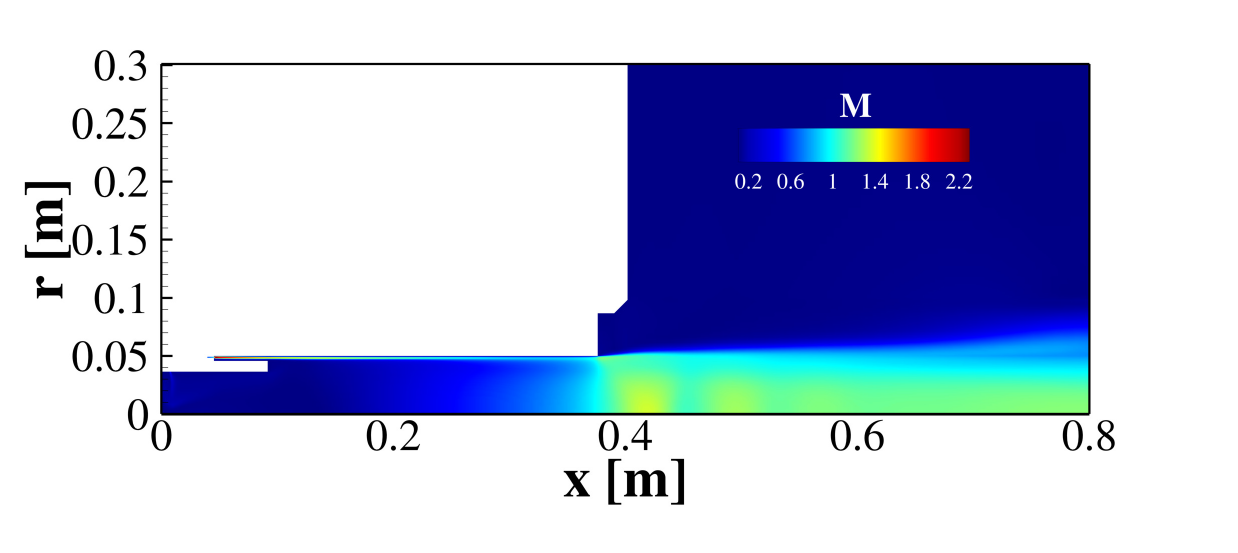}}\\
\subfloat[][\SI{200}{\kilo\watt}]{\includegraphics[trim={0 0.75cm 0 0},clip,scale=0.35]{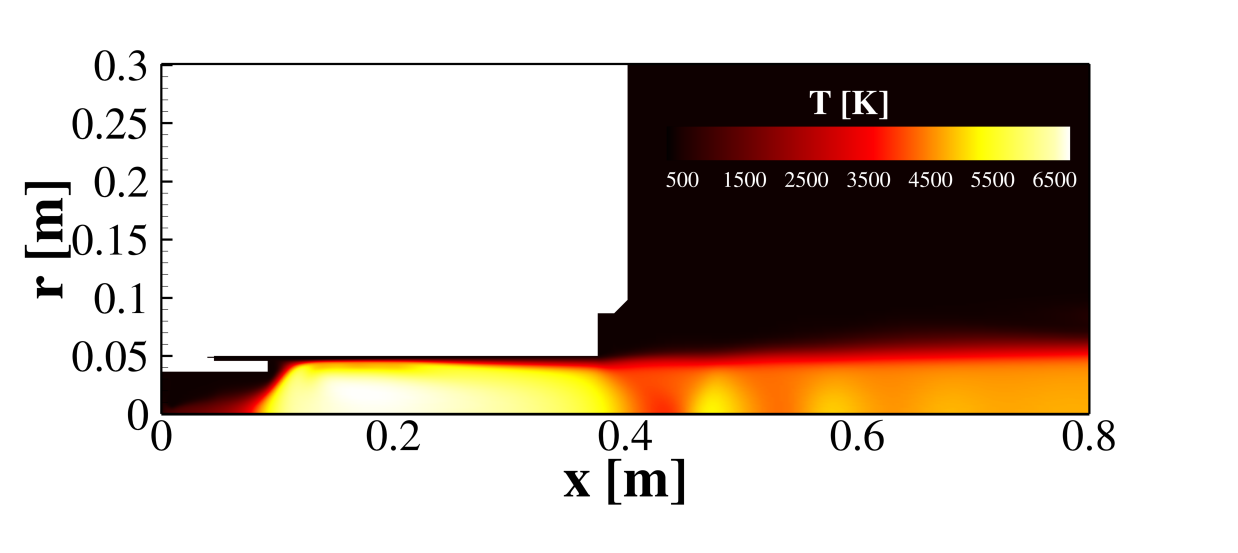}}
\subfloat[][\SI{200}{\kilo\watt}]{\includegraphics[trim={0 0.75cm 0 0},clip,scale=0.35]{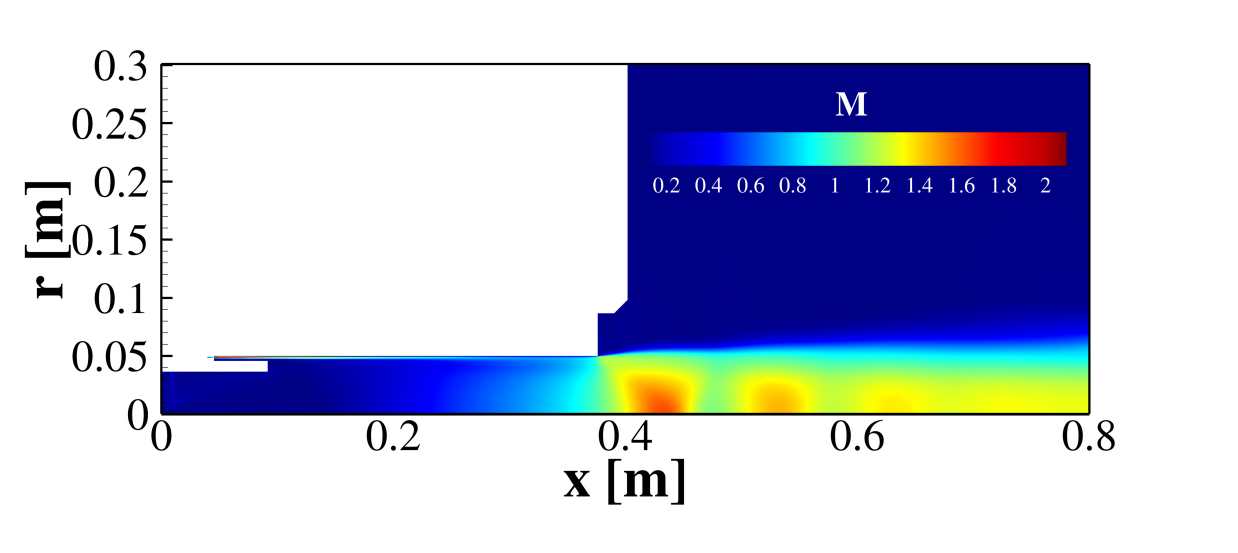}}\\
\subfloat[][\SI{250}{\kilo\watt}]{\includegraphics[trim={0 0.75cm 0 0},clip,scale=0.35]{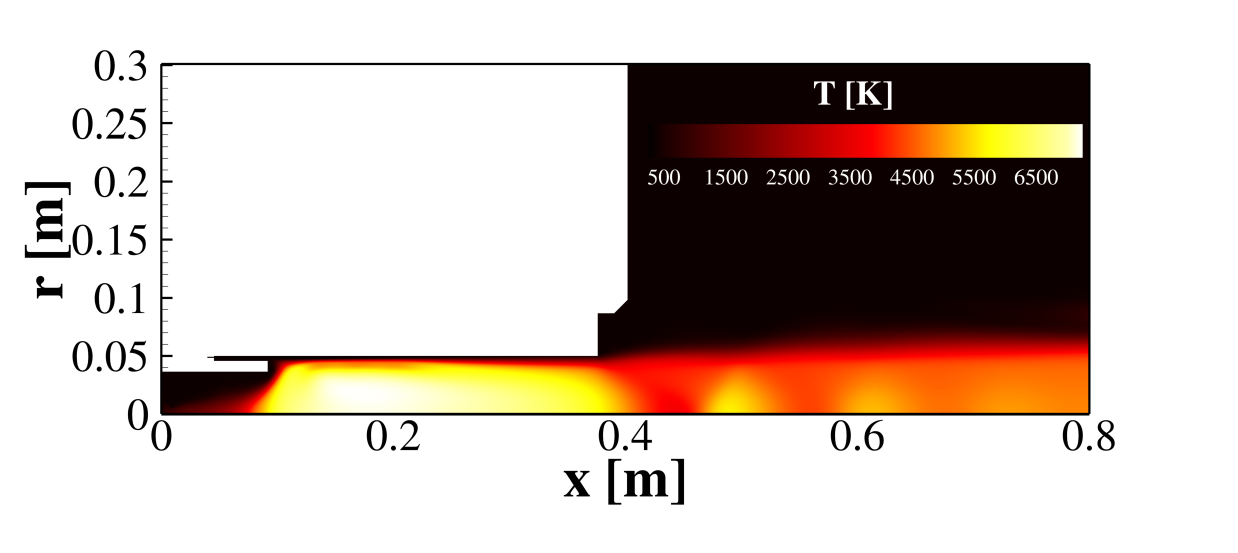}}
\subfloat[][\SI{250}{\kilo\watt}]{\includegraphics[trim={0 0.75cm 0 0},clip,scale=0.35]{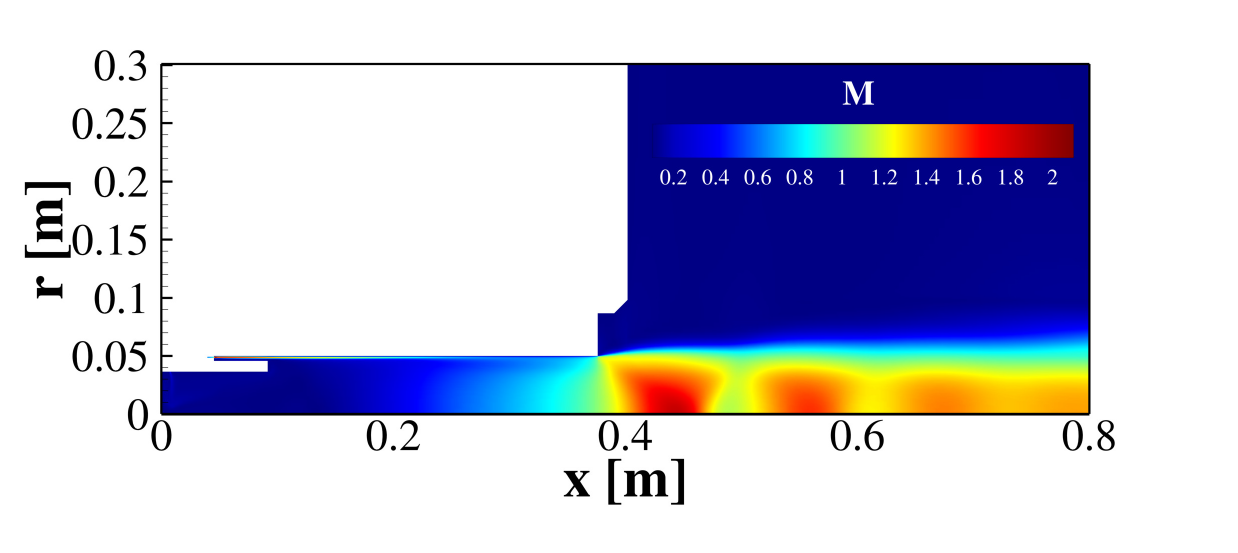}}\\
\subfloat[][\SI{300}{\kilo\watt}]{\includegraphics[trim={0 0.75cm 0 0},clip,scale=0.35]{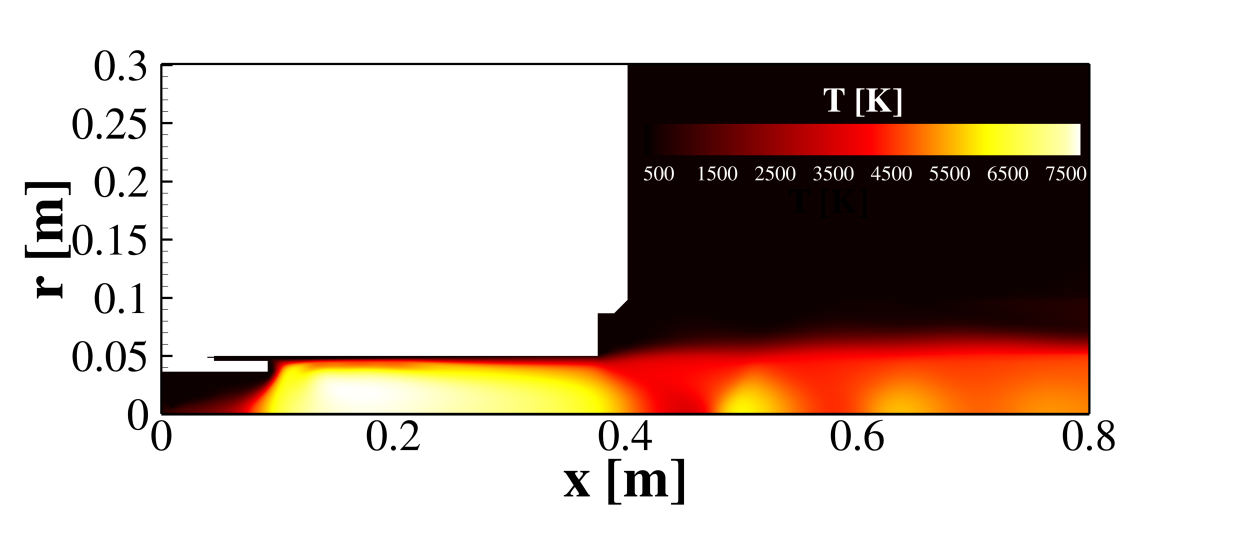}}
\subfloat[][\SI{300}{\kilo\watt}]{\includegraphics[trim={0 0.75cm 0 0},clip,scale=0.35]{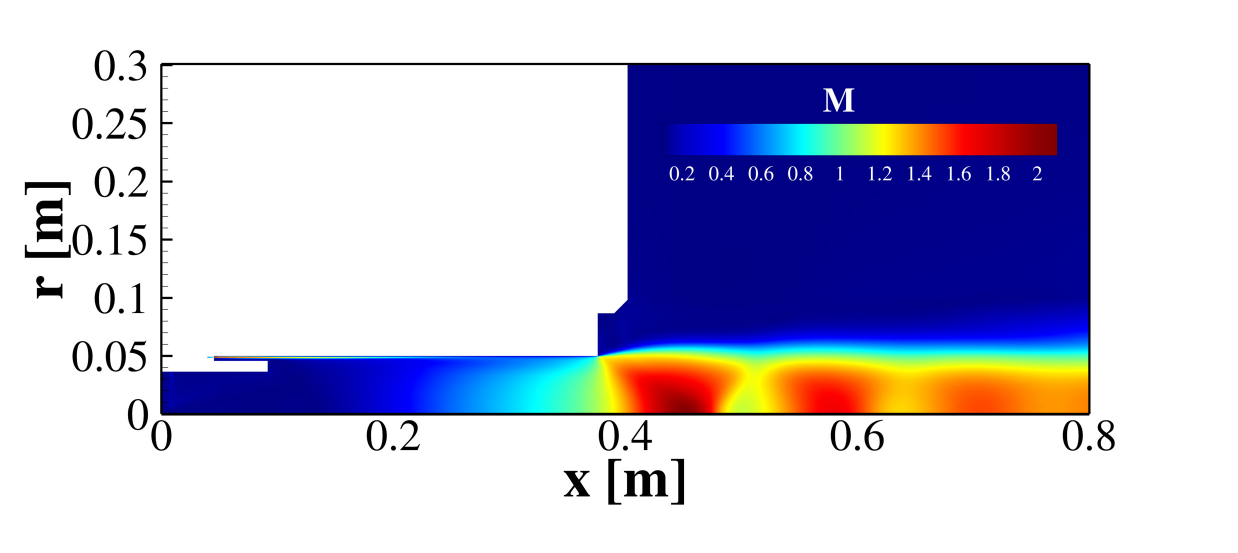}}
\caption{Heavy-particle temperature (a,c,e,g,i) and Mach number (b,d,f,h,j) distributions inside the facility at different power levels and fixed operating pressure and efficiency ($p_{\mathrm{a}} = \SI{600}{Pa}$ and $\eta = \SI{50}{\percent}$, respectively), highlighting the subsonic-supersonic transition of the plasma jet.}
\label{fig:T_M_distribution}
\end{figure}

\begin{figure}[!hbt]
\centering
\includegraphics[scale=0.6]{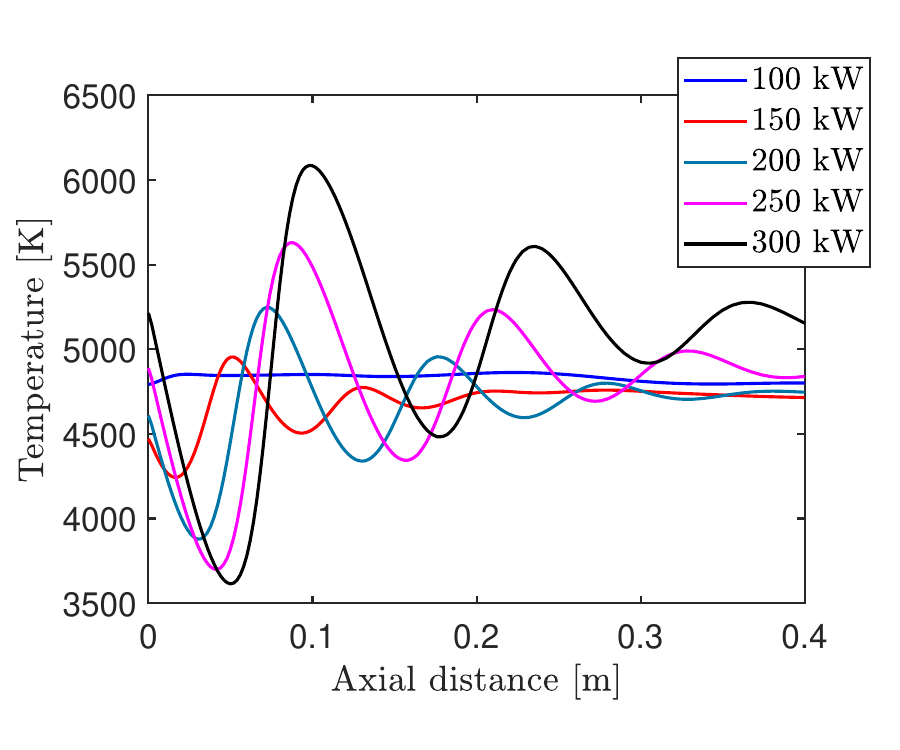}
\caption{Centerline heavy-particle temperature profiles at different power levels ($p_{\mathrm{a}} = \SI{600}{Pa}$). The axial distance is measured from the nozzle exit.}
\label{fig:T_axial_profiles}
\end{figure}

\cref{fig:T_M_distribution} reports the distributions of heavy-particle temperature and Mach number within the facility at different power levels. The Mach number contours show that the transition from subsonic to supersonic flow occurs around \SI{150}{kW}, a conclusion further corroborated by the shock-diamond structures in the temperature distribution. \cref{fig:T_axial_profiles} shows the centerline profiles of heavy-particle temperature for various power levels at \SI{600}{Pa} ambient pressure. The reported axial distance is measured from the nozzle exit. At \SI{100}{kW}, the absence of a temperature peak suggests a subsonic plasma jet. However, beginning at \SI{150}{kW}, distinct temperature peaks emerge, indicating the formation of shocks within the jet. At higher power levels, the number of peaks increases, and their positions progressively shift downstream. Additionally, both the spacing between successive shocks (\emph{i.e.,} distance between successive intensity peaks) and their intensity (reflected by peak amplitudes) grow with increasing power. These trends are consistent with the observations reported in ref. \cite{capponi2024multi}, where similar patterns were deduced from plasma light-intensity profiles. A comparison between the simulated and measured average shock spacings, presented in \cref{tab:shock_distance}, shows a good overall agreement. Given the strong correlation between radiative intensity and plasma temperature, such qualitative similarity in their profiles is expected. Thus, the numerical simulations demonstrate strong qualitative agreement with the experimental results.

It is worth mentioning that a quantitative comparison with the experimentally determined intensity profiles would require coupling the plasma dynamics and radiation. This effort is currently underway, with the ICP framework being integrated with the \textsc{MURP} radiation transport solver \cite{jo2023multi}. Moreover, ongoing efforts are focused on acquiring more quantitative plasma measurements, such as temperatures, number densities, and velocities, using advanced diagnostic techniques including Coherent Anti-Stokes Raman Scattering (CARS), Optical Emission Spectroscopy (OES), and Laser-Induced Fluorescence (LIF). These state-of-the-art diagnostics will allow for a more rigorous and quantitative validation of the simulation framework in the near future.

\subsection{Comparison against cold wall heat flux measurements}\label{sec:heat_flux_compare}
\begin{figure}[!htb]
        \centering
        \includegraphics[trim = 2.5in 0 2in 0, clip,scale=0.75]{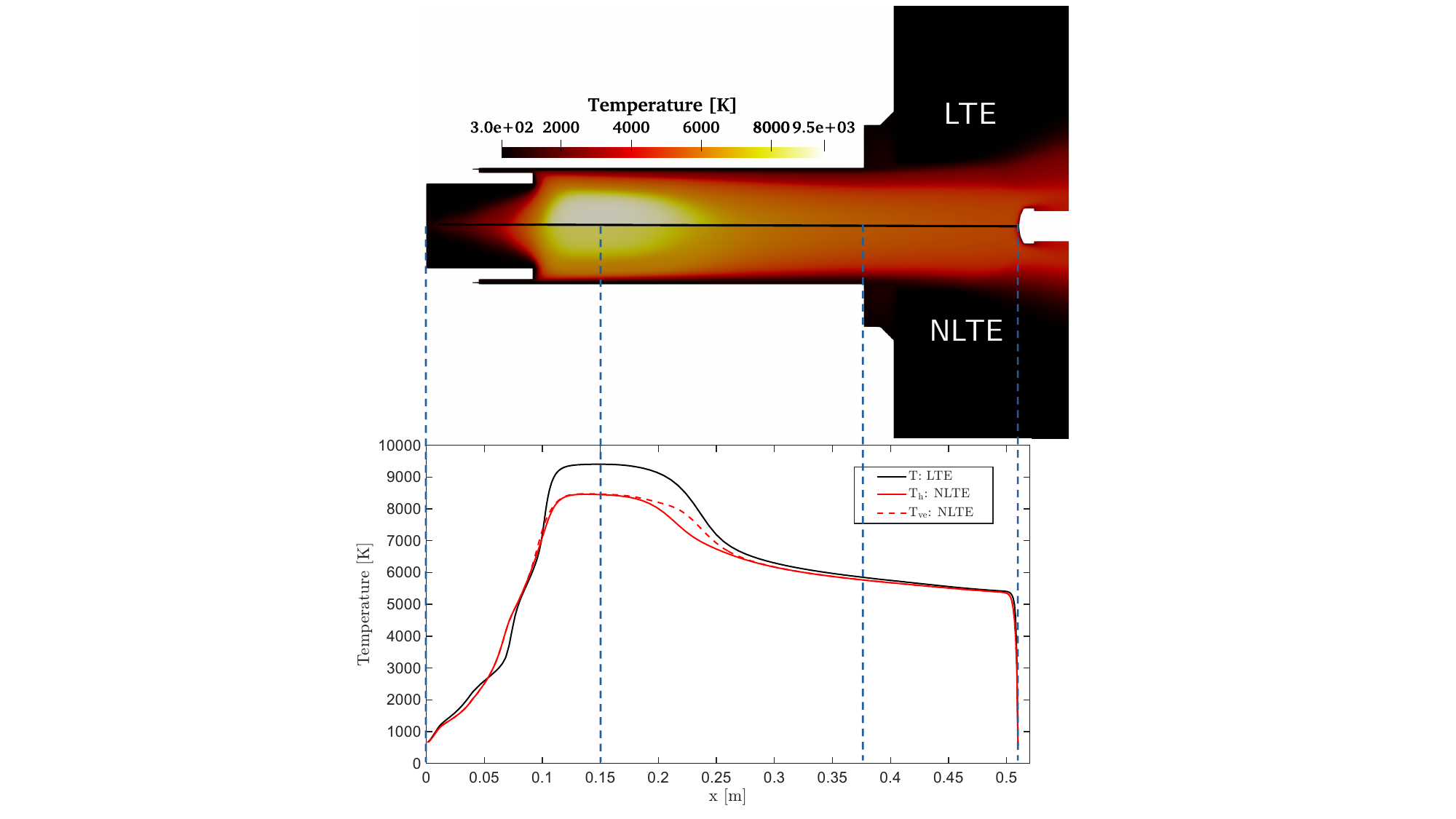}
        \caption{Comparison of the heavy-species temperature distribution and axial temperature profiles in the facility obtained using LTE and 2T NLTE simulations for Case 3 ($p_{\mathrm{a}} = \SI{5.5}{\kilo\pascal}$, $P = \SI{55}{\kilo\watt}$, $\eta = \SI{58.63}{\percent}$).} 
        \label{fig:lte_vs_nlte_contours}
    \end{figure}

This section provides a comparison of the cold wall heat flux predicted by the numerical simulations with the values measured using an in-house copper slug calorimeter probe. More details on the setup of the slug calorimeter can be found in ref. \cite{franco2024investigation}. The measurements used an isoQ \SI{30}{mm} diameter slug calorimeter (same as the TPS sample geometry discussed in \cref{sec:problem_setup}) placed at an axial distance of \SI{108}{mm} from the torch exit. \cref{tab:slug_cases_compare} provides the operating conditions for which the measurements are reported in this paper. It is worth noticing that the operating pressures in all three cases are sufficiently large to justify the use of an LTE plasma model. To verify the validity of the LTE assumption employed in this study, a two-temperature (2T) NLTE simulation was performed for Case 3, corresponding to the lowest pressure condition, and the results were compared with the baseline LTE predictions. \cref{fig:lte_vs_nlte_contours} presents a comparison of the gas (heavy-species) temperature fields obtained from both approaches. While minor discrepancies are observed within the torch, particularly in the coil region, the plasma structure in the downstream jet region remains nearly identical in both simulations. The figure also compares the axial temperature profiles obtained from the two simulations. The temperature profiles starting between x = 0 and x = \SI{0.1}{m} are nearly identical. Between x = \SI{0.1}{m} and \SI{0.3}{m}, where the influence of the RF coils is significant, the NLTE temperature profile differs from the LTE profile due to the non-equilibrium effect induced by the RF coils. The thermal non-equilibrium in this region is evident from the difference between the translational ($T_h$) and electro-vibrational ($T_{ve}$) temperatures. However, as soon as the plasma exits the torch region (around x = \SI{0.375}{m}), equilibrium between the translational and electro-vibrational modes starts prevailing, and the NLTE temperature profiles collapse with the LTE profile. This observation is consistent with several other references involving ICP studies \cite{kumar2024investigation},\cite{munafo2024self-consistent},\cite{zhang2016analysis}. Owing to the near-identical flowfields in the vicinity of the TPS sample, the predicted cold-wall heat fluxes from the two approaches are also in close agreement, with values of \SI{90.20}{W/cm^2} for the LTE case and \SI{88.10}{W/cm2^}
for the NLTE simulation (assuming a fully catalytic wall).

   \begin{table}[hbt!]
    
    \centering
    {\fontsize{10pt}{12pt}\selectfont
    \begin{tabular}{lcccccc}
    \hline\hline
    Case & Pressure & Power & Efficiency  & q [Measured] & q [Predicted] & Relative error  \\
    & (\si{\kilo\pascal}) & (\si{\kilo\watt}) & (\si{\percent}) & (\si{\watt/\centi\meter^2}) & (\si{\watt/\centi\meter^2}) & (\si{\percent}) \\
    \hline
    1 & 20 & 55 & 63.75 & 146.18 & 116.10 & 20.5 \\
    2 & 10 & 55 & 60.16 & 118.29 & 94.20 & 20.3 \\
    3 & 5.5 & 55 & 58.63 & 106.15 & 90.20 & 15.0\\
    \hline\hline
    \end{tabular}
    \caption{\label{tab:slug_cases_compare} Plasmatron X operating conditions for the cold wall heat flux measurements.}}
    \end{table}

        \begin{figure}[!htb]
        \centering
        \includegraphics[scale=0.5]{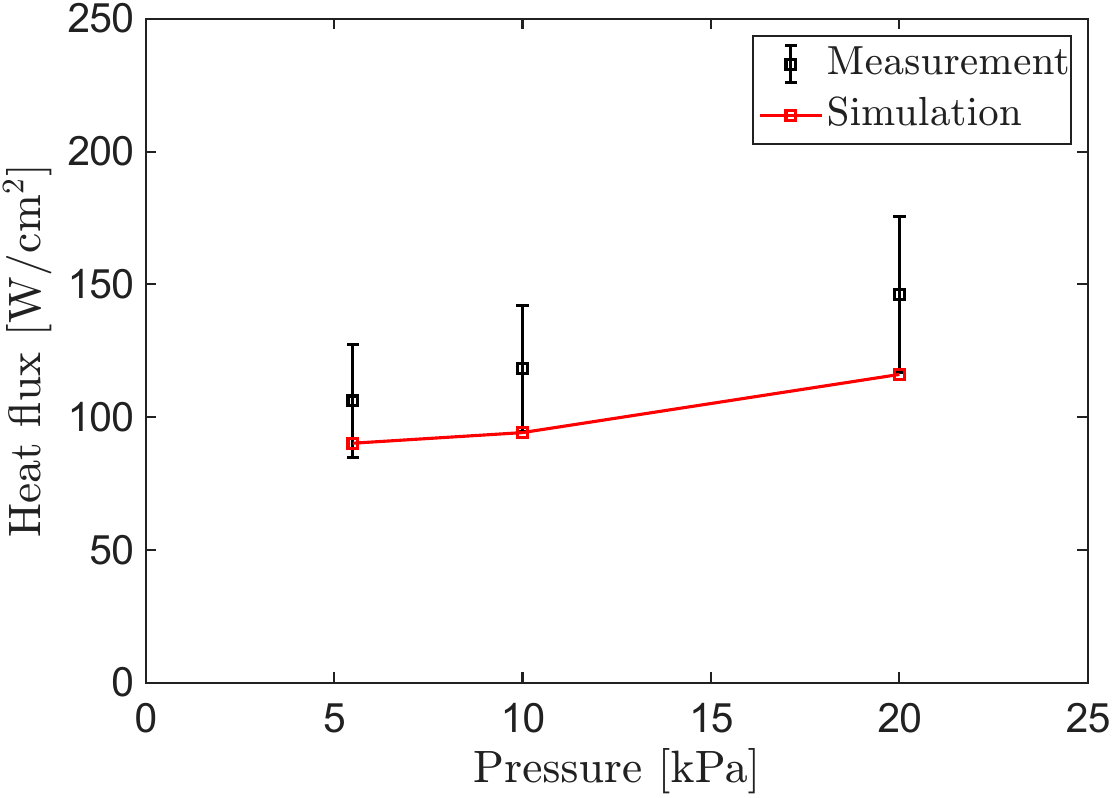}
        \caption{Stagnation cold wall heat flux at various pressure levels ($P =\SI{55}{kW}$). The measured values have been shown with 20\% error bars. } 
        \label{fig:q_vs_pressure}
    \end{figure}

\cref{fig:q_vs_pressure} shows a comparison of the cold wall stagnation heat flux at different pressures (corresponding to cases 1, 2, and 3) at \SI{55}{\kilo\watt} power. For all cases, the simulation results fall within the 20\% experimental error bars. It must be borne in mind that a \SI{20}{\percent} error is considered a good agreement as a result of the large variability to which heat flux measurements are subjected. In fact, in a slug calorimeter, the heat flux is not measured directly. Instead, it is deduced from the temporal gradient of the temperature recorded by a thermocouple mounted on the rear face of the slug. Consequently, multiple error sources must be rigorously assessed to ensure that the heat-flux estimation derived from this temperature history is reliable. For example, the inferred heat flux is sensitive to the slug diameter (as demonstrated in ref. \cite{meyers2022progress}), which introduces geometric uncertainties. Thermal losses at the interface between the slug and its holder also contribute to errors, since the theoretical model used to back-calculate heat flux assumes adiabatic conditions on all slug surfaces, except the front face exposed to the incident heat. Such idealized conditions are difficult to achieve in practice, and deviations can lead to systematic under- or over-predictions of the heat flux. In addition, the plasma jet impinging on the calorimeter typically exhibits a non-uniform spatial distribution that depends on operating conditions (\emph{e.g.}, mass flow, ambient pressure) and on the probe geometry itself. This non-uniformity can further distort the measured heat flux. Moreover, in an equilibrium boundary layer, as assumed in the numerical simulations, the surface catalycity of the sample has little influence on the heat flux because recombination occurs upstream, and the associated energy is transferred to the gas before reaching the wall. In contrast, for a non-equilibrium boundary layer, the recombination and diffusion time scales are comparable, making the heat flux strongly dependent on surface catalicity (a property that exhibits large variability). It is also important to note that absorptivity, catalycity, and the convective contribution to the heat flux all depend on the instantaneous surface temperature. Therefore, the net heat flux can vary significantly during tests with relatively long exposure durations. A detailed quantification of these error sources is given in ref. \cite{franco2024investigation}. By contrast, the heat flux obtained from numerical simulations is computed directly from the local spatial temperature gradient, bypassing the indirect estimation procedure used in experiments. Therefore, exact agreement between the two approaches is unlikely until the dominant experimental uncertainties are fully resolved and more consistent measurement–simulation comparisons are established, an area of ongoing research. Nevertheless, the reported comparisons show that the overall heat flux trends are reasonably well captured, which is encouraging and motivates further complex numerical analysis of TPS sample testing in the Plasmatron X facility.

       \begin{figure}[!htb]
        \centering
        \includegraphics[scale=0.5]{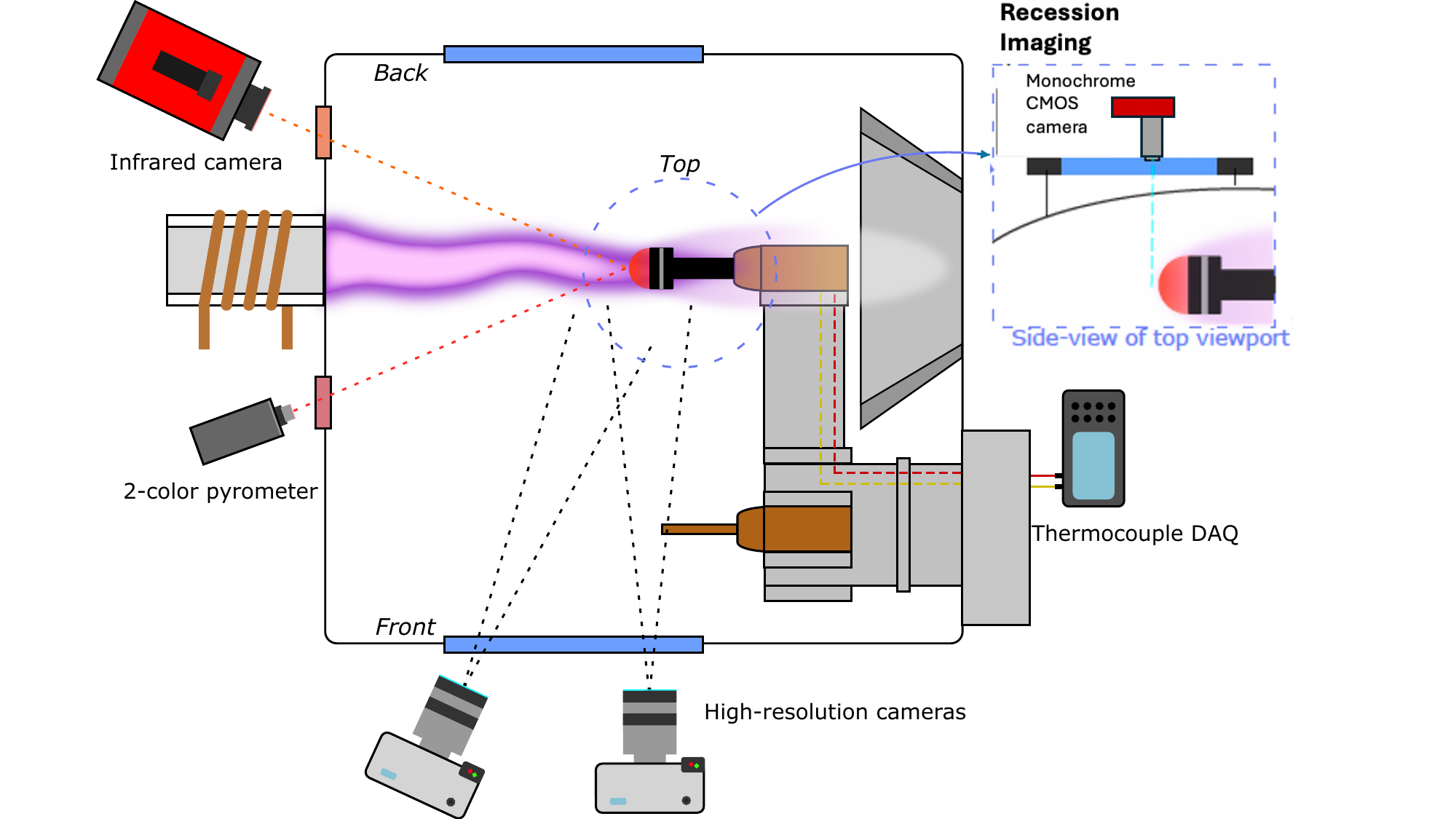}
        \caption{Schematic top-view of the Plasmatron X chamber depicting the available optical accesses and the in situ diagnostic suite \cite{oruganti5740826depth}. } 
        \label{fig:exp_setup}
    \end{figure}

\subsection{Material response modeling in Plasmatron X environment}
Coupled ablation simulations, using the framework discussed in \cref{sec:coupling_framework}, have been performed for TPS samples tested in the Plasmatron X facility and compared with the corresponding experimental data. The experiments utilized an isoQ \SI{30}{mm} graphite (Mersen 2340) sample placed at an axial distance of \SI{108}{mm} from the torch exit, for which the temporal evolution of stagnation-point surface temperature and recession was measured. The set of operating conditions remains the same as in \cref{tab:slug_cases_compare}.

\subsubsection{Experimental setup}\label{sec:exp_setup}
\begin{figure}[!htb]
        \centering
        \includegraphics[trim={0 3cm 0 3cm},clip,scale=0.45]{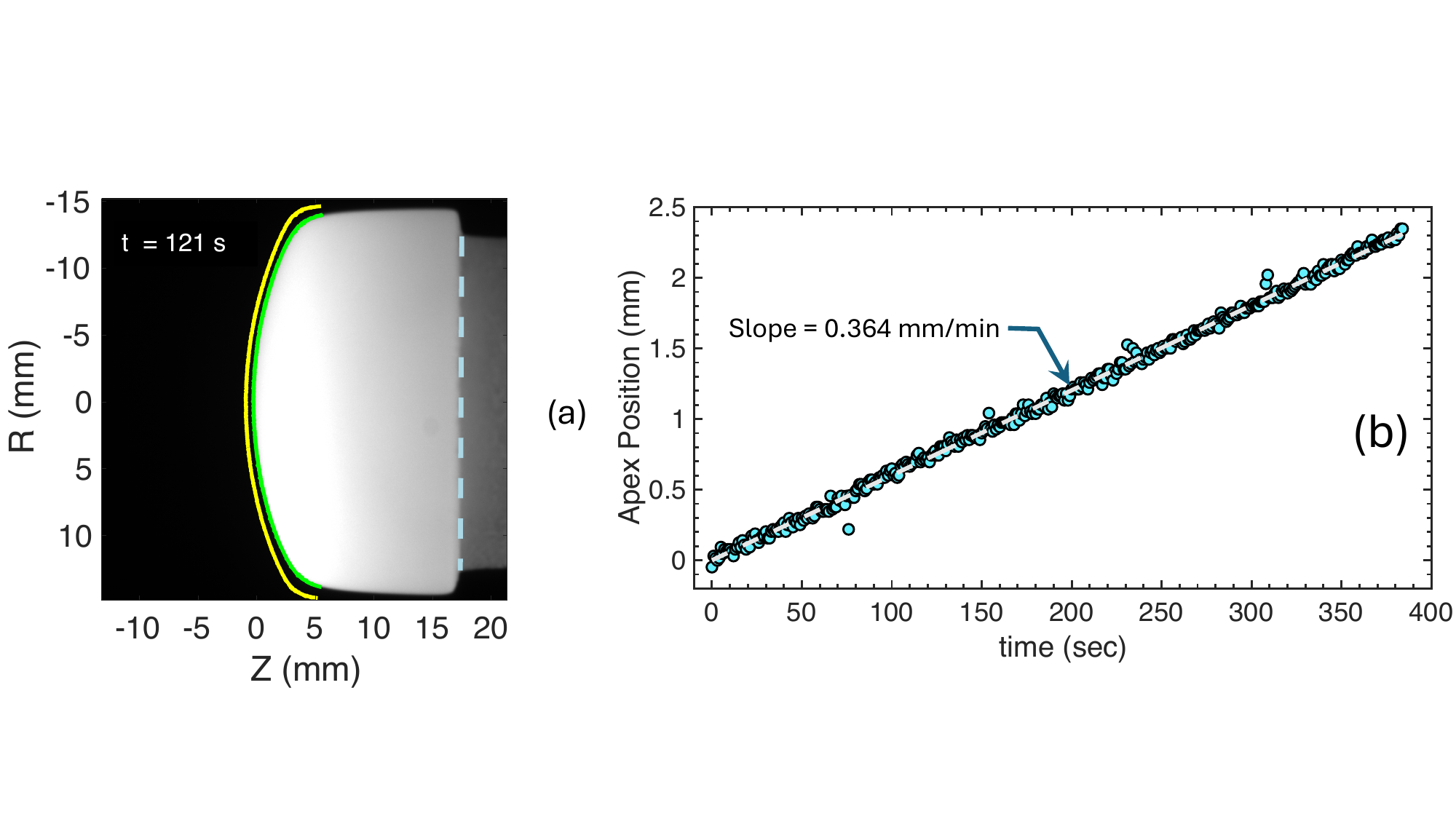}
        \caption{(a) High-resolution thermal emission image of the graphite hemisphere during ablation experiments in Plasmatron X. The yellow curve indicates the unablated sample shape, while the green curve is the updated sample edge position, and the blue dashed curve indicates the interface between the sample backface and the graphite holder. (b) Least squares fit to the sample apex position near r = 0; the slope indicates the measured ablation rate.} 
        \label{fig:recession_measurement}
    \end{figure}

\cref{fig:exp_setup} shows the simplified schematic of the Plasmatron X chamber along with the typically used instrumentation suite \cite{oruganti5740826depth}, although not all instruments were available for the present tests. 
Graphite ablation rate was characterized by tracking the edge of the hemispherical iso-Q samples in high-contrast thermal emission images acquired using a high-resolution CMOS camera with 2.2 $\mu$m pixel size at 5 MegaPixel format. The CMOS detector was mounted atop the Plasmatron X main vacuum chamber, roughly \SI{1.5}{m} from the sample location and fitted with a Nikon \SI{105}{mm} lens, which was focused at the sample midplane at f/2.8. Thermal emission from the graphite was imaged at a projected pixel size of 7.88 $\mu$m at the sample plane, with a representative image shown in \cref{fig:recession_measurement} (a). Graphite recession rate was quantified using an edge-finding routine to locate the face of the iso-Q sample during our ablation experiments. Significant thermal emission from the hemispherical samples provides high-contrast edges required for tracking of the ablating surface. Lower emission levels are observed from the graphite holder at the sample back, where thermal contact resistance from the threaded connection and, presumably, lower levels of convective heat transfer in the sample wake result in a lower holder temperature. The interface between the holder and sample was readily identified throughout each measurement, with minimal change in this reference position when the sample was at steady state temperature. The unablated surface shape in \cref{fig:recession_measurement} (a) is indicated by the yellow curve, while the ablating surface is indicated in green. Sample apex position was obtained with sub-pixel precision by fitting a quadratic polynomial to the edge surface data near r = 0. The time history of the sample apex was then fit to a least-squares line, whose slope indicated the recession rate, with the result at \SI{55}{kW} operating power and \SI{20}{kPa} static pressure shown in \cref{fig:recession_measurement} (b). Two Hietronics two-color pyrometers covering different temperature ranges are synchronized together and aligned with the stagnation point on the sample, simultaneously recording the temperature history at 10 Hz. Together, these diagnostics enabled time-resolved characterization of stagnation-point recession and surface temperature evolution under plasma exposure. More details regarding plasma diagnostics in the Plasmatron X facility, including sample mounting and preparation, can be found in References \cite{ringel2025unsteady,puttini2026orthographic,oruganti5740826depth}.

\subsubsection{Material properties}\label{sec:mat_prop}
\begin{figure}[!htb]
\centering
\subfloat[][]{\includegraphics[scale=0.5]{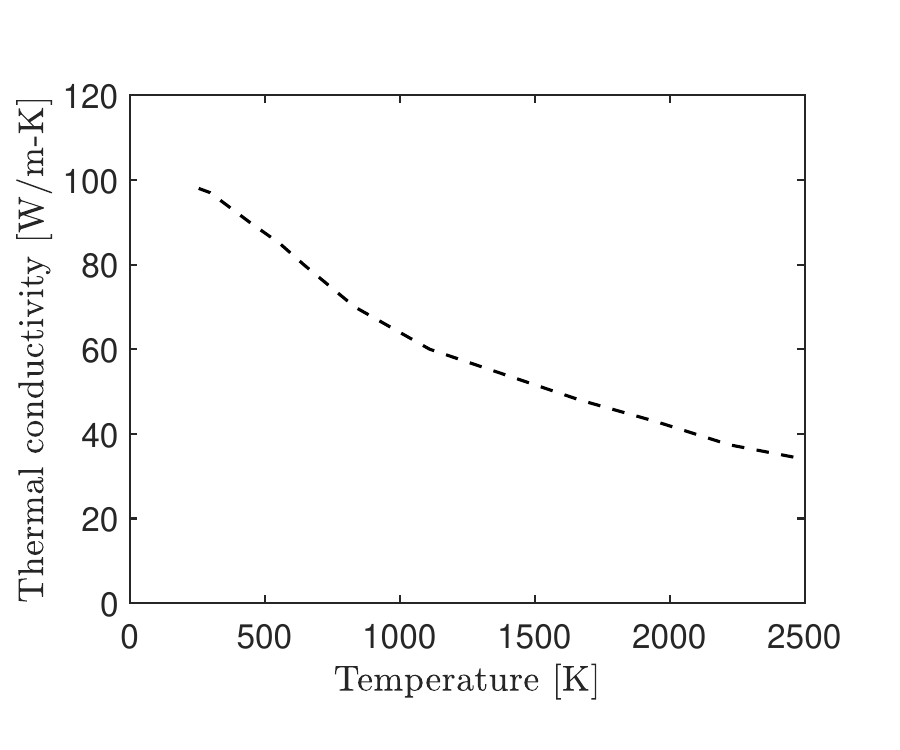}}
\subfloat[][]{\includegraphics[scale=0.5]{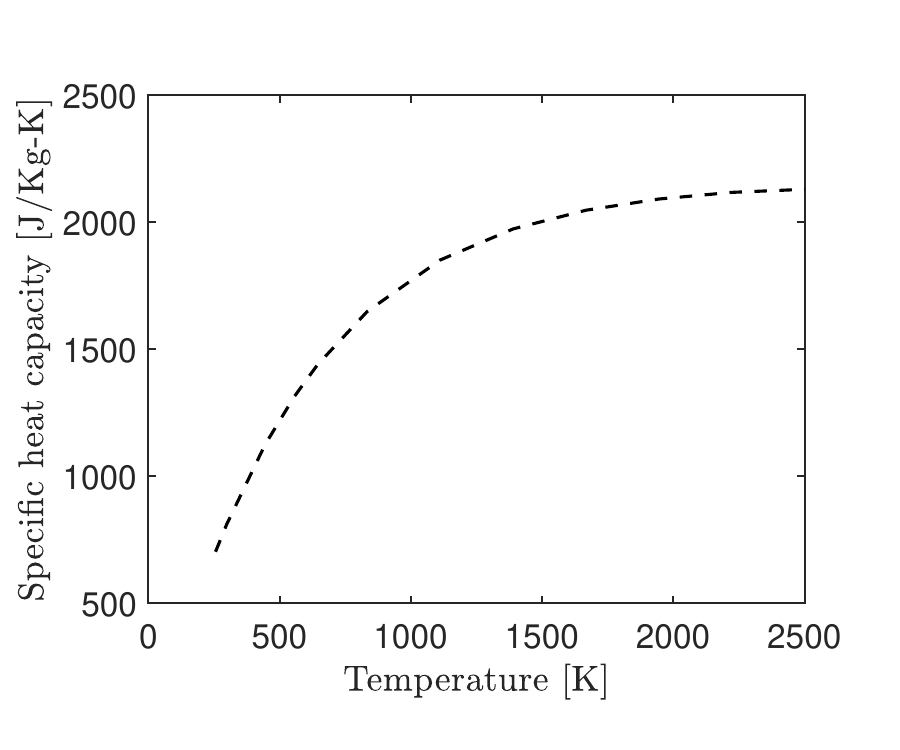}}
\caption{Thermal properties of Mersen 2160 graphite used in this work: (a) thermal conductivity,  (b) heat capacity per unit mass.}
\label{fig:mersen_properties}
\end{figure}

The graphite sample simulations in this work employ the material properties of Mersen 2160, modeled with isotropic, temperature-dependent behavior and a constant density of \SI{1.84}{\gram/\centi\meter^3}. \cref{fig:mersen_properties} shows the temperature dependence of thermal conductivity and heat capacity for the graphite sample. The surface emissivity is assumed to remain constant and equal to 0.85. It should be noted that the simulations employ material properties corresponding to a different graphite grade (Mersen 2160) than those used in the experiments (Mersen 2340), due to the unavailability of the temperature-dependent data for the latter. Nevertheless, the two grades are expected to exhibit sufficiently similar properties to justify the use of Mersen 2160 data in the simulations. \cref{tab:mersen_prop} lists the properties of two grades, showing only minor differences in various properties \cite{mersen_2160,mersen_2340}. However, uncertainties remain on how much the properties (such as thermal conductivity and specific heat capacity) may differ as the temperature of the material increases, and the effect of these uncertainties on material response has been investigated in \cref{sec:discussion_on_ablation}.

 \begin{table}[hbt!]
    
    \centering
    \begin{tabular}{lcccc}
    \hline\hline
    Property & Units & Mersen 2160 & Mersen 2340 \\
    \hline
    Density & $g/cm^3$ & 1.84 & 1.88\\
    Porosity & \% & 7 & 13 \\
    Thermal conductivity (room temperature) & W/m-K & 84 & 102 \\
    grain size & $\mu m$ & 5 & 4 \\
    \hline\hline
    \end{tabular}
    \caption{\label{tab:mersen_prop} Properties of Mersen grades.}
    \end{table}

 \begin{table}[hbt!]
    
    \centering
    \begin{tabular}{lcc}
    \hline\hline
    Time (\si{\second}) & Coupling window (\si{\second}) \\
    \hline
    0 & 0.1 \\
    5 & 1 \\
    50 & 5 \\
    \hline\hline
    \end{tabular}
    \caption{\label{tab:coupling_window} \textsc{preCICE} coupling window schedule between \textsc{HEGEL} and \textsc{CHyPS}.}
    \end{table}

\subsubsection{Comparison with numerical simulations}    
Coupled ablation simulations of the entire facility were conducted on the Frontera supercomputer at the University of Texas at Austin. The simulations used the Cascade Lake (CLX) compute nodes with an Intel Xeon Platinum 8280 CPU model, with 260 processors used for the flow solver, 40 processors for the EM solver, and 100 processors for the material solver.  

\begin{figure}[hbt!]
\centering
\includegraphics[trim={0 1in 0 1in},clip,scale=0.45]{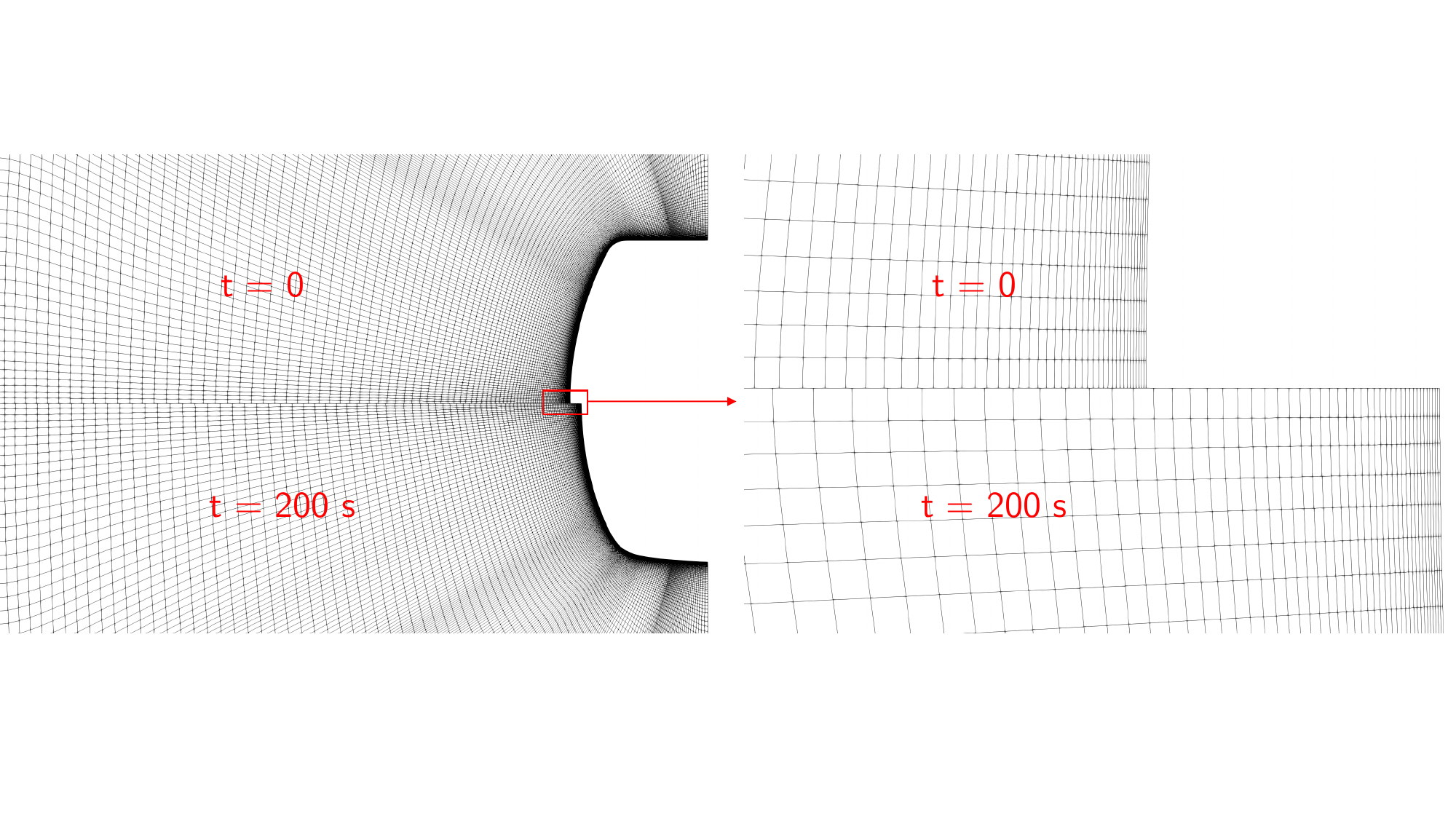}
\caption{Fluid solver mesh movement using LTI approach for Case 1 ($p_{\mathrm{a}} = \SI{20}{\kilo\watt}$, $P = \SI{55}{\kilo\watt}$, $\eta = \SI{63.75}{\percent}$).}
\label{fig:mesh_movement}
\end{figure}

Simulations were run for \SI{200}{\second} for the TPS sample, while tracking the evolution of the stagnation temperature and surface recession. \cref{tab:coupling_window} shows the coupling window schedule used for data communication between \textsc{HEGEL} and \textsc{CHyPS}, which was found to be sufficient to ensure that the results are converged with respect to the coupling window schedule as discussed in \ref{appendix:coupling_convergence}. \cref{fig:t0_vs_t200} presents the plasma and sample temperature fields at $t = \SI{0}{\second}$ and $t = \SI{200}{\second}$ in the region surrounding the sample for Case 1. Initially, the sample is uniformly at \SI{300}{\kelvin}. As the simulation progresses, the surface temperature of the sample rises steadily as a result of the intense thermal loading from the plasma. This heating leads to material ablation, wherein the surface layers of the graphite sample gradually erode as a result of sublimation and mass loss under high temperatures. The effect of this surface recession is clearly visible in the contours at $t = \SI{200}{\second}$, showing a noticeable change in geometry. Additionally, the plasma boundary layer is observed to shift slightly outward from the sample surface, a hallmark of strongly coupled flow–material interactions. This displacement occurs because the ablated gaseous products emerging from the receding surface introduce a blowing effect, which locally alters the near-wall flow and modifies the plasma–surface heat and momentum transfer characteristics. It is to be noted that the Linear Transfinite Interpolation (LTI) method updates the positions of interior mesh nodes using weighting factors based on their relative distance from the boundary. Nodes located closer to the boundary experience displacements that closely follow the boundary motion, whereas nodes farther away exhibit progressively attenuated movement. As a result, the first cell adjacent to the wall preserves its thickness throughout the simulation due to the strong influence of the nearby boundary motion. With increasing distance from the boundary, the displacement of interior nodes diminishes smoothly and eventually approaches zero far from the boundary. Hence, LTI ensures that the spatial accuracy of the heat flux remains consistent throughout the simulation. \cref{fig:mesh_movement} compares between the original mesh at t = 0 and the deformed mesh at t = \SI{200}{s} for Case 1, confirming that the near-wall cell thickness remains unchanged throughout the simulation time.

\begin{figure}[!htb]
\centering
\subfloat[][]{\includegraphics[scale=0.5]{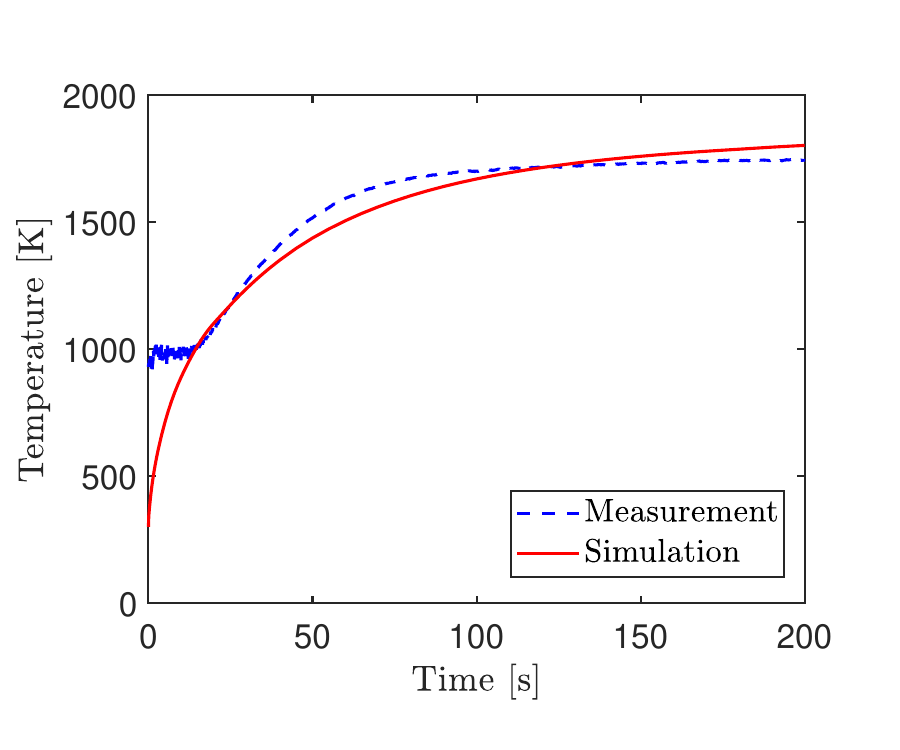}}
\subfloat[][]{\includegraphics[scale=0.5]{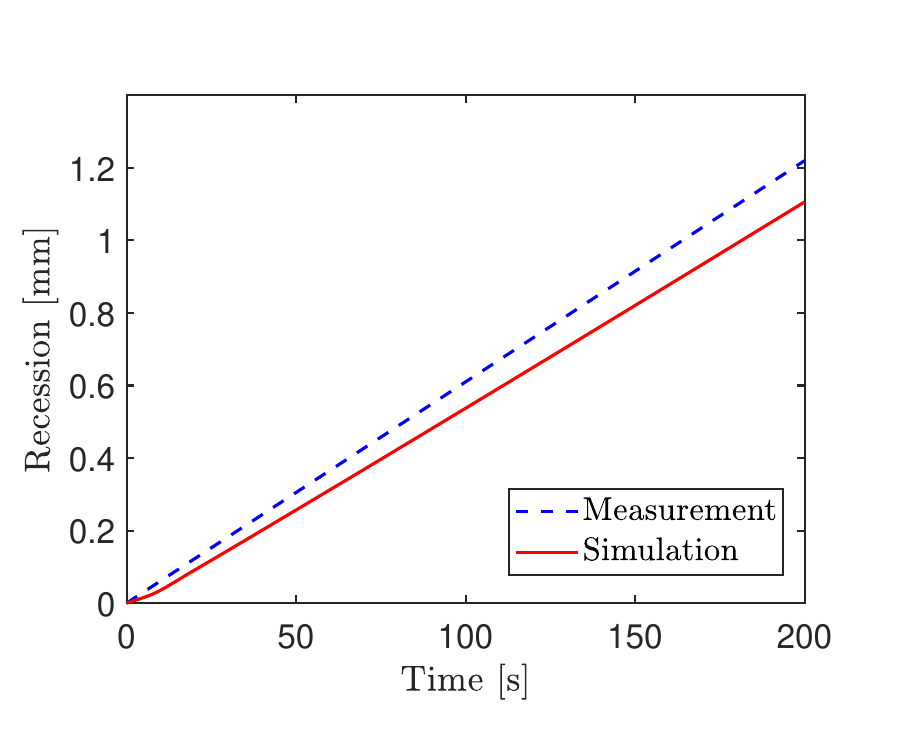}}
\caption{Comparison of the predicted and measured surface stagnation temperature and recession for Case 1 ($p_{\mathrm{a}} = \SI{20}{\kilo\pascal}$, $P = \SI{55}{kW}$, $\eta = \SI{63.75}{\percent}$).}
\label{fig:T_recession_20kPa}
\end{figure}

\begin{figure}[!htb]
\centering
\subfloat[][]{\includegraphics[scale=0.5]{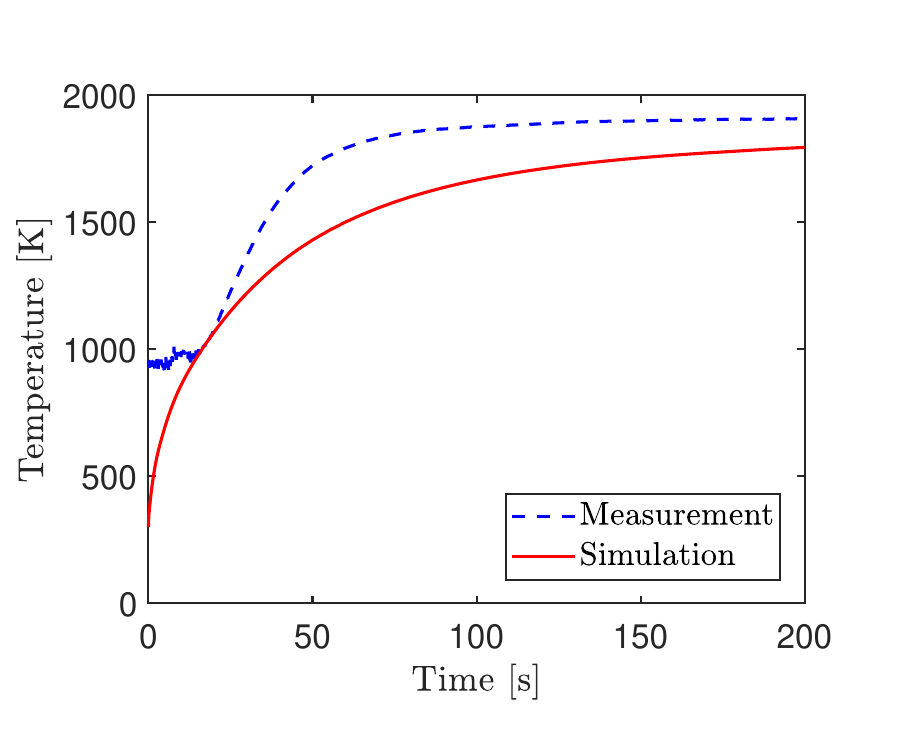}}
\subfloat[][]{\includegraphics[scale=0.5]{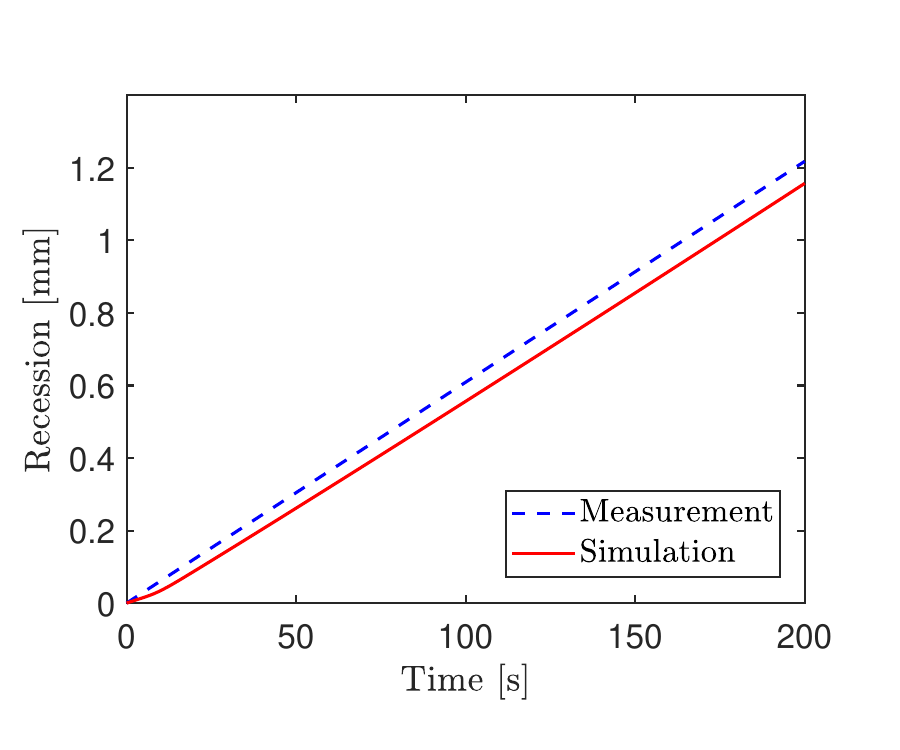}}
\caption{Comparison of the predicted and measured surface stagnation temperature and recession for Case 2 ($p_{\mathrm{a}} = \SI{10}{\kilo\pascal}$, $P = \SI{55}{\kilo\watt}$, $\eta = \SI{60.16}{\percent}$).}
\label{fig:T_recession_10kPa}
\end{figure}

\begin{figure}[!htb]
\centering
\subfloat[][]{\includegraphics[scale=0.5]{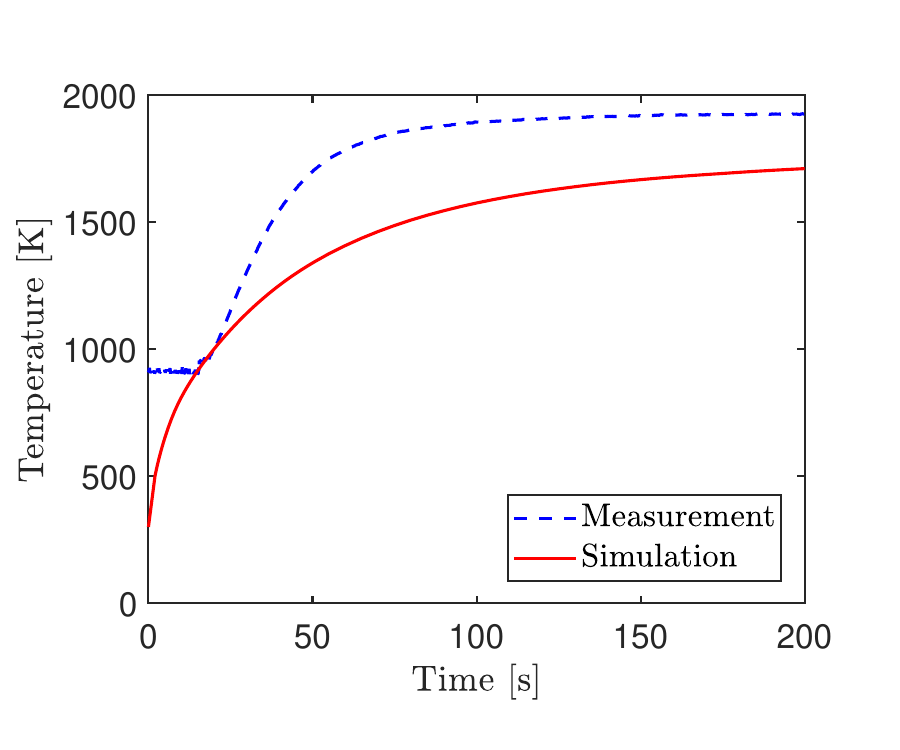}}
\subfloat[][]{\includegraphics[scale=0.5]{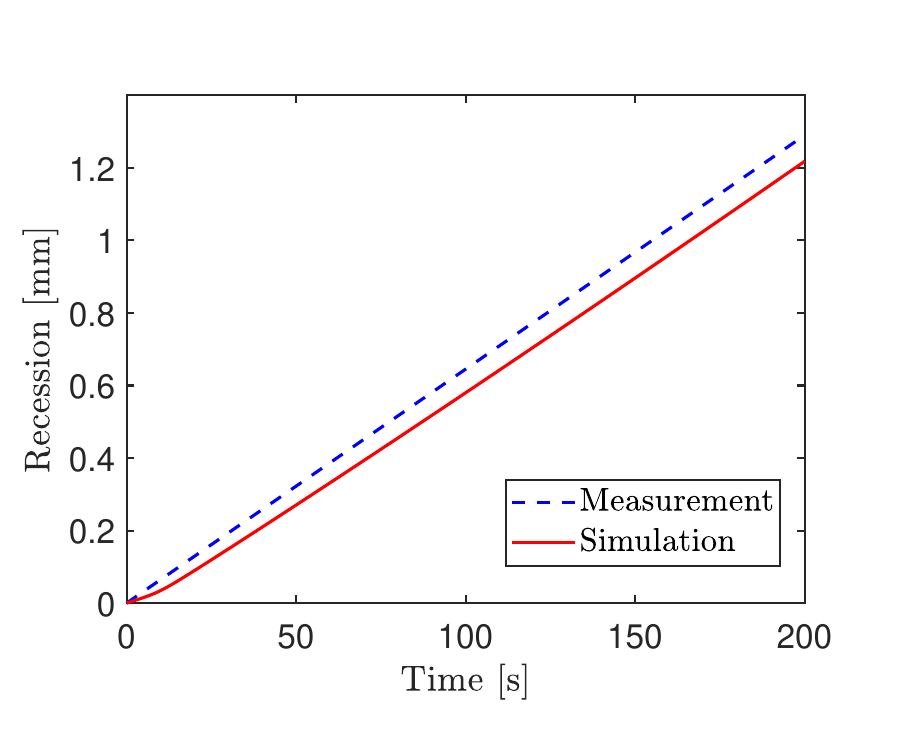}}
\caption{Comparison of the predicted and measured surface stagnation temperature and recession for Case 3 ($p_{\mathrm{a}} = \SI{5.5}{\kilo\pascal}$, $P = \SI{55}{\kilo\watt}$, $\eta = \SI{58.63}{\percent}$).}
\label{fig:T_recession_5kPa}
\end{figure}

\cref{fig:T_recession_20kPa} through \cref{fig:T_recession_5kPa} report the temporal evolution of the stagnation-point temperature and surface recession for all three cases. For Case 1, the predicted temperature evolution exhibits excellent agreement with experimental measurements throughout the entire duration. It is important to note that the pyrometers begin recording data only once the temperature exceeds approximately \SI{1000}{\kelvin}. Hence, the measured temperature curve remains flat during the initial phase. Beyond this point, the simulated temperature profile closely follows the experimental trend, with the relative deviation at the final time ($t = \SI{200}{\second}$) being only \SI{3.3}{\percent}. For cases 2 and 3, a noticeable discrepancy is observed during the early transient period, the possible source of which is addressed in the next section. Nevertheless, as the system approaches steady state, the predicted stagnation-point temperatures are in excellent agreement with the experimental data, with relative errors of \SI{5.9}{\percent} and \SI{11.2}{\percent} at $t = \SI{200}{\second}$, respectively. The surface recession histories for all three cases also demonstrate good consistency between simulations and experiments. The relative errors in the predicted steady-state recession rates remain below \SI{10}{\percent} in all cases, specifically \SI{9.83}{\percent}, \SI{4.60}{\percent}, and \SI{5.42}{\percent} for cases 1, 2, and 3, respectively. A detailed quantitative comparison of the stagnation-point temperature and surface recession rates for all three cases is given in \cref{tab:T_200s_compare,tab:disp_200s_compare}.

 \begin{table}[hbt!]
    
    \centering
    {\fontsize{10pt}{12pt}\selectfont
    \begin{tabular}{lcccccc}
    \hline\hline
    Case & Pressure & Power & Efficiency  & T [Measured] & T [Predicted] & Relative error  \\
    & (\si{\kilo\pascal}) & (\si{\kilo\watt}) & (\si{\percent}) & (\si{\kelvin}) & (\si{\kelvin}) & (\si{\percent}) \\
    \hline
    1 & 20 & 55 & 63.75 & 1745.1 & 1802.7 & 3.3 \\
    2 & 10 & 55 & 60.16 & 1908.4 & 1794.8 & 5.9 \\
    3 & 5.5 & 55 & 58.63 & 1927.3 & 1711.3 & 11.2\\
    \hline\hline
    \end{tabular}
    \caption{\label{tab:T_200s_compare} Summary of the stagnation point temperature comparison with experimental data at $t = \SI{200}{\second}$.}}
    \end{table}

     \begin{table}[hbt!]
    
    \centering
    {\fontsize{10pt}{12pt}\selectfont
    \begin{tabular}{lcccccc}
    \hline\hline
    Case & Pressure & Power & Efficiency  & Recession rate  & Recession rate  & Relative error  \\
    & (\si{\kilo\pascal}) & (\si{\kilo\watt}) & (\si{\percent}) & [Measured] (\si{\micro\meter}/s) & [Predicted] (\si{\micro\meter}/s) & (\si{\percent}) \\
    \hline
    1 & 20 & 55 & 63.75 & 6.10 & 5.50 & 9.83 \\
    2 & 10 & 55 & 60.16 & 6.08 & 5.80 & 4.60 \\
    3 & 5.5 & 55 & 58.63 & 6.45 & 6.10 & 5.42 \\
    \hline\hline
    \end{tabular}
    \caption{\label{tab:disp_200s_compare} Summary of the stagnation point recession rate comparison with experimental data. }}
    \end{table}

\begin{figure}[hbt!]
\centering
\includegraphics[trim={0 0 5cm 0},clip,scale=0.5]{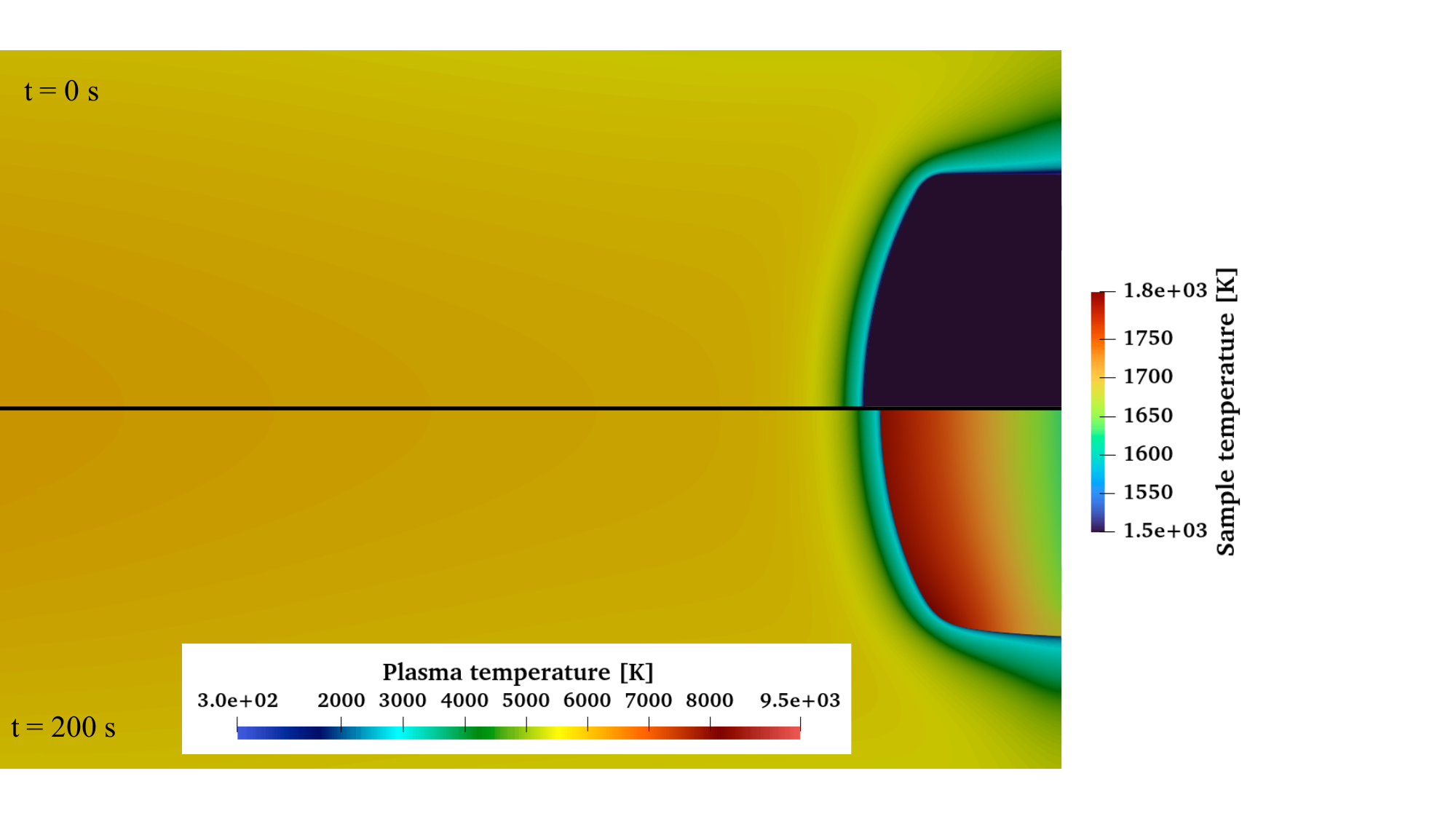}
\caption{Plasma and sample temperature fields at $t = 0$ and $t = \SI{200}{\second}$ for Case 1 ($p_{\mathrm{a}} = \SI{20}{\kilo\watt}$, $P = \SI{55}{\kilo\watt}$, $\eta = \SI{63.75}{\percent}$).}
\label{fig:t0_vs_t200}
\end{figure}

\subsection{Discussion on coupled ablation results}\label{sec:discussion_on_ablation} 
The extent of agreement observed in the ablation predictions (within a relative error of around \si{10\percent}) is particularly remarkable given the \textit{ab initio} nature of the simulation framework used in this work, which is driven exclusively by independently measured facility operating conditions such as input power, chamber pressure, and system efficiency. Crucially, the framework achieves this level of agreement without any form of parameter tuning, empirical adjustment, or ad-hoc calibration. This approach differs somewhat from practices reported in the literature, where key parameters, such as coupling efficiency or effective power, are sometimes treated as adjustable quantities and tuned to improve agreement with experimental observations \cite{mcclernan2025finite,fagnani2020investigation}. A similar reliance on experimentally prescribed boundary conditions is common in arc-jet simulations, where measured nozzle-exit flow profiles are imposed directly as inlet conditions, thereby bypassing the need to model the facility itself \cite{zibitsker2023finite}. In contrast, the present framework preserves the predictive integrity of the simulations by relying solely on physics-based models and measured inputs. By reducing reliance on such dependencies, the proposed methodology offers the potential for more predictive modeling of ICP facilities, facilitating improved assessment of material response and ablation behavior with reduced dependence on empirical inputs.

However, noticeable discrepancies arise during the early transient phase of the stagnation-point temperature evolution for cases 2 and 3, indicating that the model does not fully capture the initial thermal response. In contrast, the steady-state temperatures are predicted with relatively good agreement, exhibiting only a slight degree of underprediction. A similar trend is observed in the recession rates, where all three cases demonstrate modest underestimation compared to the reference data. These observations suggest that, while the overall predictive capability of the model is reasonable, there remain underlying sources of uncertainty that warrant a more detailed and systematic investigation. Given the multi-solver coupled nature of the present framework, which simultaneously resolves plasma flow, electromagnetic power deposition, and material response, such discrepancies are not unexpected. Small uncertainties introduced at individual modeling stages can propagate through the coupled system and modestly influence the predicted thermal response. The dominant contributors to these deviations include uncertainties in the facility coupling efficiency, limited availability of material property data, and simplified assumptions in the surface ablation model. Surface ablation in this work is modeled using an equilibrium $\mathrm{B}^{\prime}$-table formulation, which, while well-suited for steady-state conditions and high-pressure operation, may not fully capture transient finite-rate effects during the onset of ablation. A recent study on graphite ablation demonstrated that a finite-rate air–carbon ablation model provides more accurate predictions of material response compared to the equilibrium $B^{\prime}$ ablation model \cite{zibitsker2023finite}. However, the incorporation of a finite-rate formulation lies beyond the scope of the present numerical framework and is therefore identified as a key direction for future work stemming from this study. Accordingly, this subsection only examines how uncertainties in system efficiency measurements and material property characterization may affect the predicted material response.

\begin{figure}[!htb]
\centering
\subfloat[][]{\includegraphics[scale=0.4]{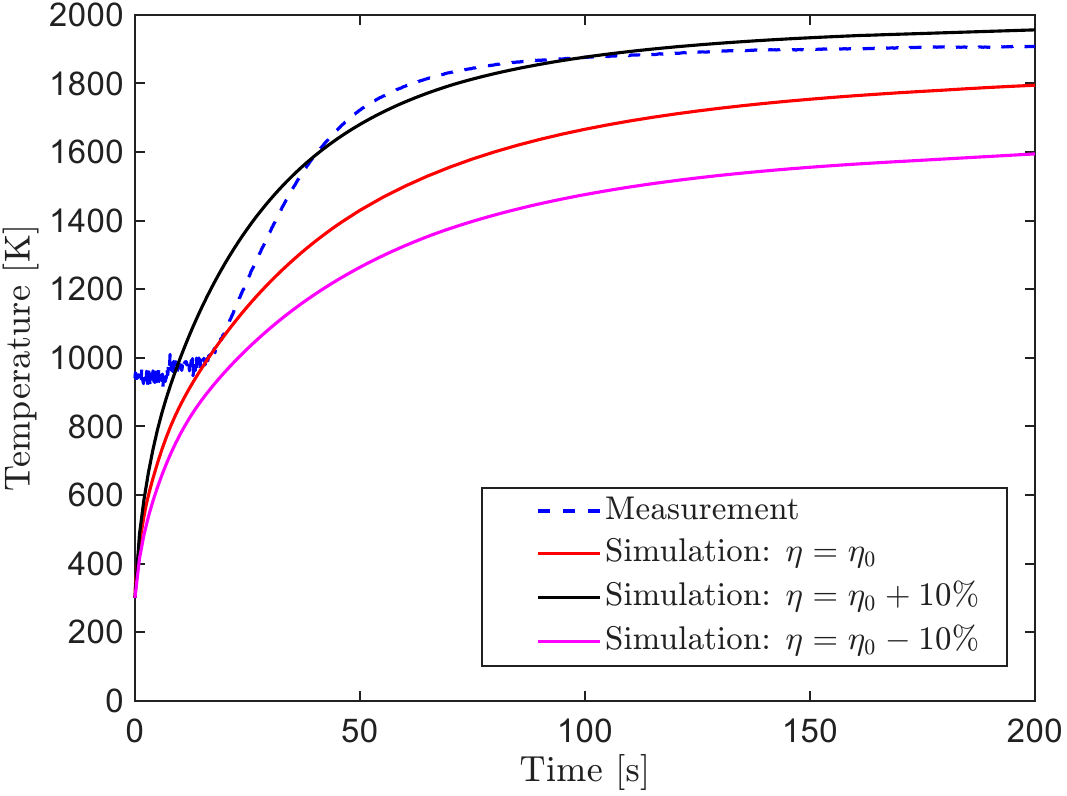}}\hspace{0.5cm}
\subfloat[][]{\includegraphics[scale=0.4]{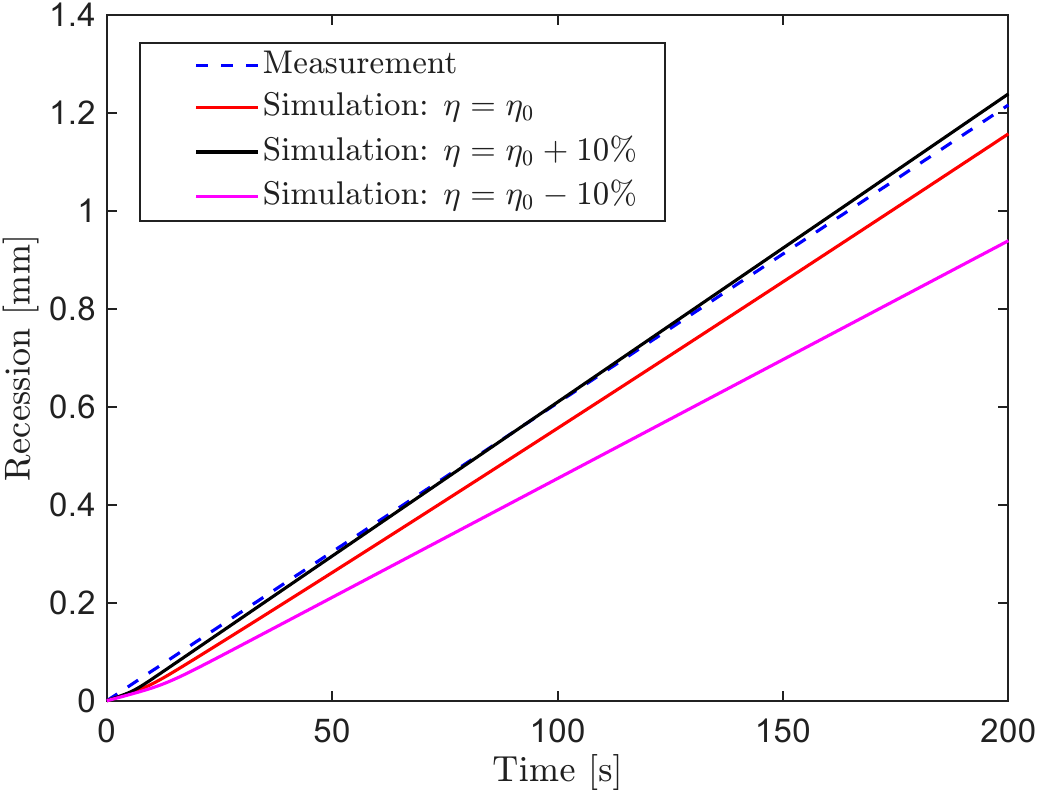}}
\caption{Comparison of the predicted and measured surface stagnation temperature and recession for Case 2 ($p_{\mathrm{a}} = \SI{10}{\kilo\pascal}$, $P = \SI{55}{kW}$, $\eta_0 = \SI{60.16}{\percent}$).}
\label{fig:T_recession_10kPa_eta}
\end{figure}

The coupling efficiency, which governs the fraction of radio-frequency (RF) generator power transferred to the plasma, is inherently difficult to quantify and may vary with operating conditions, leading to modest uncertainty in the effective plasma energy input. In particular, measured efficiencies in the Plasmatron X facility have been found to generally vary in the range of 50-70\%. To quantify the impact of uncertainty in the measured system efficiency on the predicted material response, additional simulations were performed for Case 2 (10 kPa) by varying the efficiency by $\pm10\%$ about the baseline value, $\eta_0 = 60.16\%$. \cref{fig:T_recession_10kPa_eta} presents a comparison of the surface stagnation temperature and recession for these perturbed efficiency values. The results clearly indicate that even a 10\% variation in efficiency can produce substantial changes in both surface temperature and recession behavior. Specifically, a 10\% increase in efficiency leads to an 8.99\% rise in the surface stagnation temperature (at t = \SI{200}{s}) and an 8.62\% increase in the recession rate. Conversely, a 10\% decrease in efficiency results in reductions of 11.15\% and 17.24\% in the surface temperature and recession rate, respectively. Moreover, increasing the efficiency by 10\% moved the temperature and recession values much closer to the measurements. \cref{tab:T_recession_compare_eta} summarizes the extent to which uncertainties in system efficiency propagate into the predicted material response for Case 2.

 \begin{table}[hbt!]
    
    \centering
    {\fontsize{9pt}{12pt}\selectfont
    \begin{tabular}{lcccccc}
    \hline\hline
    Efficiency  & T [Predicted] & Recession rate [Predicted] & Relative error  & Relative error\\
    (\si{\percent}) & (\si{\kelvin}) & (\si{\micro\meter}/s) &  $|\frac{T(\eta) - T(\eta_0)}{T(\eta_0)}|$ $\times 100$(\si{\percent}) & $|\frac{r(\eta) - r(\eta_0)}{r(\eta_0)}|$ $\times 100$(\si{\percent}) \\
    \hline
    $\eta_0$ & 1794.8 & 5.8 & - & - \\
    $\eta_0 + 10$ & 1956.2 & 6.3 & 8.99 & 8.62  \\
    $\eta_0 - 10$ & 1594.6 & 4.8 & 11.15 & 17.24 \\
    \hline\hline
    \end{tabular}
    \caption{\label{tab:T_recession_compare_eta} Summary of the stagnation point temperature (at $t = \SI{200}{\second}$) and recession rates for various efficiencies for Case 2 ($p_{\mathrm{a}} = \SI{10}{\kilo\pascal}$, $P = \SI{55}{kW}$, $\eta_0 = \SI{60.16}{\percent}$).}}
    \end{table}

\begin{figure}[!htb]
\centering
\subfloat[][]{\includegraphics[scale=0.4]{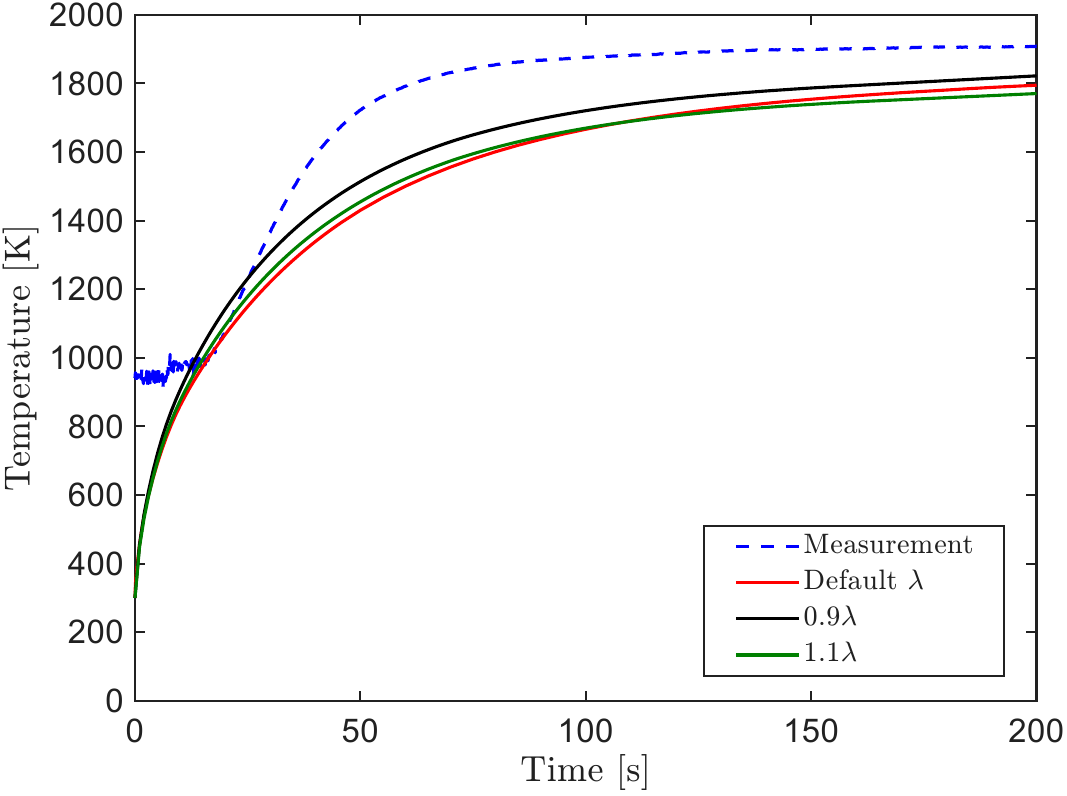}}\hspace{0.5cm}
\subfloat[][]{\includegraphics[scale=0.4]{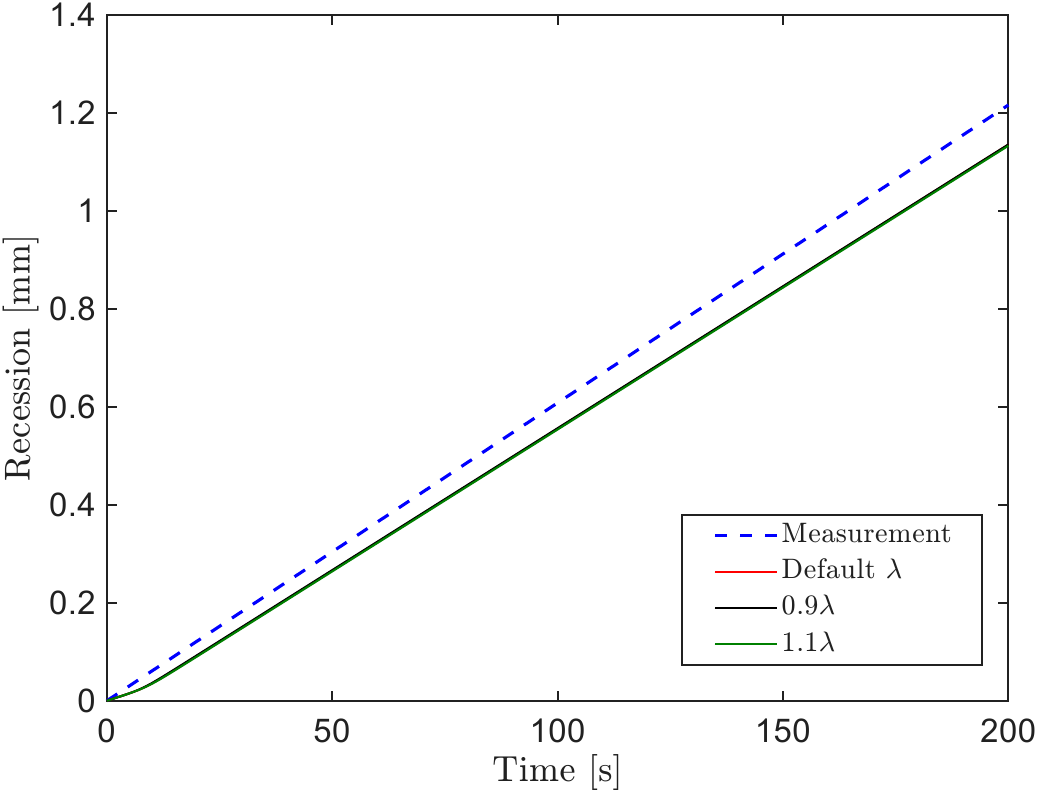}}
\caption{Comparison of the predicted and measured surface stagnation temperature and recession for Case 2 ($p_{\mathrm{a}} = \SI{10}{\kilo\pascal}$, $P = \SI{55}{kW}$, $\eta_0 = \SI{60.16}{\percent}$).}
\label{fig:T_recession_10kPa_cond}
\end{figure}

\begin{table}[hbt!]
    
    \centering
    {\fontsize{9pt}{12pt}\selectfont
    \begin{tabular}{lcccccc}
    \hline\hline
    Conductivity  & T [Predicted] & Recession rate [Predicted] & Relative error  & Relative error\\
    (W/m-K) & (\si{\kelvin}) & (\si{\micro\meter}/s) &  $|\frac{T(\lambda) - T(\lambda_0)}{T(\lambda_0)}|$ $\times 100$(\si{\percent}) & $|\frac{r(\lambda) - r(\lambda_0)}{r(\lambda_0)}|$ $\times 100$(\si{\percent}) \\
    \hline
    Default $\lambda$ & 1794.8 & 5.8 & - & - \\
    $0.9\lambda$ & 1822.1  & 5.82 & 1.52  & 0.34  \\
    $1.1\lambda$ & 1770.5 & 5.77 & 1.35 & 0.51 \\
    \hline\hline
    \end{tabular}
    \caption{\label{tab:T_recession_compare_cond} Summary of the stagnation point temperature (at $t = \SI{200}{\second}$) and recession rates for various conductivities for Case 2 ($p_{\mathrm{a}} = \SI{10}{\kilo\pascal}$, $P = \SI{55}{kW}$, $\eta_0 = \SI{60.16}{\percent}$). $\lambda_0$ refers to the default temperature-dependent conductivity.}
    }
    \end{table}

Discrepancies may also arise from the use of thermophysical property data corresponding to a closely related graphite grade rather than the exact material tested experimentally. Previous studies on POCO Graphite \cite{zibitsker2023finite} indicate that, while the specific heat capacity remains nearly identical across different graphite grades, the thermal conductivity exhibits significant variation. This observation is consistent with the fundamental nature of these properties: specific heat is largely invariant across graphite grades because it is primarily governed by the intrinsic atomic bonding and lattice structure of carbon. In contrast, thermal conductivity is highly sensitive to microstructural features and therefore varies with parameters such as grain size, porosity, etc. In particular, variations on the order of approximately 10\% in temperature-dependent thermal conductivity have been reported among different grades \cite{sheppard2002graphite}. Although graphites produced by Mersen and POCO originate from different manufacturers, the underlying trends in property variation are expected to be comparable. Consequently, different Mersen grades (e.g., 2160, 2340) are also likely to exhibit similar ranges of variation in thermal conductivity. To assess the influence of uncertainty in thermal conductivity on the predicted material response, additional simulations were performed for Case 2 (10 kPa), wherein the temperature-dependent conductivity was varied by $\pm 10\%$ about the baseline values. \cref{fig:T_recession_10kPa_cond} shows the surface stagnation temperature and recession obtained for various conductivity values. It can be seen that varying the conductivity even by 10\% doesn$^\prime$t change the surface stagnation temperature and recession history significantly. \cref{tab:T_recession_compare_cond} summarizes the extent to which uncertainties in thermal conductivity propagate into the predicted material response for Case 2. A 10\% increase in conductivity leads to a 1.35\% reduction in the surface stagnation temperature (at t = \SI{200}{s}) and a 0.51\% reduction in the recession rate. Conversely, a 10\% decrease in conductivity results in a 1.52\% increase in surface stagnation temperature and 0.34\% increase in the surface stagnation recession rate.

Addressing the above sources of uncertainty through refined facility characterization, finite-rate ablation modeling, and material-specific property calibration constitutes a natural extension of the present work.

\section{Conclusions}\label{sec:conclusions}
This work presented a loosely coupled, multiphysics framework for simulating ICP wind tunnels and predicting the thermo-chemical response of TPS materials under high-enthalpy conditions. By integrating high-fidelity solvers for the plasma, the electromagnetic field, and the material response within a partitioned coupling strategy, the framework enables end-to-end simulation of ICP wind tunnels, from the RF coils to the stagnation region of the material sample. The coupled model successfully captures key magneto-hydrodynamic features of ICP discharges, including vortex-mode recirculation, Joule-heating-driven plasma formation, and Lorentz-force-induced flow confinement. Predictions of plasma jet behavior at low pressures further demonstrate the framework’s ability to reproduce the experimentally observed transition from subsonic to supersonic flow. The framework was further validated against cold-wall heat flux measurements and graphite ablation experiments conducted at the Plasmatron X facility at the University of Illinois Urbana-Champaign. Across the three operating conditions, simulated stagnation-point heat fluxes fell within the experimental uncertainty bounds, while the flow-material coupled ablation simulations reproduced the measured temperature rise and steady-state recession rates with errors below \SI{12}{\percent} and \SI{10}{\percent}, respectively. These results confirm the robustness of the coupling strategy and its ability to capture the transient thermal response as well as the long-duration surface recession behavior of ablative materials.

The overall level of agreement obtained in this study is highly encouraging and underscores the effectiveness and robustness of the proposed approach. The modest discrepancies observed primarily reflect secondary modeling uncertainties, such as facility power-coupling characterization, simplified surface chemistry assumptions, and minor differences in available material property data. Sensitivity analyses indicate that these uncertainties can have a measurable impact on the predicted response. In particular, a $\pm10\%$ variation in system efficiency results in changes of up to 11\% and 17\% in the steady-state surface stagnation temperature and recession rate, respectively. In contrast, a similar $\pm10\%$ variation in material thermal conductivity leads to comparatively minor changes, on the order of 1.5\% in temperature and 0.5\% in recession rate.

More importantly, the framework introduced in this work establishes a fundamentally new capability for \textit{ab initio}, end-to-end simulation of high-enthalpy inductively coupled plasma wind tunnels using only independently measured facility inputs, without empirical tuning or calibration. This represents a significant advance over existing approaches and provides a rigorous, physics-based foundation for high-fidelity and predictive simulations of ICP facilities. As such, the framework enables more reliable interpretation of experimental data and offers a powerful tool for the design, analysis, and optimization of hypersonic material testing campaigns, with clear pathways for future refinements that build upon, rather than redefine, the core methodology. The current work also establishes the foundation for future developments toward tightly coupled simulations incorporating finite-rate ablation chemistry, which has been recently shown to offer significant improvements in prediction accuracy over the uncoupled fixed boundary-condition method.

Moreover, the present framework is inherently general and can be extended to predict the material response of test samples across a wide range of high-enthalpy facilities such as the arc-jets. Provided that the requisite input data for accurately modeling the plasma are available, the coupling between the plasma field and the material solver remains largely independent of the specific facility configuration.

\section*{Acknowledgments}
         This work is funded by the Vannevar Bush Faculty Fellowship OUSD(RE) Grant No: N00014-21-1-295 with M. Panesi as the Principal Investigator. The work is also supported by the Center for Hypersonics and Entry Systems Studies
(CHESS) at UIUC. Computations were performed on Frontera, an HPC resource provided by the Texas Advanced
Computing Center (TACC) at The University of Texas at Austin, on allocation CTS20006, with D. Bodony as the Principal Investigator.
    
\section*{Competing interest}
The authors declare no competing interests.

\section*{Declaration of generative AI and AI-assisted technologies in the manuscript preparation process}
During the preparation of this work, the author(s) used ChatGPT 5.3 in order to check grammatical errors and rephrase certain paragraphs to improve readability and quality of writing. After using this tool/service, the author(s) reviewed and edited the content as needed and take(s) full responsibility for the content of the published article.

\appendix
    \section{Grid convergence study for Plasmatron X torch simulations}\label{appendix:grid_convergence_torch}
\begin{figure}[hbt!]
\centering
\includegraphics[trim={0 5cm 0 5cm},clip,scale=0.45]{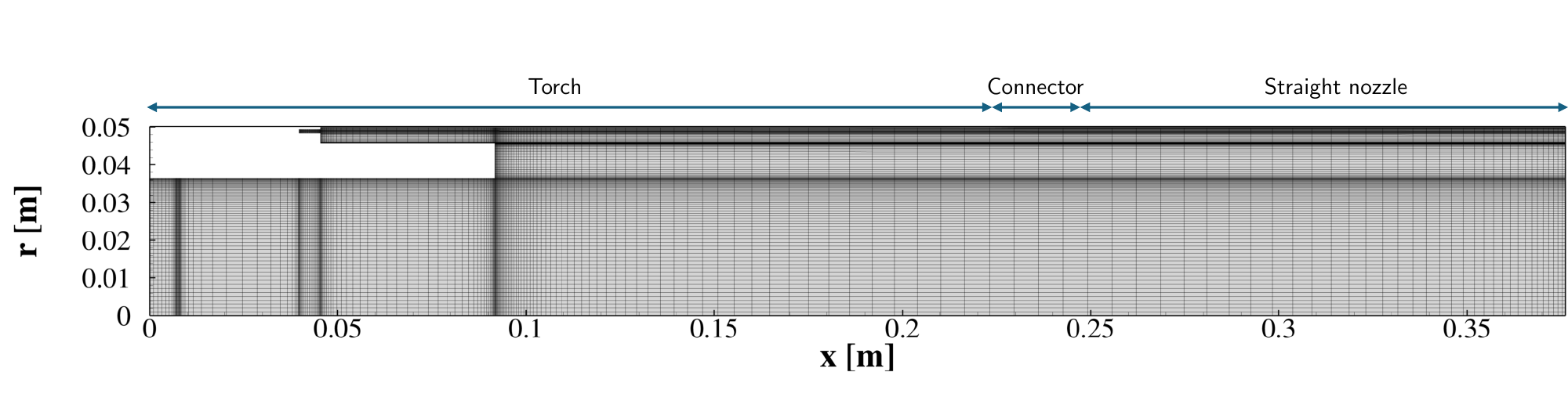}
\caption{Multi-block torch mesh (with nozzle) for the plasma solver.}
\label{fig:pX_torch_mesh}
\end{figure}

\begin{figure}[hbt!]
\centering
\subfloat{\includegraphics[scale=0.35]{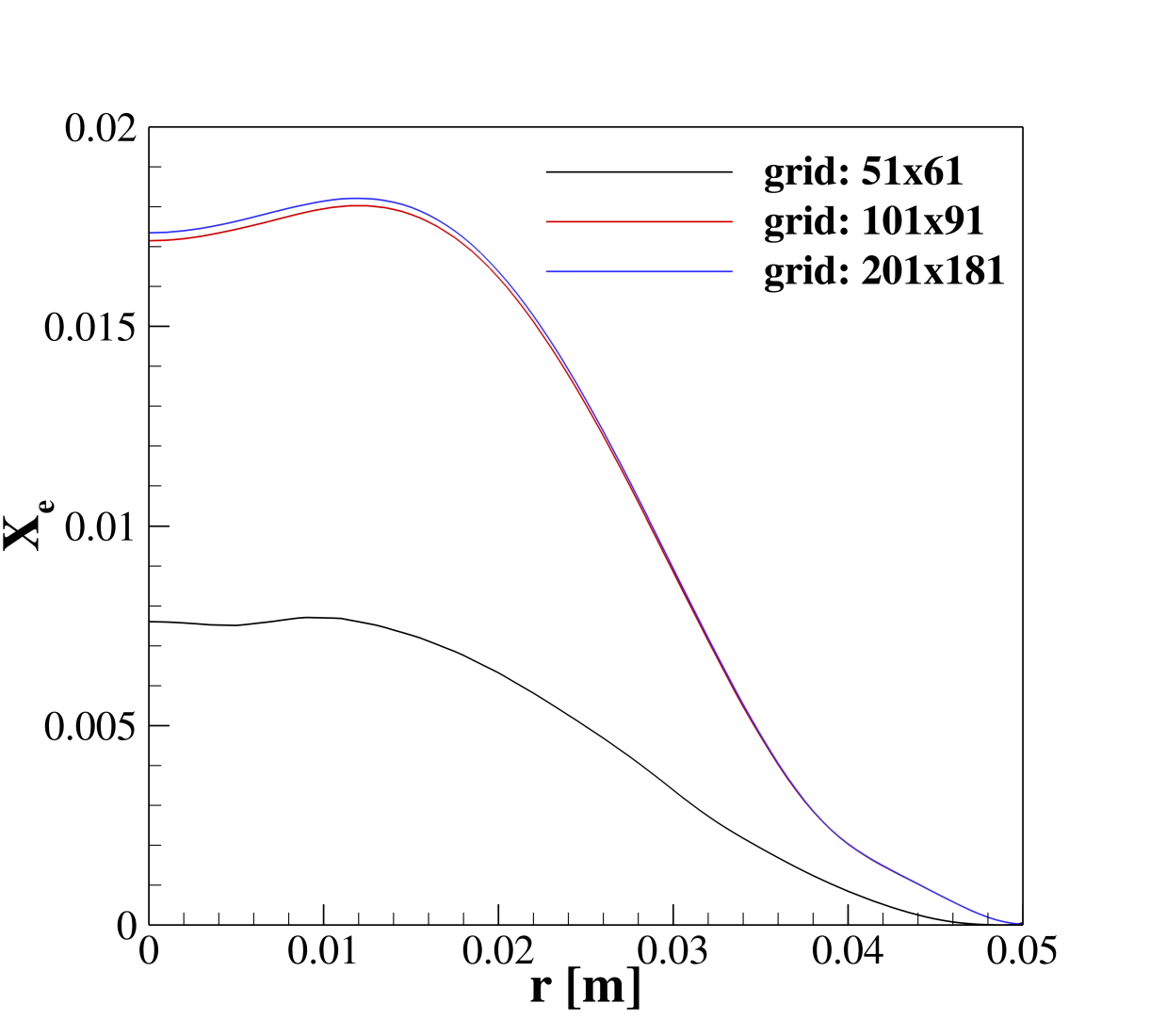}} 
\subfloat{\includegraphics[scale=0.35]{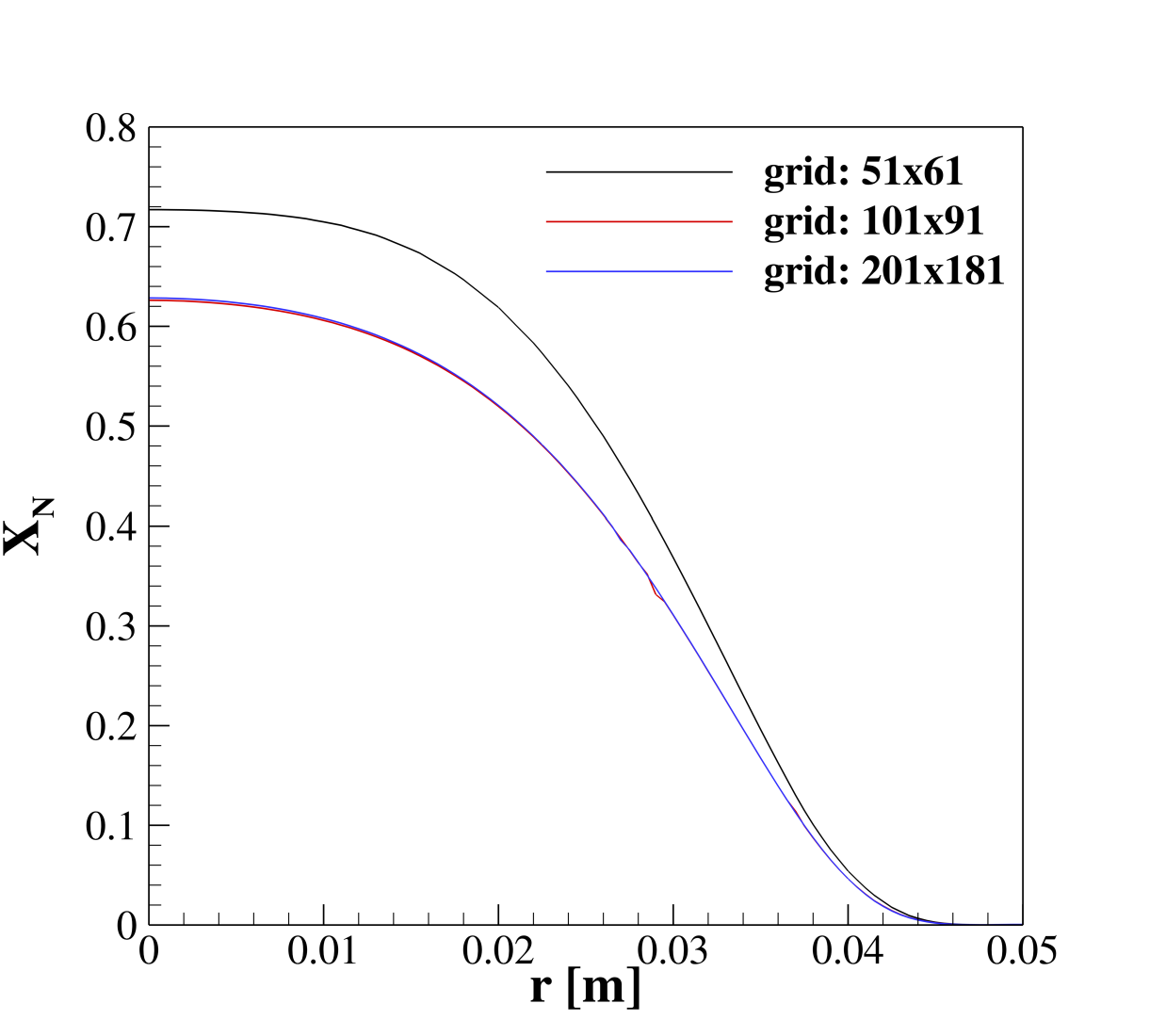}}
\\
(a) \hspace{2.6in} (b)
\subfloat{\includegraphics[scale=0.35]{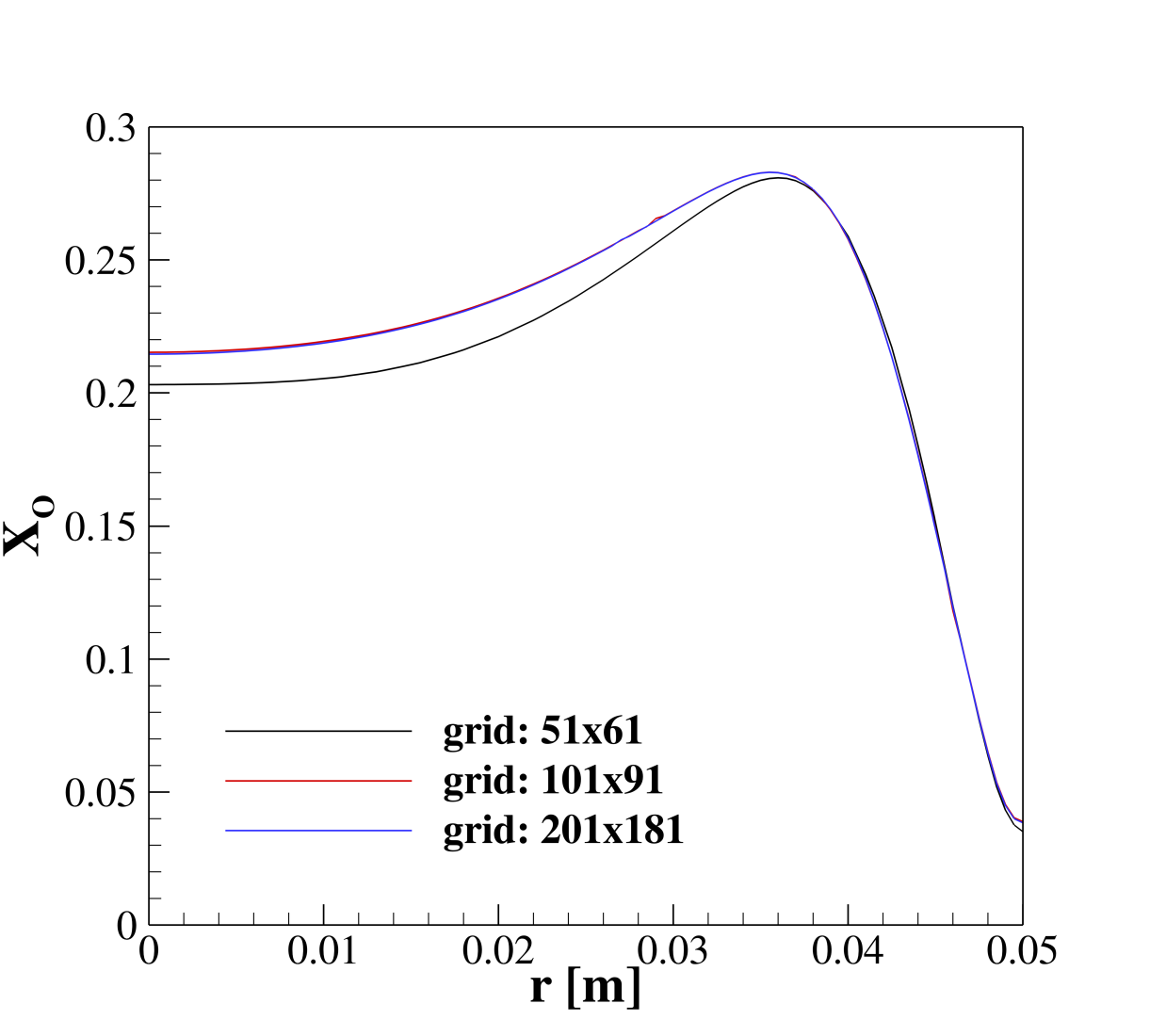}} 
\subfloat{\includegraphics[scale=0.35]{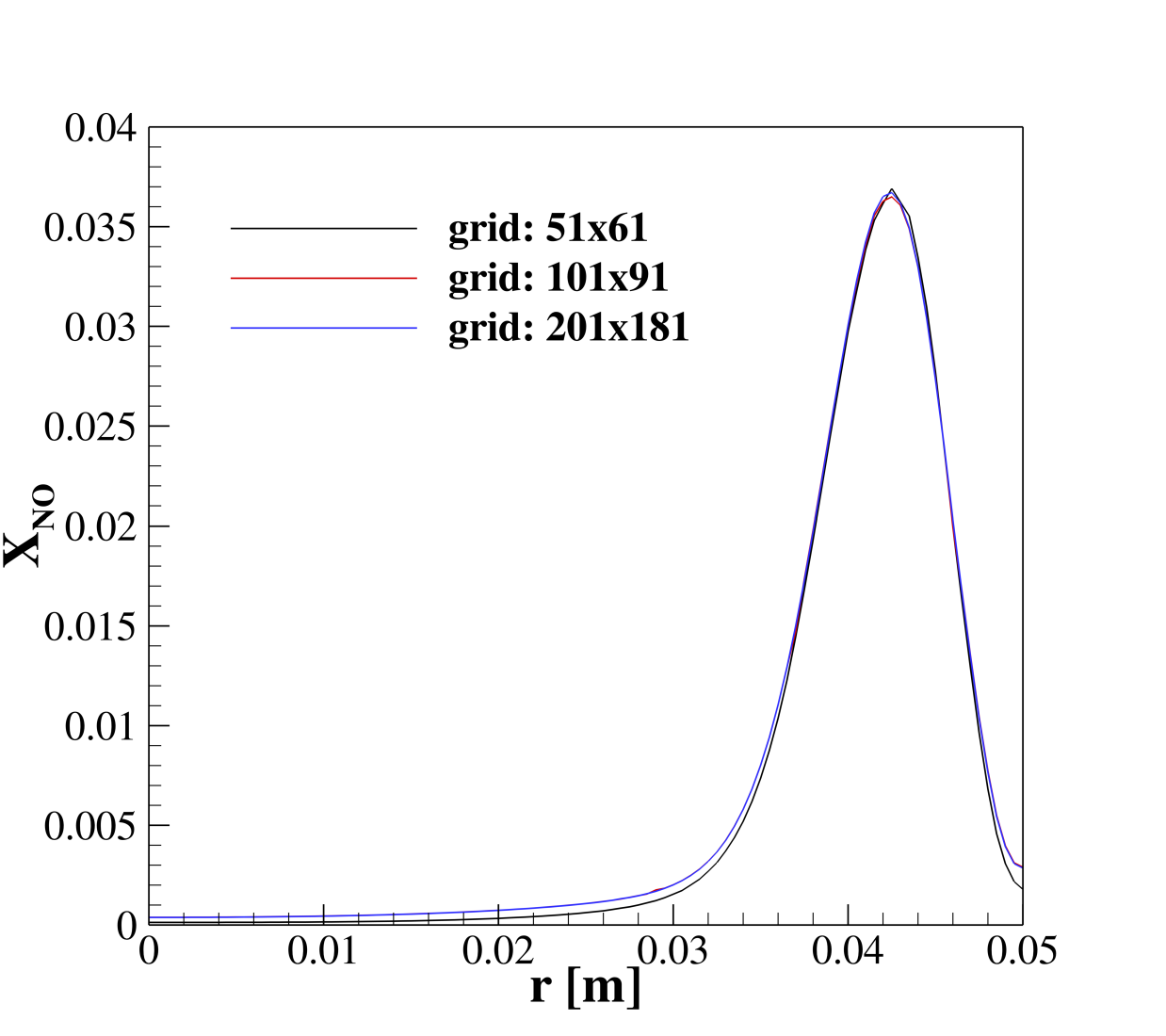}}
\\
(c) \hspace{2.6in} (d)

\caption{Profiles of mole fractions at the torch outlet. (a) Free-electrons, (b) N, (c) O, and (d) NO. ($p_{\mathrm{a}} = \SI{1000}{\pascal}$, $P = \SI{200}{\kilo\watt}$, $\eta = 50\%$). } 
\label{fig:torch_grid_refinement_conc}
\end{figure}

\begin{figure}[hbt!]
\centering
\subfloat{\includegraphics[scale=0.35]{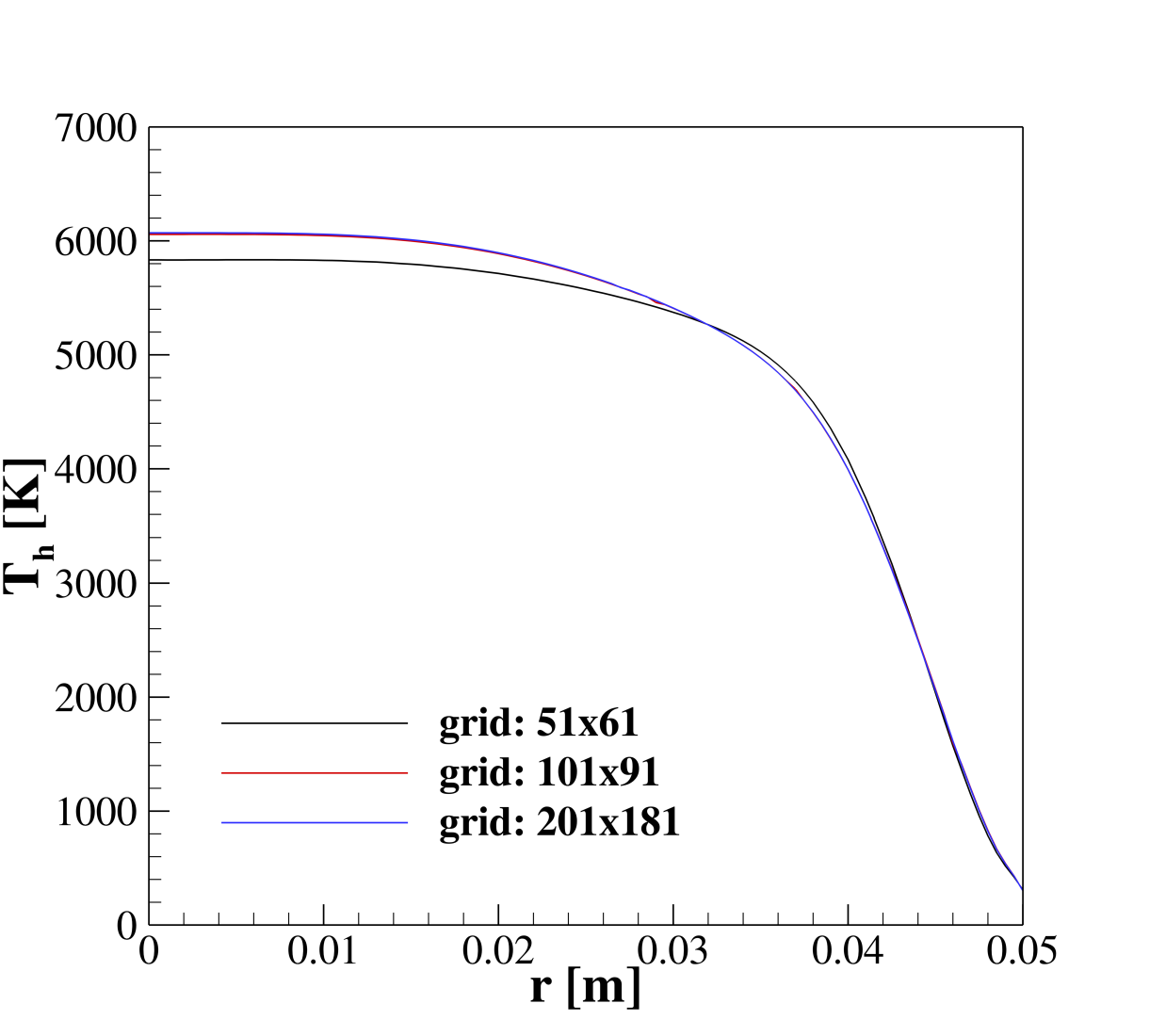}} 
\subfloat{\includegraphics[scale=0.35]{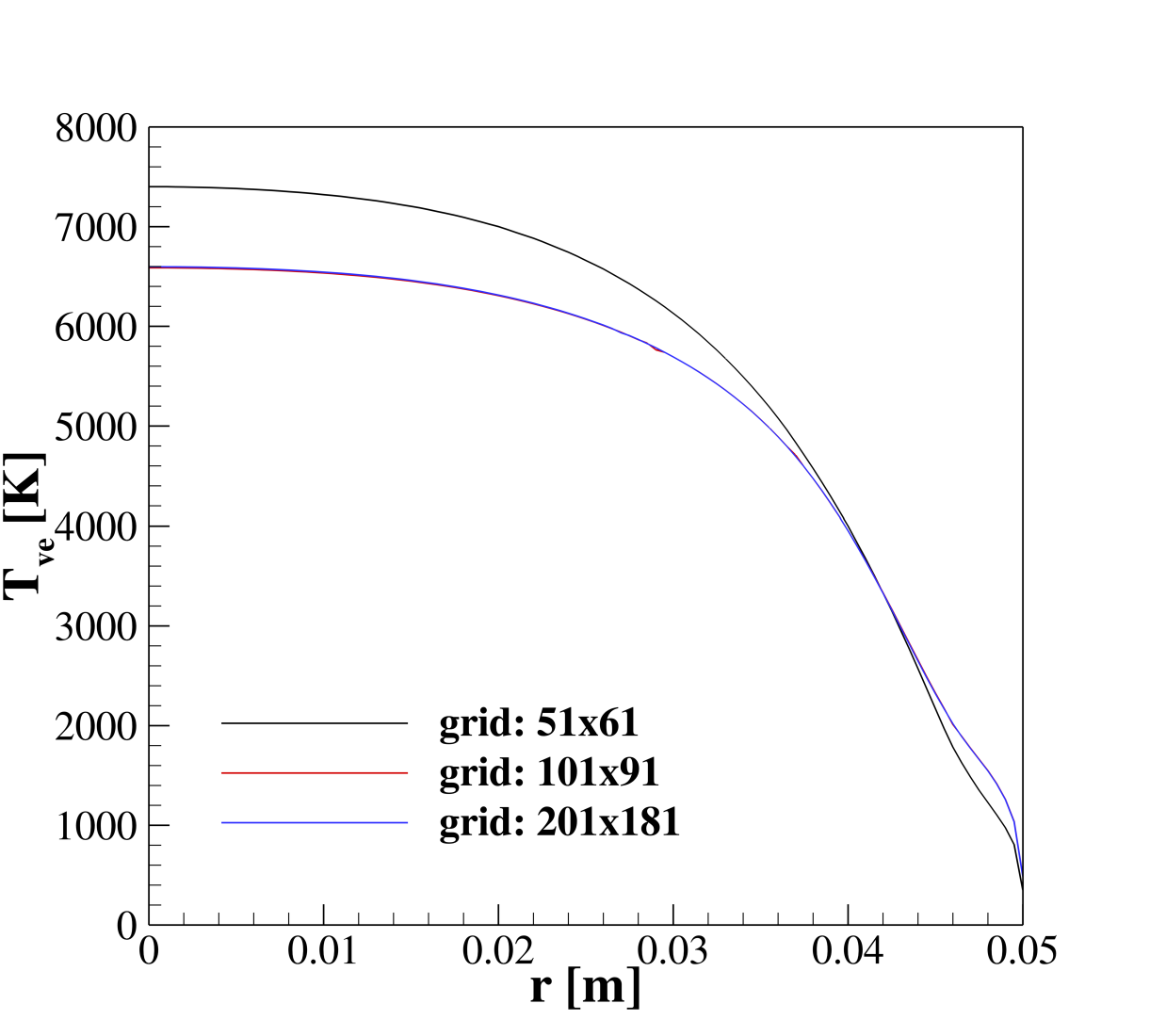}}
\\
(a) \hspace{2.6in} (b)
\subfloat{\includegraphics[scale=0.35]{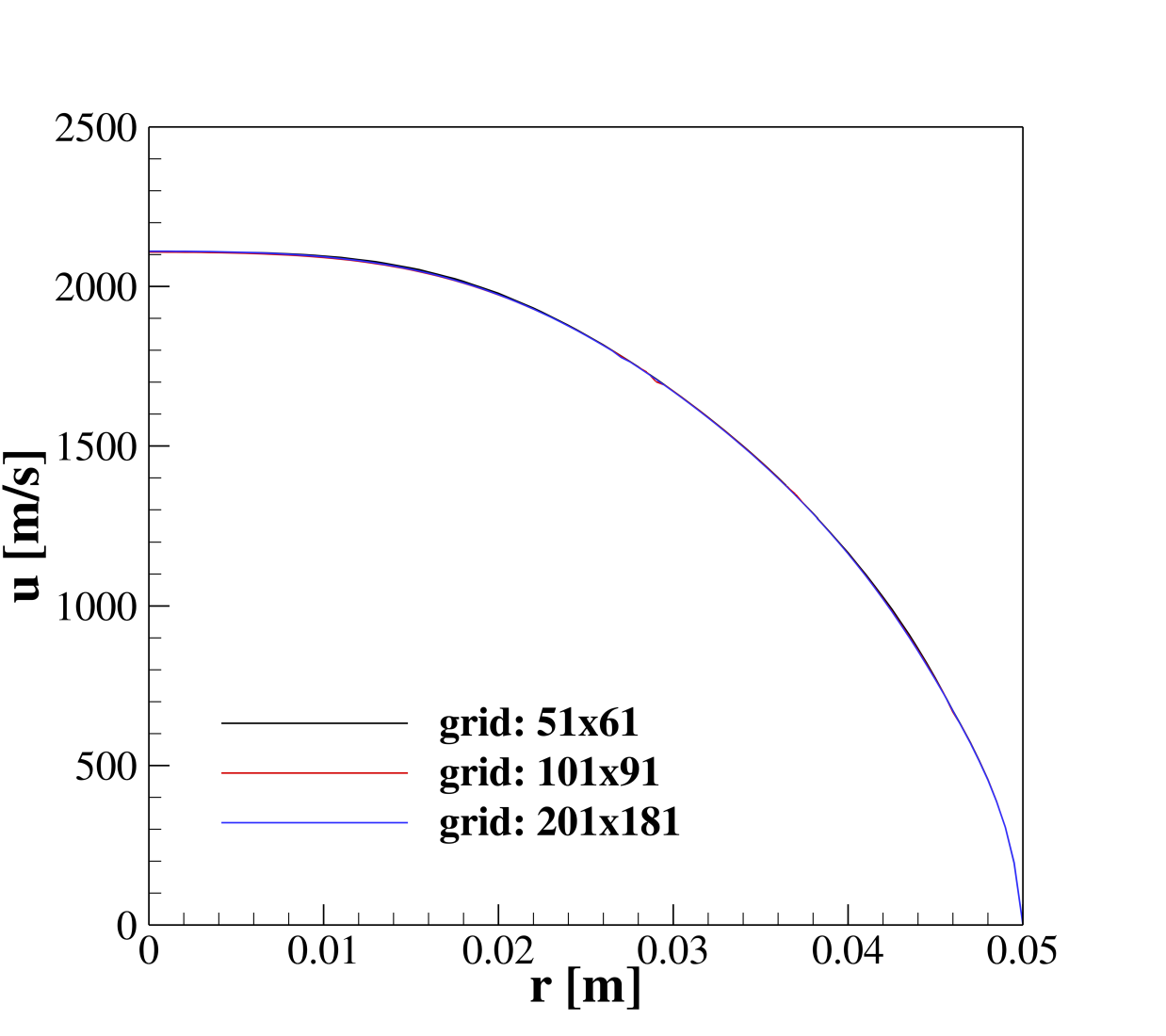}} 
\subfloat{\includegraphics[scale=0.35]{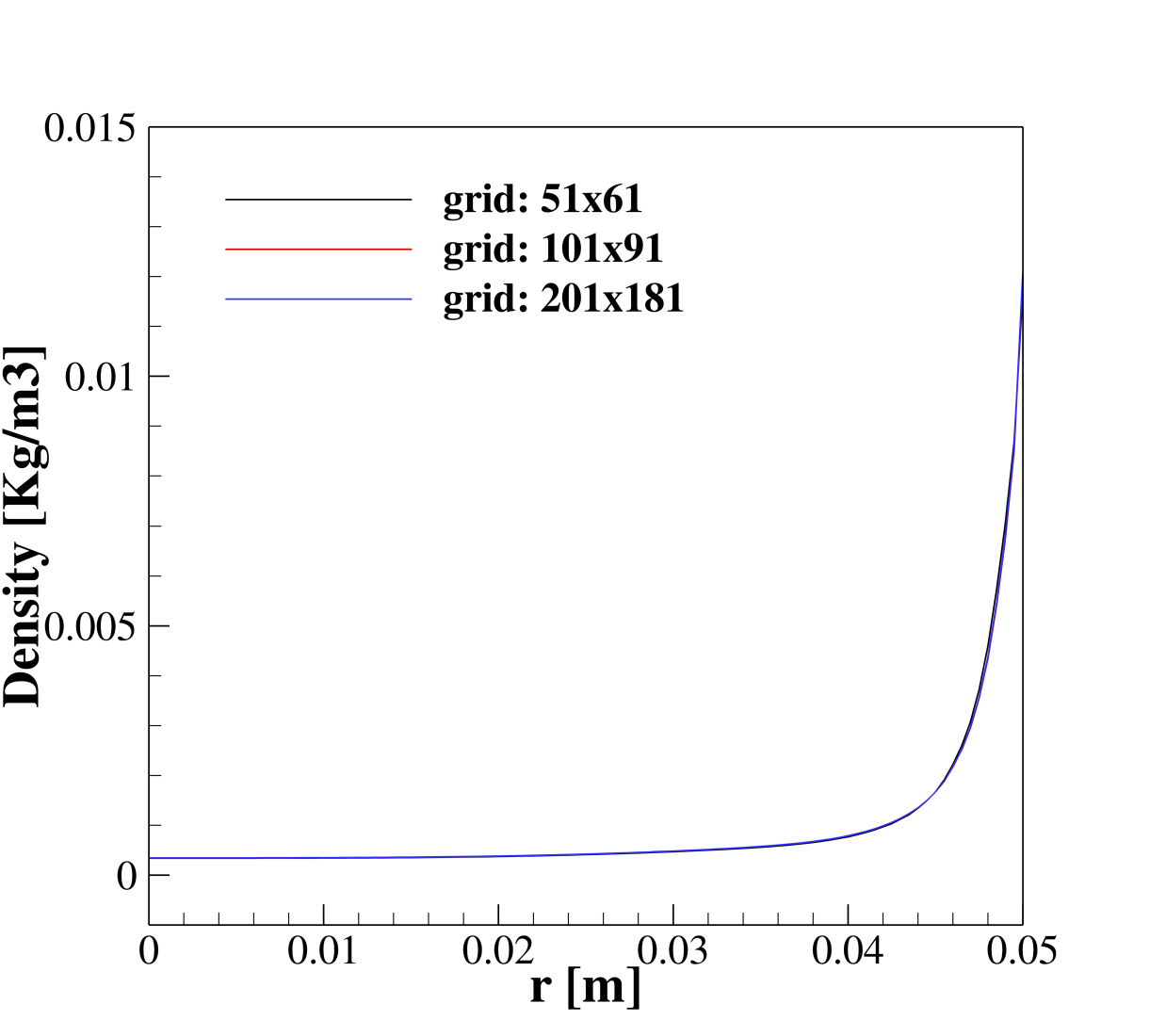}}
\\
(c) \hspace{2.6in} (d)

\caption{Profiles of various flow quantities at the torch outlet. (a) Heavy-species temperature, (b) vibronic temperature, (c) axial velocity, and (d) density. ($p_{\mathrm{a}} = \SI{1000}{\pascal}$, $P = \SI{200}{\kilo\watt}$, $\eta = 50\%$). } 
\label{fig:torch_grid_refinement_flow}

\end{figure}

A 2D axisymmetric simulation was conducted using the 2T NLTE model for the following operating conditions: central gas mass flow \SI{0.86}{g/s}, sheath gas mass flow \SI{7.13}{g/s}, pressure \SI{1}{kPa}, power \SI{200}{kW}, and efficiency $50\%$. The working gas is air modeled using an 11-species mixture with the reaction rates taken from \cite{park2001chemical}. The block-structured mesh for the torch consists of 4 blocks as shown in \cref{fig:pX_torch_mesh}. Since the plasma core is located in the biggest block (located in the induction zone), starting from x = \SI{0.0916}{m}, the grid refinement study is presented focusing on that block. The grid for the other three smaller blocks is modified accordingly to satisfy the requirements for having a point-matched multi-block mesh. Three meshes are used for the grid refinement study, consisting of the following number of nodes for the largest block: 51x61, 101x91, and 201x181. Steady-state simulations are conducted for all the meshes. \cref{fig:torch_grid_refinement_conc,fig:torch_grid_refinement_flow} show the comparison of various quantities at the torch outlet obtained from the three meshes. The results show that 101x91 mesh gives an excellent agreement with the solution obtained using the 201x181 grid. Hence, the 101x91 mesh for the torch region is used for all the simulations presented in this work.

\section{Grid convergence study for torch-chamber-sample simulations}\label{appendix:grid_convergence_with_sample}
\begin{figure}[!htb]
        \centering
        \includegraphics[scale=0.5]{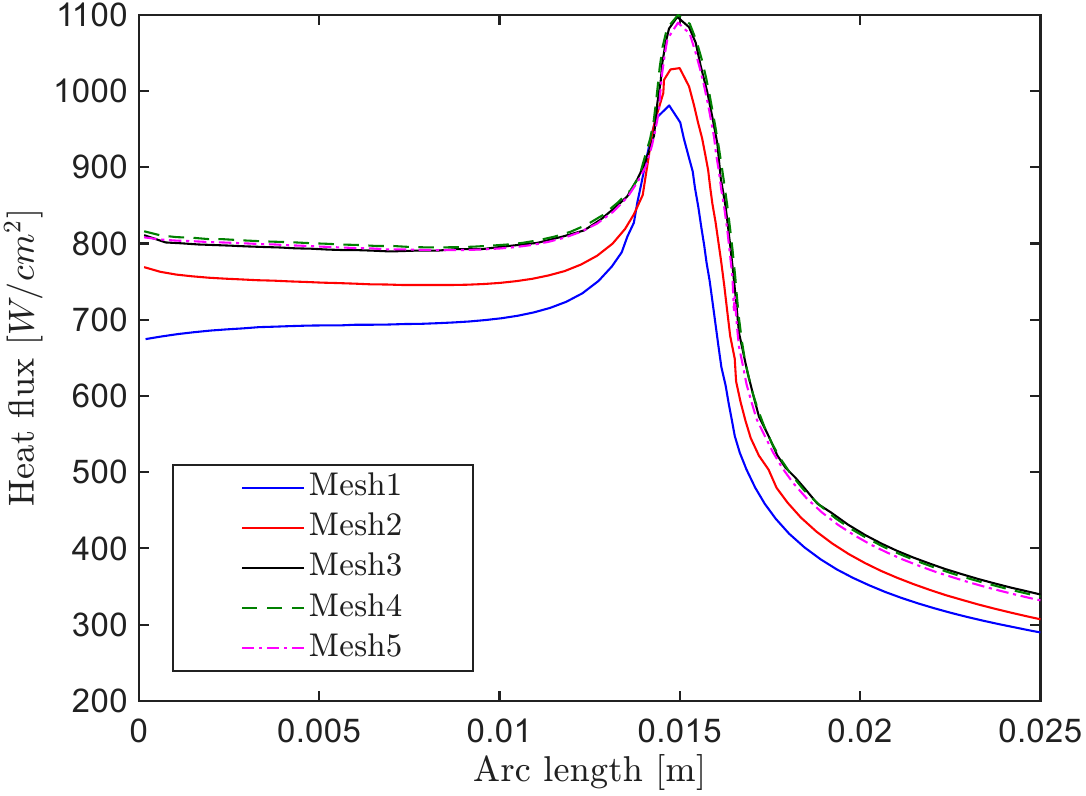}
        \caption{Heat flux profiles along the sample surface for various grids. ($p_{\mathrm{a}} = \SI{10}{\kilo\pascal}$, $P = \SI{300}{\kilo\watt}$, $\eta = 55\%$).} 
        \label{fig:grid_test}
    \end{figure}

Grid convergence study for the Plasmatron X torch-chamber-sample (with \SI{30}{mm} isoQ sample) simulations was performed using 2D axisymmetric simulations with the LTE model, under the following operating conditions: central gas mass flow \SI{0.86}{g/s}, sheath gas mass flow \SI{7.13}{g/s}, pressure \SI{10}{kPa}, power \SI{300}{kW}, and efficiency $55\%$. The mesh in the torch region was kept fixed based on the grid convergence study for the torch-only simulations as reported in \ref{appendix:grid_convergence_torch}, and only the mesh in the chamber region was changed. Five different meshes with successive refinements near the sample as well as in the chamber region were used, as tabulated in \cref{tab:q_mesh}. The first cell thickness given in the table denotes the thickness of the first cell as measured from the sample wall. \cref{fig:grid_test} compares the heat flux profiles along the sample surface obtained with various meshes. The profiles for meshes 3, 4, and 5 collapse together, indicating that mesh 3 is the coarsest mesh, which gives a grid-converged solution. Hence, mesh 3 is chosen for all the simulations presented in this work for the full facility CFD simulations.
    
    \begin{table}[!htb]
    \centering
    \begin{tabular}{lcc}
    \hline\hline
    Mesh Id & First cell thickness [m] & Number of cells \\
    \hline
    1 & 3.2e-6 & 10905 \\
    2 & 1.6e-6 & 20638 \\
    3 & 8e-7 & 58788 \\
    4 & 4e-7 & 128608 \\
    5 & 2e-7 & 285625 \\
    \hline\hline
    \end{tabular}
    \caption{\label{tab:q_mesh} \centering List of meshes used for the grid convergence study for the full facility CFD simulation.}
    \end{table}

\section{Grid convergence study for the material simulation}\label{appendix:grid_convergence_material}
\begin{figure}[!htb]
        \centering
        \includegraphics[trim = 0in 1.5in 0in 1in, clip,scale=0.45]{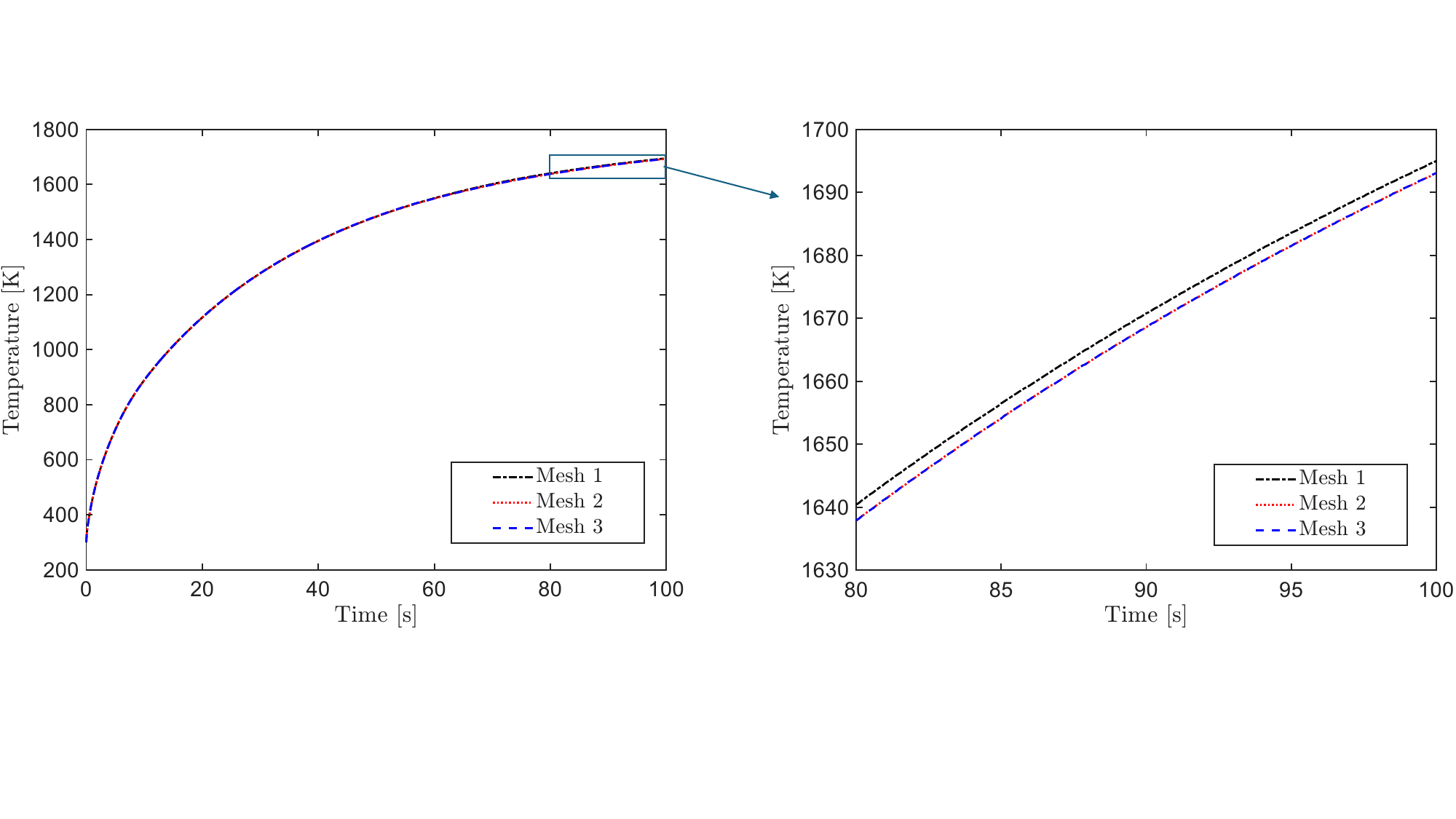}
        \caption{Surface stagnation temperature history for Case 2 ($p_{\mathrm{a}} = \SI{10}{\kilo\pascal}$, $P = \SI{55}{\kilo\watt}$, $\eta = \SI{60.16}{\percent}$).} 
        \label{fig:T_stag_chyps_mesh}
    \end{figure}

\begin{figure}[!htb]
        \centering
        \includegraphics[trim = 0in 0in 0in 0in, clip,scale=0.5]{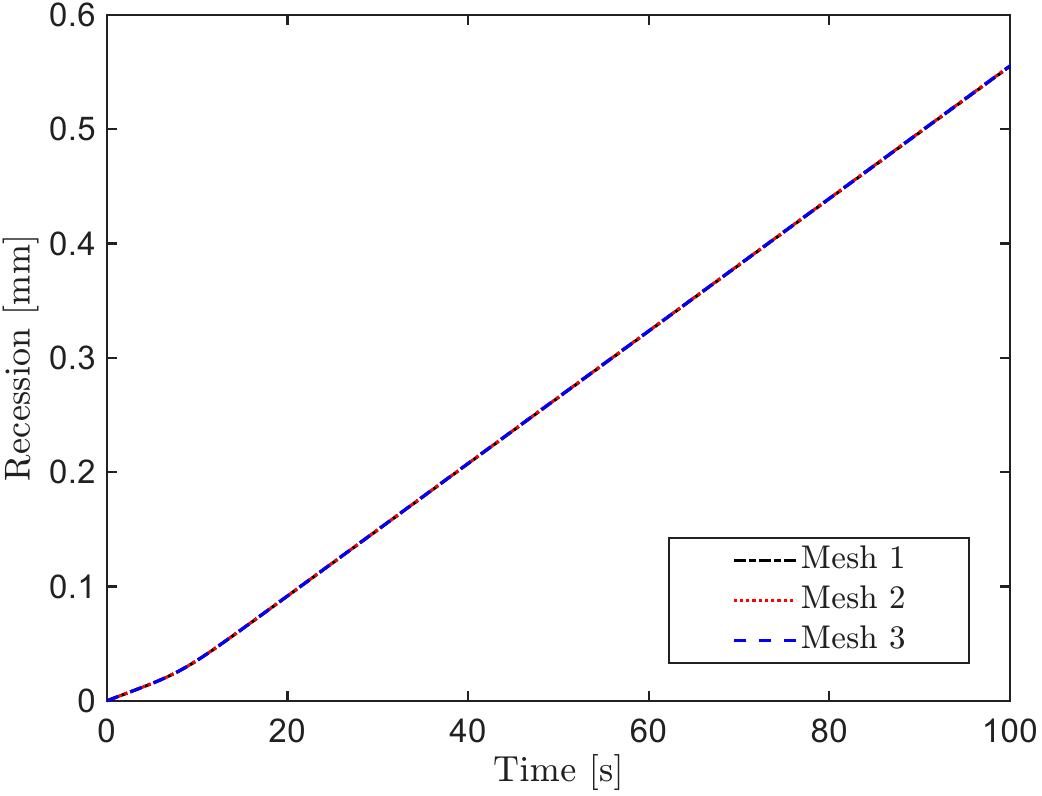}
        \caption{Surface recession history for Case 2 ($p_{\mathrm{a}} = \SI{10}{\kilo\pascal}$, $P = \SI{55}{\kilo\watt}$, $\eta = \SI{60.16}{\percent}$).} 
        \label{fig:disp_stag_chyps_mesh}
    \end{figure}

A grid convergence study for the material simulation was conducted by running coupled ablation simulations for Case 2 (\SI{10}{kPa}, \SI{55}{kW}, 60.16\%). The grid for the plasma solver was kept fixed based on the grid convergence study presented in \ref{appendix:grid_convergence_with_sample}. Three meshes for the material solver were chosen as tabulated in \cref{tab:chyps_mesh}. \cref{fig:T_stag_chyps_mesh,fig:disp_stag_chyps_mesh} show the surface stagnation temperature and recession obtained using various meshes for the material solver. While all three meshes give identical results, Mesh 2 was chosen for the simulations presented in this work for improved accuracy and consistency for the surface coupling with the CFD solver$^{\prime}$s mesh.

 \begin{table}[!htb]
    \centering
    \begin{tabular}{lcc}
    \hline\hline
    Mesh Id & First element thickness [m] & Number of elements \\
    \hline
    1 & 5e-4 & 8396 \\
    2 & 1e-4 & 33584 \\
    3 & 5e-5 & 134336 \\

    \hline\hline
    \end{tabular}
    \caption{\label{tab:chyps_mesh} \centering List of meshes used for the grid convergence study for material simulation.}
    \end{table}

    \section{Coupling frequency convergence test}\label{appendix:coupling_convergence}
    To assess that the steady-state temperature and recession rate are converged with respect to the coupling window schedule, a series of coupled ablation simulations were conducted for different coupling windows for $t \ge$ \SI{50}{s}. It is to be noted that the coupling windows selected for times $t \le$ \SI{50}{s} (\emph{i.e.,} \SI{0.1}{s} window till \SI{5}{s}, and \SI{1}{s} window till \SI{50}{s}) are already very small to capture the large gradients during initial transient. Hence, a further reduction of the coupling windows for $t \le$ \SI{50}{s} would be unnecessary and make the simulations impractical. For $t \ge$ \SI{50}{s}, three coupling windows were used: \SI{10}{s}, \SI{5}{s}, and \SI{2}{s}. Simulations were conducted for Case 1 at \SI{20}{kPa}.
    
\begin{figure}[!htb]
        \centering
        \includegraphics[trim = 0in 1.5in 0in 1in, clip,scale=0.45]{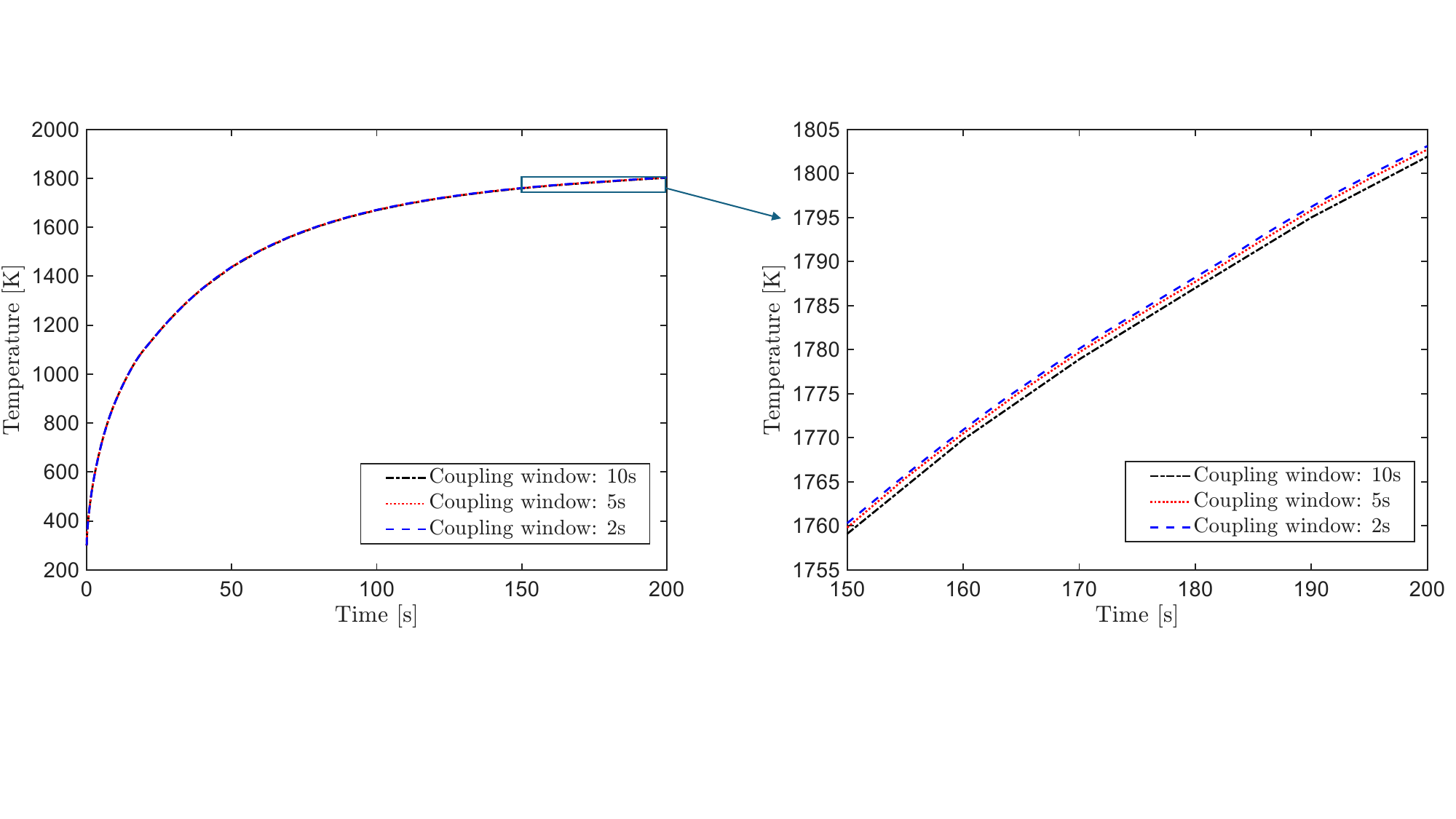}
        \caption{Surface stagnation temperature history for Case 1 ($p_{\mathrm{a}} = \SI{20}{\kilo\pascal}$, $P = \SI{55}{\kilo\watt}$, $\eta = \SI{63.75}{\percent}$).} 
        \label{fig:T_stag_coupling_window}
    \end{figure}

\begin{figure}[!htb]
        \centering
        \includegraphics[trim = 0in 0in 0in 0in, clip,scale=0.5]{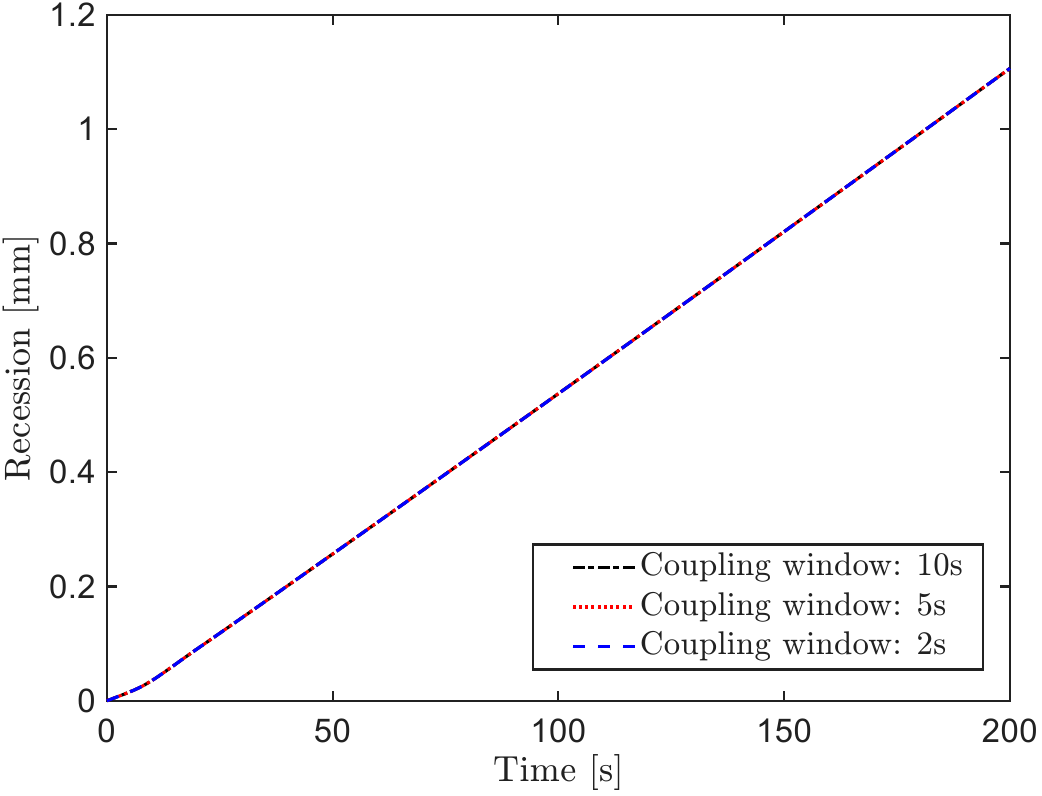}
        \caption{Surface recession history for Case 1 ($p_{\mathrm{a}} = \SI{20}{\kilo\pascal}$, $P = \SI{55}{\kilo\watt}$, $\eta = \SI{63.75}{\percent}$).} 
        \label{fig:disp_stag_coupling_window}
    \end{figure}

    \cref{fig:T_stag_coupling_window} demonstrates that the surface stagnation temperature histories obtained using different coupling window sizes exhibit near-complete overlap, indicating that even a 10 s coupling window is sufficient to achieve a solution that is effectively converged with respect to this parameter. A closer examination of the temperature profiles, however, reveals a slight deviation: the \SI{10}{s} coupling window underpredicts the temperature by approximately \SI{2}{K} at t = \SI{200}{s}, corresponding to a difference of about 0.11\% relative to the solution obtained with a \SI{2}{s} coupling window. In contrast, the temperature profile corresponding to a \SI{5}{s} coupling window remains virtually indistinguishable from that of the \SI{2}{s} case. \cref{fig:disp_stag_coupling_window} shows that the surface recession histories at the stagnation point are essentially identical for all three coupling window sizes. Based on these results, a coupling window of \SI{5}{s} is deemed sufficient and is therefore adopted for all simulations presented in this work.

\section*{References}
\bibliographystyle{aiaa}
\bibliography{bibliography}

\end{document}